\PassOptionsToPackage{x11names}{xcolor}
\documentclass[submission, Phys]{SciPost}
\usepackage{orcidlink}
\usepackage{slashed}
\hypersetup{
    colorlinks,
    linkcolor={red!50!black},
    citecolor={blue!50!black},
    urlcolor={blue!80!black}
}

\usepackage[bitstream-charter]{mathdesign}
\DeclareSymbolFont{usualmathcal}{OMS}{cmsy}{m}{n}
\DeclareSymbolFontAlphabet{\mathcal}{usualmathcal}

\definecolor{blue}{rgb}{0,0.396,0.741}
\definecolor{MathBlue}{rgb}{0.368417, 0.506779, 0.709798}
\definecolor{MathYellow}{rgb}{0.880722, 0.611041, 0.142051}
\definecolor{MathGreen}{rgb}{0.560181, 0.691569, 0.194885}
\definecolor{MathRed}{rgb}{0.922526, 0.385626, 0.209179}
\definecolor{MathViolet}{rgb}{0.528488, 0.470624, 0.701351}

\newcommand{\tr}[1]{\mathrm{tr}\left(#1\right)}

\newcommand{\eq}[1]{\eqref{eq:#1}}
\newcommand{\fig}[1]{Fig.~\ref{fig:#1}}

\newcommand{\Sec}[1]{Sec.~\ref{sec:#1}}

\newcommand{\ep}{\epsilon}
\newcommand{\nn}{\nonumber}

\newcommand{\lk}{\Big[}
\newcommand{\rk}{\Big]}

\usepackage{axodraw2}

\newcommand{\sbx}{\scalebox{0.525}}
\newcommand{\defDiag}[2]{\expandafter\newcommand%
  \csname diag-#1\endcsname{#2}}
\newcommand{\diag}[1]{\csname diag-#1\endcsname}

\newcommand{\TLfig}[1]{{\begin{array}{c}\diag{#1}\\[2ex]\text{\small #1}\end{array}}}

\newcommand{\picj}[1]{\;\parbox[c]{40pt}{\begin{picture}(40,40)(0,0)
\SetWidth{1.0}\SetScale{1.0} #1 \end{picture}}\;}
\newcommand{\picjj}[1]{\parbox[c]{0pt}{\begin{picture}(0,40)(43,0)
\SetWidth{1.0}\SetScale{1.0} #1 \end{picture}}}

\newcommand{\picbj}[1]{\;\parbox[c]{60pt}{\begin{picture}(60,40)(0,0)
\SetWidth{1.0}\SetScale{1.0} #1 \end{picture}}\;}
\newcommand{\picbjj}[1]{\parbox[c]{0pt}{\begin{picture}(0,40)(63,0)
\SetWidth{1.0}\SetScale{1.0} #1 \end{picture}}}

\def\Asc(#1,#2)(#3,#4,#5){\CArc(#1,#2)(#3,#4,#5)}
\def\Lsc(#1,#2)(#3,#4){\Line(#1,#2)(#3,#4)}

\defDiag{1}{\sbx{\picj{
\Asc(20,20)(20,-90,270)}}}

\defDiag{7}{\sbx{\picj{
\Asc(20,20)(20,0,180) 
\Asc(20,20)(20,180,360)
\Lsc(40,20)(0,20)}}}

\defDiag{51}{\sbx{\picj{
\Asc(20,20)(20,0,180)
\Asc(20,2)(27,42,138) 
\Asc(20,38)(27,222,318)
\Asc(20,20)(20,-180,0)}}}

\defDiag{62}{\sbx{\!\!\picbj{
\Asc(35,20)(20,256,76) 
\Lsc(40,39.5)(20,39.5) 
\Asc(25,20)(20,104,284)
\Lsc(30,0)(20,39.5) 
\Lsc(40,39.5)(30,0)}\!\!}}

\defDiag{63}{\sbx{\picj{
\Asc(20,20)(20,-30,90)
\Asc(20,20)(20,90,210)
\Asc(20,20)(20,210,330)
\Lsc(3,10)(20,20) 
\Lsc(20,20)(20,40) 
\Lsc(37,10)(20,20)}}}

\defDiag{841}{\sbx{\picj{
\Asc(20,20)(20,0,180)
\Asc(20,30)(22.36,-153.43,-26.57)
\Lsc(0,20)(40,20)
\Asc(20,10)(22.36,26.57,153.43)
\Asc(20,20)(20,-180,0)}}}

\defDiag{993}{\sbx{\!\!\picbj{
\Asc(35,20)(20,256,346)
\Asc(35,20)(20,-14,76) 
\Lsc(40,39.5)(20,39.5) 
\Asc(25,20)(20,104,284)
\Lsc(30,0)(20,39.5) 
\Lsc(40,39.5)(30,0)
\Asc(38,23)(24,136,250)}\!\!}}

\defDiag{952}{\sbx{\picj{
\Asc(20,20)(20,90,210)
\Asc(20,20)(20,210,330) 
\Asc(20,20)(20,-30,90) 
\Lsc(3,10)(20,40)
\Lsc(37,10)(3,10) 
\Lsc(20,40)(37,10)}}}

\defDiag{1016}{\sbx{\picj{
\Asc(20,20)(20,0,90)
\Asc(20,20)(20,-180.,-90)
\Asc(40,0)(20,90,180)
\Asc(0,40)(20,-90,0)
\Asc(0,0)(20,0,90)
\Asc(20,20)(20,-90.,0)
\Asc(20,20)(20,90,180)}}}

\defDiag{1010}{\sbx{\picbj{
\Asc(20,20)(20,90,270)
\Asc(40,20)(20,-90,90) 
\Lsc(40,40)(20,40) 
\Lsc(20,0)(40,0)
\Lsc(20,0)(20,40) 
\Lsc(40,40)(40,0) 
\Lsc(20,40)(40,0)}}}

\defDiag{1009}{\sbx{\picbj{
\Asc(40,20)(20,-120,120)
\Asc(40,20)(20,120,180) 
\Asc(40,20)(20,180,240) 
\Asc(20,20)(20,60,300)
\Asc(20,20)(20,-60,0)
\Asc(20,20)(20,0,60) 
\Lsc(40,20)(20,20)}}}

\defDiag{1020}{\sbx{\!\!\picbj{
\Asc(35,20)(20,256,346)
\Asc(35,20)(20,-14,76) 
\Lsc(40,39.5)(20,39.5) 
\Asc(25,20)(20,104,284)
\Lsc(30,0)(20,39.5) 
\Lsc(40,39.5)(35,20) 
\Lsc(35,20)(30,0)
\Lsc(34,20)(53,15)}\!\!}}

\defDiag{1011}{\sbx{\picj{
\Asc(20,20)(20,90,180)
\Asc(20,20)(20,180,270) 
\Asc(20,20)(20,270,360) 
\Asc(20,20)(20,0,90)
\Lsc(20,20)(20,40) 
\Lsc(20,20)(20,0) 
\Lsc(0,20)(20,20) 
\Lsc(40,20)(20,20)}}}

\defDiag{1022}{\sbx{\picbj{
\Lsc(20,0)(40,0)
\Lsc(20,40)(40,40)
\Asc(20,20)(20,-180,-90)
\Asc(40,20)(20,-90,90)
\Lsc(20,20)(40,20)
\Lsc(20,0)(20,20)
\Lsc(40,20)(40,0)
\Lsc(20,20)(20,40)
\Lsc(40,20)(40,40)
\Asc(20,20)(20,90,180)}}}

\defDiag{511}{\sbx{\picbj{
\Asc(20,20)(20,90,270)
\Asc(40,20)(20,-90,90) 
\Lsc(40,40)(20,40) 
\Lsc(20,0)(40,0) 
\Lsc(20,0)(20,20)
\Lsc(20,20)(40,40) 
\Lsc(20,40)(28,33) 
\Lsc(33,28)(40,20) 
\Lsc(40,20)(40,0)
\Lsc(20,20)(40,20)}}}

\defDiag{841.1.3}{\diag{841}\sbx{\picjj{\Vertex(13,20){2.5} \Vertex(27,20){2.5}}}}
\defDiag{1009.1.2}{\diag{1009}\sbx{\picbjj{\Vertex(37,30){2.5}}}}
\defDiag{1011.1.2}{\diag{1011}\sbx{\picjj{\Vertex(6,34){2.5}}}}
\defDiag{841.1.2.2.2}{\diag{841}\sbx{\picjj{\Vertex(20,0){2.5} \Vertex(20,40){2.5}}}}
\defDiag{1009.3.2}{\diag{1009}\sbx{\picbjj{\Vertex(60,20){2.5}}}}
\defDiag{1009.4.2}{\diag{1009}\sbx{\picbjj{\Vertex(30,20){2.5}}}}
\defDiag{1011.3.2}{\diag{1011}\sbx{\picjj{\Vertex(30,20){2.5}}}}

\begin{document}

\begin{center}{\Large \textbf{
General Four-Loop Beta Function for Scalar-Fermion Theories in Three Dimensions
}}\end{center}

\begin{center}
York Schröder\,\orcidlink{0000-0003-3535-4194}\textsuperscript{1$\dagger$},
Emmanuel Stamou\,\orcidlink{0000-0002-8385-6159}\textsuperscript{2$\ddagger$},
Tom Steudtner\,\orcidlink{0000-0003-1935-0417}\textsuperscript{2$\star$} and
Max Uetrecht\,\orcidlink{0000-0001-8685-2543}\textsuperscript{2$\blacklozenge$}
\end{center}

\begin{center}
{\bf 1} Centro de Ciencias Exactas, Departamento de Ciencias B\'asicas,
Universidad del B\'io-B\'io, Avenida Andr\'es Bello 720, Chill\'an, Chile
\\
{\bf 2} Fakultät Physik, Technische Universität Dortmund, D-44221 Dortmund, Germany
\\
${}^\dagger$ {\small \sf yschroder@ubiobio.cl} \;
${}^\ddagger$ {\small \sf emmanuel.stamou@tu-dortmund.de}
\\
${}^\star$ {\small \sf tom2.steudtner@tu-dortmund.de} \;
${}^\blacklozenge$ {\small \sf max.uetrecht@tu-dortmund.de}

\end{center}

\begin{center}
\today
\end{center}

\section*{Abstract}
{\bf
We present general four-loop template $\beta$-functions and anomalous field
dimensions for renormalisable scalar-fermion theories in three dimensions. By
imposing $\mathcal{N}=1$ and $\mathcal{N}=2$ supersymmetry, we obtain relations
between the template RGE coefficients, valid in any renormalisation scheme.
Directly in $d=3$, we identify a new theory with a non-trivial IR fixed point
that is under perturbative control in a large-$N$ limit.  We provide up-to-date
numerical results for all required massive tadpole master integrals up to four
loops and complement them with analytic expressions where available.
}

\vspace{10pt}
\noindent\rule{\textwidth}{1pt}
\tableofcontents\thispagestyle{fancy}
\noindent\rule{\textwidth}{1pt}
\vspace{10pt}

\section{Introduction}
\label{sec:intro}

\allowdisplaybreaks

Studying quantum systems in three dimensions is of strategic interest for many
aspects of modern physics. For instance, continuous phase transitions in
condensed-matter systems can be described as critical points in the systems
three-dimensional continuum limit. Prominent examples include scalar
$\mathrm{O}(N)$ models~\cite{Wilson:1971bg,Wilson:1971dh,Wilson:1971dc,Wilson:1973jj,
LeGuillou:1979ixc,Chetyrkin:1981jq,Gorishnii:1983gp,Kleinert:1991rg,Calabrese:2003ww,
Kleinert:1994td,Batkovich:2016jus,Kompaniets:2017yct,Adzhemyan:2019gvv,Kompaniets:2019xez,
Bednyakov:2021ojn}, and gapless Dirac electrons characterized by
the generalized Gross--Neveu universality class~\cite{Herbut:2006cs,Herbut:2009qb,
Zerf:2017zqi,Boyack:2020xpe,Herbut:2022zzw,Herbut:2023xgz,Uetrecht:2023uou,
Uetrecht:2025llz,Hawashin:2025cua}.
At high temperatures, four-dimensional quantum systems reduce effectively to three dimensions 
once the heavy Matsubara modes are 
integrated out~\cite{Ginsparg:1980ef,Farakos:1994kx,Kajantie:1995dw,Braaten:1995cm}. 
The resulting three-dimensional effective field theory governs the infrared dynamics, making its renormalisation-group flow essential for high-quality resummations~\cite{Ekstedt:2024etx,Chala:2025cya}.
Moreover, non-perturbative phenomena such as confinement
and dynamic symmetry breaking, which are complicated to capture in four dimensions, can be 
modelled in three-dimensional theories, see for instance~\cite{Appelquist:1981vg,Appelquist:1986qw,
Appelquist:1986fd,Appelquist:1988sr,Nash:1989xx,Rosenstein:1990nm}.
Furthermore, theories in three dimensions are interrelated by a web of dualities~\cite{Aharony:2014uya,Aharony:2015mjs,Seiberg:2016gmd,Aharony:2016jvv,Amoretti:2025hpi}.
Such dualities include strong-weak dualities, which make non-perturbative effects tractable.

After half a century of intensive work, three-dimensional quantum field
theories (QFTs) have remained less well-understood
than their four-dimensional counterparts in regard to their critical 
phenomena. A key reason is that in four dimensions, the renormalisation group
(RG) flow is known for each interaction of any renormalisable QFT in the perturbative region.
While perturbation theory certainly has its limits, it also enables a strong theoretical understanding whenever it is applicable.
In four dimensions, the perturbative renormalisation group equations (RGEs) of
any renormalisable QFT are readily available to high orders via template
expressions~\cite{Machacek:1983tz,Machacek:1983fi,Machacek:1984zw,Jack:1984vj,Pickering:2001aq,Luo:2002ti,Chetyrkin:2012rz,Mihaila:2012pz,Bednyakov:2012en,Bednyakov:2012rb,Chetyrkin:2013wya,Bednyakov:2013eba,Bednyakov:2013cpa,Sperling:2013eva,Sperling:2013xqa,Mihaila:2014caa,Zoller:2015tha,Chetyrkin:2016ruf,Schienbein:2018fsw,Poole:2019kcm,Poole:2019txl,Bednyakov:2021qxa,Davies:2021mnc,Steudtner:2021fzs,Jack:2024sjr,Steudtner:2024teg,Steudtner:2025blh,Henriksson:2025vyi}.

This approach of template expressions has proven useful for two main reasons.
For one, template RGEs are model independent results, which separate the
problem of computing RG flows from the difficulty of loop calculations. When
implemented in software packages such as~\cite{Fonseca:2011sy,Staub:2013tta,
Litim:2020jvl,Sartore:2020gou,Thomsen:2021ncy,Steudtner:FoRGEr},
high-loop RGEs become available even with minimal know-how or computational
power.

The second advantage is that templates allow for an exhaustive and systematic
study of perturbative RG flows. As all renormalisable QFTs are captured by template RGEs, any theorem derived from these expressions enjoys the same broad scope.
Moreover, the formalism can help to identify QFTs
where the RG flow has very rare or unique properties, and which are otherwise
difficult to identify. An example in four dimensions would be the Litim--Sannino
model~\cite{Litim:2014uca,Litim:2015iea}, which realises a perturbatively
accessible UV fixed point. In contrast, in this work, the template method will
enable us to identify a three-dimensional theory that features a novel 
IR fixed point under perturbative control at large $N$.

The main goal of this work is to bring the advantages of template RGEs into the
realm of three-dimensional QFTs. For the time being, we target gaugeless
theories and a precision up to four-loop order. Thus, the work can be
understood as an extension of the
studies of~\cite{Jack:2016utw,Fraser-Taliente:2024rql}, and a three-dimensional
equivalent of~\cite{Steudtner:2021fzs,Jack:2024sjr,Steudtner:2025blh}. We also
aim to describe known critical phenomena in a unified framework, namely the UV
fixed points in purely scalar~\cite{Townsend:1976sy,Appelquist:1981sf,
Pisarski:1982vz,Hager:2002uq,Kvedaraite:2025lgi}
and purely fermionic theories~\cite{Rosenstein:1988pt,Gat:1990xi,
deCalan:1991km,Braun:2011pp,Jakovac:2014lqa}.

This work is structured as follows: We introduce our notation and methodology
in \Sec{theory} and collect our main results in~\Sec{templates}. As an
important cross-check, we discuss the emergence of supersymmetry in~\Sec{SUSY}.
In~\Sec{FPs}, we analyse the RG flow and find hitherto unknown fixed points. 
A small outlook follows in~\Sec{conclusions}. The
App.~\ref{app:masters} holds technical details of three-dimensional
four-loop integrals, which are key ingredients for the calculations in this
work.

\section{Scalar-Fermionic Theory}\label{sec:theory}

We consider the most general three-dimensional, renormalisable QFT
of scalars and spin-$1/2$ fermions with marginal interactions in three-dimensional Minkowski space
\begin{align}\label{eq:theory}
    \mathcal{L} &= \frac12 \partial_\mu \phi_a \partial^\mu \phi_a + \frac{i}2\overline{\psi}_i \gamma^\mu\partial_\mu \psi_i - \frac14Y^{abij} \phi_a \phi_b \overline{\psi}_i  \psi_j - \frac1{6!} \eta^{abcdef} \phi_a \phi_b \phi_c \phi_d \phi_e \phi_f\,,
\end{align}
where $\phi^a$ are real scalar fields with $a=1,\,..,\,N_s$. In three
dimensions, fermions are real rank-two spinors given a suitable representation
of $\gamma$-matrices where $(\gamma^0)^\mathrm{T} = -\gamma^0$, and
$(\gamma^{1,2})^\mathrm{T} = \gamma^{1,2}$, which can be furnished via the
Pauli matrices. The fermions $\psi_i$ and $\overline{\psi}_i =
\psi_i^\dagger \gamma^0$ are enumerated by flavour indices $i=1,\,..,\,N_f$
and spinor indices are kept implicit. 
All marginal interactions are encoded by the tensor couplings $\eta$ and $Y$, that are totally symmetric in their indices connecting 
to fields of the same spin $\eta^{abcdef} = \eta^{(abcdef)}$ and $Y^{abij} = Y^{(ab)(ij)}$. 
In the following, we condense our notation of fermion chains via
\begin{equation}
    Y^{a_1 b_1 i_1 i_2} Y^{a_2 b_2 i_2 i_3} \dots Y^{a_n b_n i_n i_{n+1}} = Y^{a_1 b_1 a_2 b_2 \dots a_n b_n}\,,
\end{equation}
where the fermion indices are suppressed altogether. This formalism allows us
to compute RGEs up to four-loop order in a very general manner. We utilise
dimensional regularisation to $d =
3-2\epsilon$~\cite{Bollini:1972bi,Bollini:1972ui}, and the modified minimal
subtraction ($\overline{\mathrm{MS}}$)
scheme~\cite{tHooft:1973mfk,Bardeen:1978yd}. The technique of infrared
rearrangement with a common mass parameter~\cite{Chetyrkin:1997fm} allows us to
separate UV from IR divergences while only computing tadpole integrals. The
computation is conducted via the \texttt{MaRTIn}
framework~\cite{Brod:2024zaz,Kuipers:2012rf,Nogueira:1991ex}, which we have
extended to four loop-order. It leverages the \texttt{FMFT}
package~\cite{Pikelner:2017tgv} for efficient reductions of integrals to a
basis of masters given in~\cite{Czakon:2004bu}. These master integrals are
computed numerically using the algorithm of~\cite{Laporta:2000dsw} to high
precision. Where possible, analytic results are extracted using the integer
relation algorithm
\texttt{PSLQ}~\cite{FergusonBailey1992,FergusonBaileyArno1999}. This is
sufficient to retrieve all template RGEs in analytic form. Details are
relegated to App.~\ref{app:masters}, which serves as an update
to~\cite{Schroder:2003kb}. As a result, the template RGEs are obtained as
contractions of the coupling tensors $Y^{ab}$ and $\eta^{abcdef}$, which at
this stage are fully model independent but may be cumbersome to interpret in terms of a 
specific model. In this work, extracting specific RGEs from template expressions is achieved with the in-house software~\texttt{FoRGEr}~\cite{Steudtner:FoRGEr},
which we plan to make openly available in the future.

\section{Template RGEs}\label{sec:templates}

In this section, we collect the template expressions for scalar and fermionic
field anomalous dimensions \begin{equation} \gamma_{\phi,\psi} = -
  \frac{\mathrm{d} \log \sqrt{Z}_{\phi,\psi}}{\mathrm{d} \log \mu} =
  \sum_{\ell=1}^\infty \frac{\gamma_{\phi,\psi}^{(2\ell)}}{(4\pi)^{2\ell}}\,,
\end{equation} where $\phi^\text{(bare)}_{a} = (\sqrt{Z}_\phi)_{ab} \phi_b $ and
$\psi^\text{(bare)}_{i} = (\sqrt{Z}_\psi)_{ij} \psi_j $ and $\mu$ is the $\overline{\text{MS}}$ renormalisation scale. Moreover, we provide $\beta$-functions for quartic Yukawa and scalar sextic couplings 
\begin{align}
    \beta_{Y}^{ab} &= \frac{\mathrm{d} Y^{ab}}{\mathrm{d} \log \mu} = \gamma_\psi Y^{ab} + Y^{ab}\gamma_\psi^\mathrm{T} + \gamma_\phi^{ac} Y^{bc} +  \gamma_\phi^{bc} Y^{ac} + \hat{\beta}^{ab}_Y\,,\\
    \beta_{\eta}^{abcdef} &= \frac{\mathrm{d} \eta^{abcdef}}{\mathrm{d} \log \mu} = \mathcal{S}_{6}\,\gamma_\phi^{ag} \eta^{gbcdef} + \hat{\beta}^{abcdef}_\eta\,,\label{eq:beta-eta}
\end{align}
in terms of their respective vertex corrections
\begin{equation}
    \hat{\beta}_X = \sum_{\ell=1}^\infty \frac{\hat{\beta}_X^{(2\ell)}}{(4\pi)^{2\ell}}
\end{equation}
up to four loops ($\ell = 2$). 
These RGEs are derived from the overall RG-invariance of the bare couplings
\begin{equation}
    {Y}^{ab}_{\text{(bare)}} = \mu^{2\epsilon} \left( {Y}^{ab} + \delta {Y}^{ab}\right) \,,\quad
    {\eta}^{abcdef}_{\text{(bare)}} = \mu^{4\epsilon} \left( \eta^{abcdef} + \delta \eta^{abcdef} \right)\,,
\end{equation}
where $\delta {Y}^{ab}$ and $\delta \eta^{abcdef}$ are counterterms, and the powers of the RG scale $\mu$ render the renormalised couplings marginal in $d=3-2\epsilon$ dimensions.

The ansatz for the RGEs incorporates the fact that
odd-dimensional RGEs only receive contributions at even loop orders, starting
at two loops. For ease of notation, we make heavy use of the symmetric
permutation symbol $\mathcal{S}_{n}$, which indicates that $n$ inequivalent
permutations of the external indices (both scalar and fermionic) are to be
added. For instance in \eq{beta-eta}, $\mathcal{S}_6$ indicates that the leg
correction consists of six terms, where $a$ is exchanged with the other
external indices: 
\begin{equation}
    \mathcal{S}_{6}\,\gamma_\phi^{ag} \eta^{gbcdef} = \gamma_\phi^{ag} \eta^{gbcdef} + \gamma_\phi^{bg} \eta^{gacdef} + \gamma_\phi^{cg} \eta^{gabdef}+ \gamma_\phi^{dg} \eta^{gabcef}+ \gamma_\phi^{eg} \eta^{gabcdf}+ \gamma_\phi^{fg} \eta^{gabcde}\,,
\end{equation}
whereas other permutations among the indices $b$--$f$ in the expression on the
left-hand-side are not considered as they do not generate new terms owing to
the symmetry of $\eta$.

Note that the results on $\beta_Y$ and $\beta_\eta$ presented here
also allow for the automatised extraction of the RGEs of the
operators $\phi \bar{\psi} \psi$ and $\bar{\psi} \psi$ (via $\beta_Y$) as well as
$\phi^5$, $\phi^4$, $\phi^3$, $\phi^2$ and $\phi^1$ (via $\beta_\eta$) by utilising
the dummy-field technique~\cite{Martin:1993zk,Luo:2002ti,Schienbein:2018fsw}. Also, these RGEs are implemented in 
\texttt{FoRGEr}~\cite{Steudtner:FoRGEr}.

\subsection{Anomalous Field Dimensions}

The general shape of the two- and four-loop scalar field anomalous dimensions are 
\begin{align}
	\gamma_\phi^{ab\,(2)} &= \gamma_{\phi,1}^{(2)} \,\tr{Y^{acbc}} \,,\\[.5em]
    \gamma_\phi^{ab\,(4)} &= \gamma_{\phi,1}^{(4)} \, \eta^{acdefg} \eta^{bcdefg} 
    + \gamma_{\phi,2}^{(4)} \, \tr{Y^{acde}} \tr{Y^{bcde}}  + \gamma_{\phi,3}^{(4)} \, \tr{Y^{acde}} \tr{Y^{bdce}}  \notag \\
    &\phantom{=\,} +  \gamma_{\phi,4}^{(4)} \, \tr{Y^{acbd}} \tr{Y^{cede}} + \gamma_{\phi,5}^{(4)} \, \tr{Y^{acbcdede}} + \gamma_{\phi,6}^{(4)} \, \tr{Y^{acbdcede}} \notag \\
    &\phantom{=\,} +  \gamma_{\phi,7}^{(4)} \, \tr{Y^{acbddece}} +  \gamma_{\phi,8}^{(4)} \, \tr{Y^{acdebcde}} +  \gamma_{\phi,9}^{(5)} \, \tr{Y^{accebdde}} \,. \label{eq:gamma-s}
\end{align}
Here we have listed all possible contractions excluding contributions from 
tadpole diagrams as they do not
contribute to the anomalous dimensions. In the $\overline{\mathrm{MS}}$
scheme, we find the explicit coefficients 
\begin{align*}
	 \gamma^{(2)}_{\phi,1} & = \tfrac{1}{12} \,, &
	 \gamma^{(4)}_{\phi,1} & = \tfrac{1}{1440} \,, &
	 \gamma^{(4)}_{\phi,2} & = -\tfrac{\pi^2}{192} \,, &
	 \gamma^{(4)}_{\phi,3} & = -\tfrac{1}{144} \,, &
	 \gamma^{(4)}_{\phi,4} & = -\tfrac{11}{432} \,, \\
	 \gamma^{(4)}_{\phi,5} & = -\tfrac{5}{216} \,, &
	 \gamma^{(4)}_{\phi,6} & = -\tfrac{1}{18} \,, &
	 \gamma^{(4)}_{\phi,7} & = -\tfrac{\pi^2}{96} \,, &
	 \gamma^{(4)}_{\phi,8} & = \tfrac{1}{72} \,, &
	 \gamma^{(4)}_{\phi,9} & = -\tfrac{1}{72} \,. \tag{\stepcounter{equation}\theequation}
\end{align*}
Analogously, the fermion anomalous dimension includes the contractions 
\begin{align}
	\gamma_\psi^{(2)} &= \gamma_{\psi,1}^{(2)} \,Y^{abab} \,,\\[.5em]
    \gamma_\psi^{(4)} &= \gamma_{\psi,1}^{(4)} \, \tr{Y^{abcd}} Y^{acbd} + \gamma_{\psi,2}^{(4)} \, \tr{Y^{abcd}} Y^{abcd} + \gamma_{\psi,3}^{(4)} \, \tr{Y^{abac}} Y^{bdcd} \notag \\
    &\phantom{=\,} + \gamma_{\psi,4}^{(4)}\, Y^{abcdcdab} + \gamma_{\psi,5}^{(4)}\, Y^{abbccdda} + \gamma_{\psi,6}^{(4)}\, Y^{abcdbdac} + \gamma_{\psi,7}^{(4)}\, Y^{abcdabcd}\notag \\
    &\phantom{=\,} + \gamma_{\psi,8}^{(4)}\, Y^{abbcaddc}\ \label{eq:gamma-f}
\end{align}
up to four-loop order, which take the explicit values 
\begin{align*}
	 \gamma^{(2)}_{\psi,1} & = \tfrac{1}{12} \,, &
	 \gamma^{(4)}_{\psi,1} & = -\tfrac{1}{36} \,, &
	 \gamma^{(4)}_{\psi,2} & = -\tfrac{\pi^2}{192} \,, &
	 \gamma^{(4)}_{\psi,3} & = -\tfrac{1}{27} \,, &
	 \gamma^{(4)}_{\psi,4} & = -\tfrac{1}{216} \,, \\
	 \gamma^{(4)}_{\psi,5} & = -\tfrac{\pi^2}{96} \,, &
	 \gamma^{(4)}_{\psi,6} & = \tfrac{1}{36} \,, &
	 \gamma^{(4)}_{\psi,7} & = \tfrac{1}{72} \,, &
	 \gamma^{(4)}_{\psi,8} & = -\tfrac{1}{72}
     \tag{\stepcounter{equation}\theequation}
\end{align*}
in the $\overline{\text{MS}}$ scheme. 

It is worth noting that both anomalous
dimensions in~\eq{gamma-s} and \eq{gamma-f} are manifestly symmetric under
exchange of their external indices. 
In general, anomalous dimension might also contain non-symmetric contributions, inherited from antisymmetric parts of the field strength renormalisations $\sqrt{Z}_{\phi,\psi}$. 
The scalar and fermion kinetic terms 
\begin{equation}
    \mathcal{L} = \tfrac12 Z_{\phi}^{ab} \partial_\mu \phi_a \partial^\mu \phi_b + \tfrac{i}2 Z_{\phi}^{ij} \overline{\psi}_i \gamma^\mu \partial_\mu \psi_j + \dots
\end{equation}
have renormalisation constants $Z_{\phi}^{ab}$ and $Z_{\psi}^{ij}$ that are
symmetric under $(a\leftrightarrow b)$ and $(i\leftrightarrow j)$,
respectively, and fixed by the renormalisation of two-point functions. However,
there is an ambiguity in defining the field strength tensors
$\sqrt{Z}_{\phi,\psi}$ such that $Z_{\phi,\psi} = \sqrt{Z}_{\phi,\psi} \cdot
\sqrt{Z}_{\phi,\psi}$, which is related to their antisymmetric part. This
ambiguity has also been encountered in four dimensional theories, see,
e.g.~\cite{Bednyakov:2012en,Fortin:2012hn,Herren:2017uxn,Steudtner:2024teg,Steudtner:2025blh}.

The ambiguity leads to subtleties in theories with
hidden flavour symmetries. Such kind of QFTs contain several scalars or
fermions that have identical quantum numbers, yet do not transform together under a global
symmetry. In this case a rotation among the fields is encoded in the theory,
but involves a redefinition of couplings. 
As a result, zeros in the RGEs of these couplings do not indicate fixed points with conformal symmetry.
Instead, $\beta$-functions require
a shift proportional to antisymmetric leg corrections in order to preserve
conformality~\cite{Fortin:2012hn,Herren:2021yur,Pannell:2025ajf}. In four dimensions, such
corrections start occurring from three-loop order. We eventually expect the
same phenomena to also occur in three dimensions, though they are still absent
at four loops. 

\subsection{Quartic Yukawa Vertex}

We now turn to the vertex corrections of the quartic Yukawa coupling. At
two-loop order there are only a few contractions 
\begin{align}\label{eq:chain2L}
	\hat{\beta}_Y^{(2) ab} &= \beta_{Y,1}^{(2)}\,\tr{Y^{abcd}} Y^{cd} + \beta_{Y,2}^{(2)}\,\tr{Y^{acbd}} Y^{cd} + \beta_{Y,3}^{(2)}\,Y^{cdabcd} + \beta_{Y,4}^{(2)}\,\mathcal{S}_{2} Y^{accddb} \notag\\
    &\phantom{=\,} 
    + \beta_{Y,5}^{(2)}\,\mathcal{S}_{4} Y^{acbdcd}\,,
\end{align}
which do not feature the sextic interactions $\eta$ at all. We find the
explicit values
\begin{align*}
	 \beta^{(2)}_{Y,1} & = 0 \,, &
	 \beta^{(2)}_{Y,2} & = \tfrac{1}{2} \,, &
	 \beta^{(2)}_{Y,3} & = \tfrac{1}{2} \,, &
	 \beta^{(2)}_{Y,4} & = 0 \,, &
	 \beta^{(2)}_{Y,5} & = \tfrac{1}{2} \,. 
     \tag{\stepcounter{equation}\theequation}
\end{align*}

At four-loop order, we find a more complicated shape with a total number of $132$ coefficients,
which we have categorised by the number of sextic interactions and fermion
traces
\begin{align}
	\hat{\beta}_Y^{(4) ab} &= \hat{\beta}_{Y,\eta^2}^{(4) ab} + \hat{\beta}_{Y,\eta^1}^{(4) ab} + \hat{\beta}_{Y,Y^2Y^2Y}^{(4) ab} +\hat{\beta}_{Y,Y^4Y}^{(4) ab} +  \hat{\beta}_{Y,Y^3Y^2}^{(4) ab} + \hat{\beta}_{Y,Y^2Y^3}^{(4) ab} +\hat{\beta}_{Y,Y^5}^{(4) ab}\,.
\end{align}
The first two terms include contractions with the scalar sextic coupling 
\begin{align}
    \hat{\beta}_{Y,\eta^2}^{(4) ab}  &= 
    \beta_{Y,1}^{(4)}\,\eta^{acefgh} \eta^{bdefgh} Y^{cd} 
    + \beta_{Y,2}^{(4)}\,\eta^{abefgh}\eta^{cdefgh} Y^{cd} \,, \\[.5em]
    \hat{\beta}_{Y,\eta^1}^{(4) ab}  &=
      \beta_{Y,3}^{(4)}\,\eta^{abcdef} Y^{cgdefg} 
     + \beta_{Y,4}^{(4)}\,\mathcal{S}_2\,\eta^{abcdef} Y^{cdegfg}
     + \beta_{Y,5}^{(4)}\,\mathcal{S}_2\,\eta^{acdefg} Y^{cdbefg}\notag\\
    &\phantom{=\,} 
     + \beta_{Y,6}^{(4)}\,\mathcal{S}_4\,\eta^{acdefg} Y^{bcdefg}
    + \beta_{Y,7}^{(4)}\, \eta^{abcdef} \tr{Y^{egfg}} Y^{cd}\notag\\
    &\phantom{=\,} 
    + \beta_{Y,8}^{(4)}\, \eta^{abcefg} \tr{Y^{defg}} Y^{cd}
    + \beta_{Y,9}^{(4)}\,\mathcal{S}_2\,\eta^{acdefg} \tr{Y^{befg}} Y^{cd} \,.
\end{align}
The next terms have both external fermion lines connected to the same Yukawa,
while also featuring two closed fermion traces
\begin{align}\label{eq:chain4L}
    \hat{\beta}_{Y,Y^2Y^2Y}^{(4) ab}  &=
      \beta_{Y,10}^{(4)}\,\tr{Y^{abef}} \tr{Y^{cdef}} Y^{cd} 
    + \beta_{Y,11}^{(4)}\,\tr{Y^{abef}} \tr{Y^{cedf}} Y^{cd}\notag\\
    &\phantom{=\,} 
    + \beta_{Y,12}^{(4)}\,\tr{Y^{abce}} \tr{Y^{dfef}} Y^{cd} 
    + \beta_{Y,13}^{(4)}\,\tr{Y^{aebf}} \tr{Y^{cdef}} Y^{cd} \notag\\
    &\phantom{=\,} 
    + \beta_{Y,14}^{(4)}\,\tr{Y^{aebf}} \tr{Y^{cedf}} Y^{cd} 
    + \beta_{Y,15}^{(4)}\,\mathcal{S}_2\,\tr{Y^{acbe}} \tr{Y^{dfef}} Y^{cd} \notag\\
    &\phantom{=\,} 
    + \beta_{Y,16}^{(4)}\,\tr{Y^{acef}} \tr{Y^{bdef}} Y^{cd} 
    + \beta_{Y,17}^{(4)}\,\tr{Y^{aecf}} \tr{Y^{bfde}} Y^{cd} \notag\\
    &\phantom{=\,} 
    + \beta_{Y,18}^{(4)}\,\tr{Y^{aecf}} \tr{Y^{bfde}} Y^{cd} 
    + \beta_{Y,19}^{(4)}\,\mathcal{S}_2 \tr{Y^{acef}} \tr{Y^{bedf}} Y^{cd} \,,
\end{align}
or one such trace
\begin{align}
    \hat{\beta}_{Y,Y^4Y}^{(4) ab}  &=
      \beta_{Y,20}^{(4)}\,\tr{Y^{abefcdef}} Y^{cd} 
    + \beta_{Y,21}^{(4)}\,\tr{Y^{abcdefef}} Y^{cd} 
    + \beta_{Y,22}^{(4)}\,\tr{Y^{abceeffd}} Y^{cd}\notag\\
    &\phantom{=\,} 
    + \beta_{Y,23}^{(4)}\,\tr{Y^{abcedfef}} Y^{cd} 
    + \beta_{Y,24}^{(4)}\,\tr{Y^{aebecfdf}} Y^{cd}
    + \beta_{Y,25}^{(4)}\,\mathcal{S}_2\,\tr{Y^{aebfcdef}} Y^{cd} \notag\\
    &\phantom{=\,} 
    + \beta_{Y,26}^{(4)}\,\tr{Y^{aebfcedf}} Y^{cd}  
    + \beta_{Y,27}^{(4)}\,\tr{Y^{aebfcfde}} Y^{cd}
    + \beta_{Y,28}^{(4)}\,\mathcal{S}_2\,\tr{Y^{acbedfef}} Y^{cd} \notag\\
    &\phantom{=\,} 
    + \beta_{Y,29}^{(4)}\,\mathcal{S}_2\,\tr{Y^{acbeefdf}} Y^{cd} 
    + \beta_{Y,30}^{(4)}\,\tr{Y^{acbdefef}} Y^{cd}
    + \beta_{Y,31}^{(4)}\,\tr{Y^{aecdbfef}} Y^{cd}\notag\\
    &\phantom{=\,} 
    + \beta_{Y,32}^{(4)}\,\tr{Y^{acefbdef}} Y^{cd}
    + \beta_{Y,33}^{(4)}\,\tr{Y^{aecfbedf}} Y^{cd}
    + \beta_{Y,34}^{(4)}\,\tr{Y^{aecebfdf}} Y^{cd}\notag\\
    &\phantom{=\,} 
    + \beta_{Y,35}^{(4)}\,\mathcal{S}_2\,\tr{Y^{acdebfef}} Y^{cd} \,.
\end{align}
Note that contributions $\propto\tr{Y^3}$ or higher odd numbers of Yukawas are
not a priori excluded in three dimensions. Such terms emerge 
in diagrams that contain traces over an odd number of $\gamma$ matrices, 
which, in contrast to four dimensions, do not vanish in three dimensions. 
Rather, they give expressions containing fully antisymmetric Levi--Civita
tensors, e.g.
\begin{equation}\label{eq:3-gamma}
  \mathrm{tr}(\gamma^\mu \gamma^\nu \gamma^\rho) = -2 i \,\epsilon^{\mu\nu\rho}\,. 
\end{equation} 
This excludes many possible contractions. After integration of the fermion loop
with an odd number of $\gamma$ matrices, 
one of the loop momenta is spent and the remaining three need to
be contracted with the Levi--Civita tensor for it not to vanish due to the
antisymmetry. Thus, only the contractions 
\begin{align}
    \hat{\beta}_{Y,Y^3Y^2}^{(4) ab} &=
    \beta_{Y,36}^{(4)} \,\tr{Y^{abcdef}} Y^{cdef} 
    + \beta_{Y,37}^{(4)} \,\tr{Y^{abcedf}} Y^{cdef}
    + \beta_{Y,38}^{(4)} \,\tr{Y^{acbedf}} Y^{cdef}\notag\\
    &\phantom{=\,} 
    + \beta_{Y,39}^{(4)} \,\mathcal{S}_2\,\tr{Y^{acbdef}} Y^{cdef} 
    + \beta_{Y,40}^{(4)} \,\mathcal{S}_4\,\tr{Y^{acdfef}} Y^{bcde}  \notag\\
    &\phantom{=\,} 
    + \beta_{Y,41}^{(4)} \,\mathcal{S}_4\,\tr{Y^{adcfef}} Y^{bcde}
    + \beta_{Y,42}^{(4)} \,\mathcal{S}_4\,\tr{Y^{afcdef}} Y^{bcde} \notag\\
    &\phantom{=\,} 
    + \beta_{Y,43}^{(4)} \,\mathcal{S}_4\,\tr{Y^{afdecf}} Y^{bcde}
\end{align}
do not vanish. There are further contractions with a single fermion loop 
\begin{align}
    \hat{\beta}_{Y,Y^2Y^3}^{(4) ab}  &= 
    \beta_{Y,44}^{(4)} \, \tr{Y^{abcd}} Y^{efcdef}
    + \beta_{Y,45}^{(4)} \, \tr{Y^{abcd}} Y^{ceefdf}\notag\\
    &\phantom{=\,} 
    + \beta_{Y,46}^{(4)} \, \mathcal{S}_2\,\tr{Y^{abcd}} Y^{cedfef} 
    + \beta_{Y,47}^{(4)} \, \tr{Y^{acbd}} Y^{efcdef}\notag\\
    &\phantom{=\,} 
    + \beta_{Y,48}^{(4)} \, \mathcal{S}_2 \, \tr{Y^{acbd}} Y^{ceefdf}
    + \beta_{Y,49}^{(4)} \, \mathcal{S}_2 \, \tr{Y^{acbd}} Y^{cedfef}\notag\\
    &\phantom{=\,} 
    + \beta_{Y,50}^{(4)} \, \mathcal{S}_4 \, \tr{Y^{acde}} Y^{bcdfef}
    + \beta_{Y,51}^{(4)} \, \mathcal{S}_4 \, \tr{Y^{acde}} Y^{bdcfef}\notag\\
    &\phantom{=\,} 
    + \beta_{Y,52}^{(4)} \, \mathcal{S}_4 \, \tr{Y^{acde}} Y^{bfcdef}
    + \beta_{Y,53}^{(4)} \, \mathcal{S}_4 \, \tr{Y^{acde}} Y^{bfcfde}\notag\\
    &\phantom{=\,} 
    + \beta_{Y,54}^{(4)} \, \mathcal{S}_4 \, \tr{Y^{acde}} Y^{bdefcf}
    + \beta_{Y,55}^{(4)} \, \mathcal{S}_4 \, \tr{Y^{acde}} Y^{bfefcd}\notag\\
    &\phantom{=\,} 
    + \beta_{Y,56}^{(4)} \, \mathcal{S}_4 \, \tr{Y^{acde}} Y^{bfdecf}
    + \beta_{Y,57}^{(4)} \, \mathcal{S}_2 \, \tr{Y^{acde}} Y^{dfbcef}\notag\\
    &\phantom{=\,} 
    + \beta_{Y,58}^{(4)} \, \mathcal{S}_4 \, \tr{Y^{acde}} Y^{dfbfce}
    + \beta_{Y,59}^{(4)} \, \mathcal{S}_4 \, \tr{Y^{acde}} Y^{debfef}\notag\\
    &\phantom{=\,} 
    + \beta_{Y,60}^{(4)} \, \mathcal{S}_4 \, \tr{Y^{acde}} Y^{dfbecf}
    + \beta_{Y,61}^{(4)} \, \mathcal{S}_4 \, \tr{Y^{cdef}} Y^{acbdef} \notag\\
    &\phantom{=\,} 
    + \beta_{Y,62}^{(4)} \, \mathcal{S}_4 \, \tr{Y^{cdef}} Y^{acbedf}
    + \beta_{Y,63}^{(4)} \, \mathcal{S}_4 \, \tr{Y^{cede}} Y^{afbfcd} \notag\\
    &\phantom{=\,} 
    + \beta_{Y,64}^{(4)} \, \mathcal{S}_4 \, \tr{Y^{cede}} Y^{acbfdf}
    + \beta_{Y,65}^{(4)} \, \mathcal{S}_4 \, \tr{Y^{cede}} Y^{afbcdf} \notag\\
    &\phantom{=\,} 
    + \beta_{Y,66}^{(4)} \, \mathcal{S}_2 \, \tr{Y^{cdef}} Y^{acdebf}
    + \beta_{Y,67}^{(4)} \, \mathcal{S}_2 \, \tr{Y^{cdef}} Y^{acefbd} \notag\\
    &\phantom{=\,} 
    + \beta_{Y,68}^{(4)} \, \mathcal{S}_2 \, \tr{Y^{cede}} Y^{afcdbf}
    + \beta_{Y,69}^{(4)} \, \mathcal{S}_4 \, \tr{Y^{cede}} Y^{acdfbf} \notag\\
    &\phantom{=\,} 
    + \beta_{Y,70}^{(4)} \, \tr{Y^{cdef}} Y^{cdabef}
    + \beta_{Y,71}^{(4)} \, \tr{Y^{cdef}} Y^{ceabdf} \notag\\
    &\phantom{=\,} 
    + \beta_{Y,72}^{(4)} \, \tr{Y^{cede}} Y^{cfabdf}\,,
\end{align}
as well as a final family with no closed fermion loop at all 
\begin{align}
    \hat{\beta}_{Y,Y^5}^{(4) ab}  &= 
     \beta_{Y,73}^{(4)} \, \mathcal{S}_4 \, Y^{acbdcedfef}
     + \beta_{Y,74}^{(4)} \, \mathcal{S}_4 \, Y^{acbddecfef}
     + \beta_{Y,75}^{(4)} \, \mathcal{S}_4 \, Y^{acbdceefdf} \notag\\
    &\phantom{=\,} 
    + \beta_{Y,76}^{(4)} \, \mathcal{S}_4 \, Y^{acbddeefcf}
    + \beta_{Y,77}^{(4)} \, \mathcal{S}_4 \, Y^{acbdefcedf}
    + \beta_{Y,78}^{(4)} \, \mathcal{S}_4 \, Y^{acbdefdecf}  \notag\\
    &\phantom{=\,} 
    + \beta_{Y,79}^{(4)} \, \mathcal{S}_4 \, Y^{acbdefcdef}
    + \beta_{Y,80}^{(4)} \, \mathcal{S}_4 \, Y^{acbdefefcd}
    + \beta_{Y,81}^{(4)} \, \mathcal{S}_4 \, Y^{acdebcdfef} \notag\\
    &\phantom{=\,} 
    + \beta_{Y,82}^{(4)} \, \mathcal{S}_4 \, Y^{accdbedfef}
    + \beta_{Y,83}^{(4)} \, \mathcal{S}_4 \, Y^{accdbeefdf}
    + \beta_{Y,84}^{(4)} \, \mathcal{S}_4 \, Y^{acdebdcfef} \notag\\
    &\phantom{=\,} 
    + \beta_{Y,85}^{(4)} \, \mathcal{S}_4 \, Y^{acdebdefcf}
    + \beta_{Y,86}^{(4)} \, \mathcal{S}_4 \, Y^{acdebfcdef}
    + \beta_{Y,87}^{(4)} \, \mathcal{S}_4 \, Y^{acdebffdce} \notag\\
    &\phantom{=\,} 
    + \beta_{Y,88}^{(4)} \, \mathcal{S}_4 \, Y^{acdebfdecf}
    + \beta_{Y,89}^{(4)} \, \mathcal{S}_4 \, Y^{acdebfcfde}
    + \beta_{Y,90}^{(4)} \, \mathcal{S}_4 \, Y^{accddebfef} \notag\\
    &\phantom{=\,} 
    + \beta_{Y,91}^{(4)} \, \mathcal{S}_4 \, Y^{accdefbdef}
    + \beta_{Y,92}^{(4)} \, \mathcal{S}_4 \, Y^{accdefbedf}
    + \beta_{Y,93}^{(4)} \, \mathcal{S}_4 \, Y^{acdecfbdef} \notag\\
    &\phantom{=\,} 
    + \beta_{Y,94}^{(4)} \, \mathcal{S}_4 \, Y^{acdecdbfef}
    + \beta_{Y,95}^{(4)} \, \mathcal{S}_4 \, Y^{acdecfbfde}
    + \beta_{Y,96}^{(4)} \, \mathcal{S}_4 \, Y^{acdedfbcef} \notag\\
    &\phantom{=\,} 
    + \beta_{Y,97}^{(4)} \, \mathcal{S}_4 \, Y^{acdeefbdcf}
    + \beta_{Y,98}^{(4)} \, \mathcal{S}_4 \, Y^{acdedfbfce}
    + \beta_{Y,99}^{(4)} \, \mathcal{S}_4 \, Y^{acdedebfcf} \notag\\
    &\phantom{=\,} 
    + \beta_{Y,100}^{(4)} \, \mathcal{S}_2 \, Y^{accddeefbf}
    + \beta_{Y,101}^{(4)} \, \mathcal{S}_2 \, Y^{acdeefcfbd}
    + \beta_{Y,102}^{(4)} \, \mathcal{S}_2 \, Y^{acdecfdebf} \notag\\
    &\phantom{=\,} 
    + \beta_{Y,103}^{(4)} \, \mathcal{S}_4 \, Y^{accdefdfbe}
    + \beta_{Y,104}^{(4)} \, \mathcal{S}_4 \, Y^{acdecfefbd}
    + \beta_{Y,105}^{(4)} \, \mathcal{S}_4 \, Y^{accdefefbd} \notag\\
    &\phantom{=\,} 
    + \beta_{Y,106}^{(4)} \, \mathcal{S}_2 \, Y^{acdedfefbc}
    + \beta_{Y,107}^{(4)} \, \mathcal{S}_4 \, Y^{cdacbedfef}
    + \beta_{Y,108}^{(4)} \, \mathcal{S}_4 \, Y^{cdacbeefdf} \notag\\
    &\phantom{=\,} 
    + \beta_{Y,109}^{(4)} \, \mathcal{S}_4 \, Y^{cdaebcdfef}
    + \beta_{Y,110}^{(4)} \, \mathcal{S}_4 \, Y^{cdaebcefdf}
    + \beta_{Y,111}^{(4)} \, \mathcal{S}_4 \, Y^{cdaebecfdf} \notag\\
    &\phantom{=\,} 
    + \beta_{Y,112}^{(4)} \, \mathcal{S}_4 \, Y^{cdaebfcdef}
    + \beta_{Y,113}^{(4)} \, \mathcal{S}_4 \, Y^{cdaebfcedf}
    + \beta_{Y,114}^{(4)} \, \mathcal{S}_4 \, Y^{cdaebfcfde}  \notag\\
    &\phantom{=\,} 
    + \beta_{Y,115}^{(4)} \, \mathcal{S}_4 \, Y^{cdaebfefcd}
    + \beta_{Y,116}^{(4)} \, \mathcal{S}_2 \, Y^{cdacdebfef}
    + \beta_{Y,117}^{(4)} \, \mathcal{S}_4 \, Y^{cdacefbdef} \notag\\
    &\phantom{=\,} 
    + \beta_{Y,118}^{(4)} \, \mathcal{S}_4 \, Y^{cdacefbedf}
    + \beta_{Y,119}^{(4)} \, \mathcal{S}_2 \, Y^{cdaecfbdef}
    + \beta_{Y,120}^{(4)} \, \mathcal{S}_2 \, Y^{cdaecfbedf} \notag\\
    &\phantom{=\,} 
    + \beta_{Y,121}^{(4)} \, \mathcal{S}_4 \, Y^{cdaeefbcdf}
    + \beta_{Y,122}^{(4)} \, \mathcal{S}_2 \, Y^{cdaeefbfcd}
    + \beta_{Y,123}^{(4)} \, \mathcal{S}_2 \, Y^{cdabcedfef} \notag\\
    &\phantom{=\,}  
    + \beta_{Y,124}^{(4)} \, \mathcal{S}_2 \, Y^{cdabceefdf}
    + \beta_{Y,125}^{(4)} \, \mathcal{S}_2 \, Y^{cdabefcedf} 
    + \beta_{Y,126}^{(4)} \, \mathcal{S}_2 \, Y^{cdabefcdef} \notag\\
    &\phantom{=\,} 
    + \beta_{Y,127}^{(4)} \, \mathcal{S}_2 \, Y^{cdabefefcd} 
    + \beta_{Y,128}^{(4)} \,  Y^{cdefabefcd}
    + \beta_{Y,129}^{(4)} \,  Y^{cdefabcdef}\notag\\
    &\phantom{=\,} 
    + \beta_{Y,130}^{(4)} \,  Y^{cdefabcedf} 
    + \beta_{Y,131}^{(4)} \,  Y^{cedeabcfdf}
    + \beta_{Y,132}^{(4)} \,  Y^{cedeabdfcf}\,.
\end{align}

In the $\overline{\text{MS}}$ scheme, we find the explicit coefficients
\begin{align*}
     \beta^{(4)}_{Y,1} & = \tfrac{1}{48} \,, &
	 \beta^{(4)}_{Y,2} & = 0 \,, &
	 \beta^{(4)}_{Y,3} & = -\tfrac{1}{2} \,, &
	 \beta^{(4)}_{Y,4} & = -\tfrac{1}{4} \,, &
	 \beta^{(4)}_{Y,5} & = -\tfrac{1}{4} \,, \\
	 \beta^{(4)}_{Y,6} & = -\tfrac{1}{8} \,, &
	 \beta^{(4)}_{Y,7} & = 0 \,, &
	 \beta^{(4)}_{Y,8} & = 0 \,, &
	 \beta^{(4)}_{Y,9} & = 0 \,, &
	 \beta^{(4)}_{Y,10} & = 0 \,, \\
	 \beta^{(4)}_{Y,11} & = 0 \,, &
	 \beta^{(4)}_{Y,12} & = 0 \,, &
	 \beta^{(4)}_{Y,13} & = 0 \,, &
	 \beta^{(4)}_{Y,14} & = \tfrac{1}{8} \,, &
	 \beta^{(4)}_{Y,15} & = -\tfrac{1}{12} \,, \\
	 \beta^{(4)}_{Y,16} & = -\tfrac{\pi^2}{32} \,, &
	 \beta^{(4)}_{Y,17} & = \tfrac{1}{8} \,, &
	 \beta^{(4)}_{Y,18} & = 0 \,, &
	 \beta^{(4)}_{Y,19} & = -\tfrac{1}{4} \,, &
	 \beta^{(4)}_{Y,20} & = 0 \,, \\
	 \beta^{(4)}_{Y,21} & = 0 \,, &
	 \beta^{(4)}_{Y,22} & = 0 \,, &
	 \beta^{(4)}_{Y,23} & = -\tfrac{\pi^2}{8} \,, &
	 \beta^{(4)}_{Y,24} & = -\tfrac{1}{2} \,, &
	 \beta^{(4)}_{Y,25} & = 0 \,, \\
	 \beta^{(4)}_{Y,26} & = -\tfrac{\pi^2}{8} \,, &
	 \beta^{(4)}_{Y,27} & = -\tfrac{\pi^2}{16} \,, &
	 \beta^{(4)}_{Y,28} & = -\tfrac{3}{4} \,, &
	 \beta^{(4)}_{Y,29} & = -\tfrac{\pi^2}{16} \,, &
	 \beta^{(4)}_{Y,30} & = -\tfrac{1}{4} \,, \\
	 \beta^{(4)}_{Y,31} & = 0 \,, &
	 \beta^{(4)}_{Y,32} & = -\tfrac{1}{4} \,, &
	 \beta^{(4)}_{Y,33} & = -\tfrac{1}{2} \,, &
	 \beta^{(4)}_{Y,34} & = -\tfrac{1}{4} \,, &
	 \beta^{(4)}_{Y,35} & = -\tfrac{3}{4} \,, \\
	 \beta^{(4)}_{Y,36} & = 0 \,, &
	 \beta^{(4)}_{Y,37} & = 0 \,, &
	 \beta^{(4)}_{Y,38} & = 0 \,, &
	 \beta^{(4)}_{Y,39} & = 0 \,, &
	 \beta^{(4)}_{Y,40} & = 0 \,, \\
	 \beta^{(4)}_{Y,41} & = 0 \,, &
	 \beta^{(4)}_{Y,42} & = 0 \,, &
	 \beta^{(4)}_{Y,43} & = 0 \,, &
	 \beta^{(4)}_{Y,44} & = 0 \,, &
	 \beta^{(4)}_{Y,45} & = 0 \,, \\
	 \beta^{(4)}_{Y,46} & = 0 \,, &
	 \beta^{(4)}_{Y,47} & = 0 \,, &
	 \beta^{(4)}_{Y,48} & = -\tfrac{\pi^2}{32} \,, &
	 \beta^{(4)}_{Y,49} & = 0 \,, &
	 \beta^{(4)}_{Y,50} & = -\tfrac{1}{4} \,, \\
	 \beta^{(4)}_{Y,51} & = -\tfrac{1}{8} \,, &
	 \beta^{(4)}_{Y,52} & = -\tfrac{3}{8} \,, &
	 \beta^{(4)}_{Y,53} & = 0 \,, &
	 \beta^{(4)}_{Y,54} & = -\tfrac{\pi^2}{32} \,, &
	 \beta^{(4)}_{Y,55} & = 0 \,, \\
	 \beta^{(4)}_{Y,56} & = -\tfrac{\pi^2}{32} \,, &
	 \beta^{(4)}_{Y,57} & = -\tfrac{1}{2} \,, &
	 \beta^{(4)}_{Y,58} & = -\tfrac{1}{4} \,, &
	 \beta^{(4)}_{Y,59} & = -\tfrac{\pi^2}{32} \,, &
	 \beta^{(4)}_{Y,60} & = -\tfrac{\pi^2}{16} \,, \\
	 \beta^{(4)}_{Y,61} & = -\tfrac{\pi^2}{32} \,, &
	 \beta^{(4)}_{Y,62} & = -\tfrac{1}{8} \,, &
	 \beta^{(4)}_{Y,63} & = 0 \,, &
	 \beta^{(4)}_{Y,64} & = -\tfrac{1}{12} \,, &
	 \beta^{(4)}_{Y,65} & = -\tfrac{1}{8} \,, \\
	 \beta^{(4)}_{Y,66} & = -\tfrac{\pi^2}{32} \,, &
	 \beta^{(4)}_{Y,67} & = 0 \,, &
	 \beta^{(4)}_{Y,68} & = 0 \,, &
	 \beta^{(4)}_{Y,69} & = 0 \,, &
	 \beta^{(4)}_{Y,70} & = -\tfrac{\pi^2}{32} \,, \\
	 \beta^{(4)}_{Y,71} & = 0 \,, &
	 \beta^{(4)}_{Y,72} & = -\tfrac{1}{6} \,, &
	 \beta^{(4)}_{Y,73} & = -\tfrac{1}{8} \,, &
	 \beta^{(4)}_{Y,74} & = 0 \,, &
	 \beta^{(4)}_{Y,75} & = -\tfrac{\pi^2}{32} \,, \\
	 \beta^{(4)}_{Y,76} & = -\tfrac{\pi^2}{32} \,, &
	 \beta^{(4)}_{Y,77} & = \tfrac{1}{8} \,, &
	 \beta^{(4)}_{Y,78} & = 0 \,, &
	 \beta^{(4)}_{Y,79} & = \tfrac{1}{8} \,, &
	 \beta^{(4)}_{Y,80} & = -\tfrac{1}{24} \,, \\
	 \beta^{(4)}_{Y,81} & = -\tfrac{1}{4} \,, &
	 \beta^{(4)}_{Y,82} & = 0 \,, &
	 \beta^{(4)}_{Y,83} & = 0 \,, &
	 \beta^{(4)}_{Y,84} & = \tfrac{1}{8} \,, &
	 \beta^{(4)}_{Y,85} & = -\tfrac{1}{4} \,, \\
	 \beta^{(4)}_{Y,86} & = \tfrac{1}{2}-\tfrac{\pi^2}{16} \,, &
	 \beta^{(4)}_{Y,87} & = -\tfrac{1}{8} \,, &
	 \beta^{(4)}_{Y,88} & = -\tfrac{1}{8} \,, &
	 \beta^{(4)}_{Y,89} & = \tfrac{1}{8} \,, &
	 \beta^{(4)}_{Y,90} & = 0 \,, \\
	 \beta^{(4)}_{Y,91} & = 0 \,, &
	 \beta^{(4)}_{Y,92} & = 0 \,, &
	 \beta^{(4)}_{Y,93} & = -\tfrac{\pi^2}{16} \,, &
	 \beta^{(4)}_{Y,94} & = -\tfrac{1}{8} \,, &
	 \beta^{(4)}_{Y,95} & = -\tfrac{\pi^2}{32} \,, \\
	 \beta^{(4)}_{Y,96} & = -\tfrac{1}{4} \,, &
	 \beta^{(4)}_{Y,97} & = -\tfrac{3}{8} \,, &
	 \beta^{(4)}_{Y,98} & = -\tfrac{\pi^2}{32} \,, &
	 \beta^{(4)}_{Y,99} & = -\tfrac{1}{12} \,, &
	 \beta^{(4)}_{Y,100} & = 0 \,, \\
	 \beta^{(4)}_{Y,101} & = 0 \,, &
	 \beta^{(4)}_{Y,102} & = -\tfrac{\pi^2}{32} \,, &
	 \beta^{(4)}_{Y,103} & = 0 \,, &
	 \beta^{(4)}_{Y,104} & = -\tfrac{\pi^2}{32} \,, &
	 \beta^{(4)}_{Y,105} & = 0 \,, \\
	 \beta^{(4)}_{Y,106} & = 0 \,, &
	 \beta^{(4)}_{Y,107} & = 0 \,, &
	 \beta^{(4)}_{Y,108} & = -\tfrac{\pi^2}{32} \,, &
	 \beta^{(4)}_{Y,109} & = \tfrac{1}{2}-\tfrac{\pi^2}{16} \,, &
	 \beta^{(4)}_{Y,110} & = -\tfrac{\pi^2}{16} \,, \\
	 \beta^{(4)}_{Y,111} & = -\tfrac{1}{4} \,, &
	 \beta^{(4)}_{Y,112} & = \tfrac{1}{2}-\tfrac{\pi^2}{16} \,, &
	 \beta^{(4)}_{Y,113} & = 1-\tfrac{\pi^2}{6} \,, &
	 \beta^{(4)}_{Y,114} & = \tfrac{1}{2}-\tfrac{\pi^2}{16} \,, &
	 \beta^{(4)}_{Y,115} & = -\tfrac{1}{4} \,, \\
	 \beta^{(4)}_{Y,116} & = 0 \,, &
	 \beta^{(4)}_{Y,117} & = 0 \,, &
	 \beta^{(4)}_{Y,118} & = -\tfrac{1}{4} \,, &
	 \beta^{(4)}_{Y,119} & = 1-\tfrac{\pi^2}{6} \,, &
	 \beta^{(4)}_{Y,120} & = -1\,, \\
	 \beta^{(4)}_{Y,121} & = -\tfrac{\pi^2}{16} \,, &
	 \beta^{(4)}_{Y,122} & = -\tfrac{\pi^2}{32} \,, &
	 \beta^{(4)}_{Y,123} & = 0 \,, &
	 \beta^{(4)}_{Y,124} & = -\tfrac{\pi^2}{16} \,, &
	 \beta^{(4)}_{Y,125} & = 0 \,, \\
	 \beta^{(4)}_{Y,126} & = 0 \,, &
	 \beta^{(4)}_{Y,127} & = -\tfrac{1}{12} \,, &
	 \beta^{(4)}_{Y,128} & = -\tfrac{1}{4} \,, &
	 \beta^{(4)}_{Y,129} & = \tfrac{1}{4} \,, &
	 \beta^{(4)}_{Y,130} & = 1-\tfrac{\pi^2}{8} \,, \\
	 \beta^{(4)}_{Y,131} & = \tfrac{1}{4} \,, &
	 \beta^{(4)}_{Y,132} & = -\tfrac{\pi^2}{16} \,.
     \tag{\stepcounter{equation}\theequation}
\end{align*}

Especially noteworthy is the vanishing of the coefficients
$\beta^{(4)}_{Y,36}$--$\beta^{(4)}_{Y,43}$, which are the ones featuring traces over an 
odd number of $\gamma$ matrices.
Resolving such traces via relations such as~\eq{3-gamma} generates Levi--Civita tensors, which are only well-defined in three, but not $(3-2\varepsilon)$ spacetime dimensions.
In this calculation, we have employed the Breitenlohner-Maison-'t Hooft-Veltman scheme (BMHV)~\cite{tHooft:1972tcz,Breitenlohner:1977hr}, which separates all Lorentz indices into $3$- and $(-2\varepsilon)$-dimensional ranges and treats the Levi--Civita tensor as manifestly three-dimensional.
While the BMHV scheme enables a consistent application of dimensional regularisation, it potentially introduces complications such as the need for evanescent counterterms, see e.g.~\cite{Belusca-Maito:2023wah} for a review. However, as all terms $\beta^{(4)}_{Y,36}$--$\beta^{(4)}_{Y,43}$ vanish, no such phenomena arise when computing the RGEs.
In fact, the four-loop Yukawa vertex corrections are the only place where such traces occur in this work. Thus, all other RGEs are compatible with a na\"ive treatment of dimensional regularisation.  Overall though, our results are derived in the consistent BMHV scheme without being affected from its technical complications.

\subsection{Sextic Vertex}

Finally, we turn towards the scalar sextic vertex corrections. The two-loop expression 
\begin{align}
	\hat{\beta}_\eta^{(2) abcdef} &= \beta_{\eta,1}^{(2)}\,\mathcal{S}_{10}\,\eta^{abcghi}\eta^{defghi} + \beta_{\eta,2}^{(2)}\,\mathcal{S}_{15}\,\tr{Y^{abgh}} \eta^{cdefgh} \notag\\
    &\phantom{=\,} + \beta_{\eta,3}^{(2)}\,\mathcal{S}_{15}\,\tr{Y^{agbh}} \eta^{cdefgh} + \beta_{\eta,4}^{(2)}\,\mathcal{S}_{90}\,\tr{Y^{abcdegfg}} \notag\\
    &\phantom{=\,} + \beta_{\eta,5}^{(2)}\,\mathcal{S}_{45}\,\tr{Y^{abcgdefg}}
\end{align}
with the explicit coefficients
\begin{align*}
     \beta^{(2)}_{\eta,1} & = \tfrac{1}{6} \,, &
	 \beta^{(2)}_{\eta,2} & = 0 \,, &
	 \beta^{(2)}_{\eta,3} & = \tfrac{1}{2} \,, &
	 \beta^{(2)}_{\eta,4} & = -1\,, &
	 \beta^{(2)}_{\eta,5} & = -2 \,. 
     \tag{\stepcounter{equation}\theequation}
\end{align*}
The four-loop corrections consist of several terms 
\begin{align}
	\hat{\beta}_\eta^{(4) abcdef} &= 
    \hat{\beta}_{\eta,\eta^3}^{(4) abcdef} 
    + \hat{\beta}_{\eta,\eta^3}^{(4) abcdef} 
    + \hat{\beta}_{\eta,\eta^2 Y^2}^{(4) abcdef}
    + \hat{\beta}_{\eta,\eta  Y^2 Y^2}^{(4) abcdef}
    + \hat{\beta}_{\eta,\eta  Y^4}^{(4) abcdef} \notag \\
    &\phantom{=\,} 
    + \hat{\beta}_{\eta,Y^2 Y^2 Y^2}^{(4) abcdef}
    + \hat{\beta}_{\eta,Y^4 Y^2}^{(4) abcdef}
    + \hat{\beta}_{\eta,Y^6}^{(4) abcdef}\,,
\end{align}
which read 
\begin{align}
	\hat{\beta}_{\eta,\eta^3}^{(4) abcdef} &= 
       \beta_{\eta,1}^{(4)}\,\mathcal{S}_{15}\,\eta^{abghij}\eta^{ghijkl} \eta^{cdefkl} 
    + \beta_{\eta,2}^{(4)}\,\mathcal{S}_{15}\,\eta^{aghijk}\eta^{bhijkl} \eta^{cdefkg}\notag\\
    &\phantom{=\,} 
    + \beta_{\eta,3}^{(4)}\,\mathcal{S}_{10}\,\eta^{abcghi}\eta^{ghijkl} \eta^{defjkl}
    + \beta_{\eta,4}^{(4)}\,\mathcal{S}_{60}\,\eta^{aghijk}\eta^{bcghil} \eta^{defjkl} \notag\\
    &\phantom{=\,} 
    + \beta_{\eta,5}^{(4)}\,\mathcal{S}_{15}\,\eta^{abghij}\eta^{cdijkl} \eta^{efklgh} \,,
\end{align}

\begin{align}
	\hat{\beta}_{\eta,\eta^2 Y^2}^{(4) abcdef} &= 
       \beta_{\eta,6}^{(4)}\,\mathcal{S}_{10}\,\eta^{abcghk}\eta^{defijl} \tr{Y^{ghij}}
     + \beta_{\eta,7}^{(4)}\,\mathcal{S}_{10}\,\eta^{abcghk}\eta^{defijl} \tr{Y^{gihj}} \notag\\
    &\phantom{=\,} 
    + \beta_{\eta,8}^{(4)}\,\mathcal{S}_{10}\,\eta^{abcgij}\eta^{defhij} \tr{Y^{gkhk}}
    + \beta_{\eta,9}^{(4)}\,\mathcal{S}_{15}\,\eta^{abghij}\eta^{cdefgh} \tr{Y^{ikjk}}  \notag\\
    &\phantom{=\,} 
    + \beta_{\eta,10}^{(4)}\,\mathcal{S}_{15}\,\eta^{abghij}\eta^{cdefgk} \tr{Y^{khij}}
    + \beta_{\eta,11}^{(4)}\,\mathcal{S}_{30}\,\eta^{abcdgh}\eta^{eghijk} \tr{Y^{eijk}}\notag\\
    &\phantom{=\,} 
    + \beta_{\eta,12}^{(4)}\,\mathcal{S}_{60}\,\eta^{abcghi}\eta^{deghjk} \tr{Y^{fijk}}
    + \beta_{\eta,13}^{(4)}\,\mathcal{S}_{60}\,\eta^{abcghi}\eta^{deghjk} \tr{Y^{fkij}}\notag\\
    &\phantom{=\,} 
    + \beta_{\eta,14}^{(4)}\,\mathcal{S}_{60}\,\eta^{abcghi}\eta^{dghijk} \tr{Y^{efjk}}
    + \beta_{\eta,15}^{(4)}\,\mathcal{S}_{60}\,\eta^{abcghi}\eta^{dghijk} \tr{Y^{ejfk}} \notag\\
    &\phantom{=\,} 
    + \beta_{\eta,16}^{(4)}\,\mathcal{S}_{45}\,\eta^{abgijk}\eta^{cdhijk} \tr{Y^{efgh}}
    + \beta_{\eta,17}^{(4)}\,\mathcal{S}_{90}\,\eta^{abgijk}\eta^{cdhijk} \tr{Y^{egfh}} \,,
\end{align}

\begin{align}
	\hat{\beta}_{\eta,\eta  Y^2 Y^2}^{(4) abcdef} &= 
       \beta_{\eta,18}^{(4)}\,\mathcal{S}_{45}\,\eta^{abghij}\tr{Y^{cdgh}} \tr{Y^{efij}}
     + \beta_{\eta,19}^{(4)}\,\mathcal{S}_{45}\,\eta^{abcghk}\tr{Y^{cgdh}} \tr{Y^{eifj}}
     \notag\\ &\phantom{=\,} 
     + \beta_{\eta,20}^{(4)}\,\mathcal{S}_{90}\,\eta^{abcghk}\tr{Y^{cdgh}} \tr{Y^{eifj}}
     + \beta_{\eta,21}^{(4)}\,\mathcal{S}_{60}\,\eta^{abcghi}\tr{Y^{deij}} \tr{Y^{fghj}}
     \notag\\ &\phantom{=\,}
     + \beta_{\eta,22}^{(4)}\,\mathcal{S}_{60}\,\eta^{abcghi}\tr{Y^{deij}} \tr{Y^{fjgh}}
     + \beta_{\eta,24}^{(4)}\,\mathcal{S}_{120}\,\eta^{abcghi}\tr{Y^{diej}} \tr{Y^{fghj}}
     \notag\\ &\phantom{=\,}
     + \beta_{\eta,24}^{(4)}\,\mathcal{S}_{120}\,\eta^{abcghi}\tr{Y^{diej}} \tr{Y^{fjgh}}
     + \beta_{\eta,25}^{(4)}\,\mathcal{S}_{15}\,\eta^{abcdgh}\tr{Y^{ghij}} \tr{Y^{efij}}
     \notag\\ &\phantom{=\,}
     + \beta_{\eta,26}^{(4)}\,\mathcal{S}_{15}\,\eta^{abcdgh}\tr{Y^{gihj}} \tr{Y^{efij}}
     + \beta_{\eta,27}^{(4)}\,\mathcal{S}_{15}\,\eta^{abcdgh}\tr{Y^{giij}} \tr{Y^{efhj}}
     \notag\\ &\phantom{=\,}
     + \beta_{\eta,28}^{(4)}\,\mathcal{S}_{15}\,\eta^{abcdgh}\tr{Y^{ghij}} \tr{Y^{eifj}}
     + \beta_{\eta,29}^{(4)}\,\mathcal{S}_{15}\,\eta^{abcdgh}\tr{Y^{gihj}} \tr{Y^{eifj}}
     \notag\\ &\phantom{=\,}
     + \beta_{\eta,30}^{(4)}\,\mathcal{S}_{30}\,\eta^{abcdgh}\tr{Y^{giij}} \tr{Y^{ehfj}}
     + \beta_{\eta,31}^{(4)}\,\mathcal{S}_{15}\,\eta^{abcdgh}\tr{Y^{eijg}} \tr{Y^{fijh}}
     \notag\\ &\phantom{=\,}
     + \beta_{\eta,32}^{(4)}\,\mathcal{S}_{15}\,\eta^{abcdgh}\tr{Y^{eijg}} \tr{Y^{fjih}}
     + \beta_{\eta,33}^{(4)}\,\mathcal{S}_{15}\,\eta^{abcdgh}\tr{Y^{egij}} \tr{Y^{fhij}}
     \notag\\ &\phantom{=\,}
     + \beta_{\eta,34}^{(4)}\,\mathcal{S}_{30}\,\eta^{abcdgh}\tr{Y^{eijg}} \tr{Y^{fhij}}\,,
\end{align}

\begin{align}
	\hat{\beta}_{\eta,\eta  Y^4}^{(4) abcdef} &=  
      \beta_{\eta,35}^{(4)}\,\mathcal{S}_{45}\,\eta^{abghij} \tr{Y^{cdefghij}}
    + \beta_{\eta,36}^{(4)}\,\mathcal{S}_{45}\,\eta^{abghij} \tr{Y^{cdghefij}}
    \notag\\ &\phantom{=\,}
    + \beta_{\eta,37}^{(4)}\,\mathcal{S}_{90}\,\eta^{abghij} \tr{Y^{cdegfhij}}
    + \beta_{\eta,38}^{(4)}\,\mathcal{S}_{180}\,\eta^{abghij} \tr{Y^{cgdhefij}}
    \notag\\ &\phantom{=\,}
    + \beta_{\eta,39}^{(4)}\,\mathcal{S}_{45}\,\eta^{abghij} \tr{Y^{cgdheifj}}
    + \beta_{\eta,40}^{(4)}\,\mathcal{S}_{60}\,\eta^{abcghi} \tr{Y^{defghjij}}
    \notag\\ &\phantom{=\,}
    + \beta_{\eta,41}^{(4)}\,\mathcal{S}_{60}\,\eta^{abcghi} \tr{Y^{degjfhij}}
    + \beta_{\eta,42}^{(4)}\,\mathcal{S}_{60}\,\eta^{abcghi} \tr{Y^{defjghij}}
    \notag\\ &\phantom{=\,}
    + \beta_{\eta,43}^{(4)}\,\mathcal{S}_{60}\,\eta^{abcghi} \tr{Y^{defjgjhi}}
    + \beta_{\eta,44}^{(4)}\,\mathcal{S}_{60}\,\eta^{abcghi} \tr{Y^{degjfjhi}}
    \notag\\ &\phantom{=\,}
    + \beta_{\eta,45}^{(4)}\,\mathcal{S}_{60} \,\eta^{abcghi} \tr{Y^{gjdhejfi}}
    + \beta_{\eta,46}^{(4)}\,\mathcal{S}_{120}\,\eta^{abcghi} \tr{Y^{gjdjehfi}}
    \notag\\ &\phantom{=\,}
    + \beta_{\eta,47}^{(4)}\,\mathcal{S}_{60} \,\eta^{abcghi} \tr{Y^{ghdjeifj}}
    + \beta_{\eta,48}^{(4)}\,\mathcal{S}_{120}\,\eta^{abcghi} \tr{Y^{ghdiejfj}}
    \notag\\ &\phantom{=\,}
    + \beta_{\eta,49}^{(4)}\,\mathcal{S}_{15} \,\eta^{abcdgh} \tr{Y^{efghijij}}
    + \beta_{\eta,50}^{(4)}\,\mathcal{S}_{15} \,\eta^{abcdgh} \tr{Y^{efijghij}}
    \notag\\ &\phantom{=\,}
    + \beta_{\eta,51}^{(4)}\,\mathcal{S}_{15} \,\eta^{abcdgh} \tr{Y^{efgihjij}}
    + \beta_{\eta,52}^{(4)}\,\mathcal{S}_{15} \,\eta^{abcdgh} \tr{Y^{efgiijhj}}
    \notag\\ &\phantom{=\,}
    + \beta_{\eta,53}^{(4)}\,\mathcal{S}_{15} \,\eta^{abcdgh} \tr{Y^{egijfhij}}
    + \beta_{\eta,54}^{(4)}\,\mathcal{S}_{15} \,\eta^{abcdgh} \tr{Y^{eigjfihj}}
    \notag\\ &\phantom{=\,}
    + \beta_{\eta,55}^{(4)}\,\mathcal{S}_{15} \,\eta^{abcdgh} \tr{Y^{eiijfjgh}}
    + \beta_{\eta,56}^{(4)}\,\mathcal{S}_{15} \,\eta^{abcdgh} \tr{Y^{eigifjhj}}
    \notag\\ &\phantom{=\,}
    + \beta_{\eta,57}^{(4)}\,\mathcal{S}_{30} \,\eta^{abcdgh} \tr{Y^{eghifjij}}
    + \beta_{\eta,58}^{(4)}\,\mathcal{S}_{15} \,\eta^{abcdgh} \tr{Y^{egfhijij}}
    \notag\\ &\phantom{=\,}
    + \beta_{\eta,59}^{(4)}\,\mathcal{S}_{15} \,\eta^{abcdgh} \tr{Y^{eifigjhj}}
    + \beta_{\eta,60}^{(4)}\,\mathcal{S}_{15} \,\eta^{abcdgh} \tr{Y^{eifjgihj}}
    \notag\\ &\phantom{=\,}
    + \beta_{\eta,61}^{(4)}\,\mathcal{S}_{15} \,\eta^{abcdgh} \tr{Y^{eifjgjhi}}
    + \beta_{\eta,62}^{(4)}\,\mathcal{S}_{30} \,\eta^{abcdgh} \tr{Y^{eifjgjhi}}
    \notag\\ &\phantom{=\,}
    + \beta_{\eta,63}^{(4)}\,\mathcal{S}_{30} \,\eta^{abcdgh} \tr{Y^{egfihjij}}
    + \beta_{\eta,64}^{(4)}\,\mathcal{S}_{30} \,\eta^{abcdgh} \tr{Y^{egfiijhj}}\,,
\end{align}

\begin{align}
	\hat{\beta}_{\eta,Y^2 Y^2 Y^2}^{(4) abcdef} &=  
      \beta_{\eta,65}^{(4)}\,\mathcal{S}_{15}\,\tr{Y^{abgh}} \tr{Y^{cdhi}} \tr{Y^{efgi}}
      \notag\\ &\phantom{=\,}
      + \beta_{\eta,66}^{(4)}\,\mathcal{S}_{90}\,\tr{Y^{agbh}} \tr{Y^{cdhi}} \tr{Y^{efgi}}
      \notag\\ &\phantom{=\,}
      + \beta_{\eta,67}^{(4)}\,\mathcal{S}_{180}\,\tr{Y^{abgh}} \tr{Y^{chdi}} \tr{Y^{egfi}}
      \notag\\ &\phantom{=\,}
      + \beta_{\eta,68}^{(4)}\,\mathcal{S}_{120}\,\tr{Y^{agbh}} \tr{Y^{chdi}} \tr{Y^{egfi}}\,,
\end{align}

\begin{align}
	\hat{\beta}_{\eta,Y^4 Y^2}^{(4) abcdef} &=  
      \beta_{\eta,69}^{(4)}\,\mathcal{S}_{45}\,\tr{Y^{abgh}} \tr{Y^{cdgiefhi}} 
      + \beta_{\eta,70}^{(4)}\,\mathcal{S}_{45}\,\tr{Y^{abgh}} \tr{Y^{cdefgihi}}
      \notag\\ &\phantom{=\,}
      + \beta_{\eta,71}^{(4)}\,\mathcal{S}_{45}\,\tr{Y^{agbh}} \tr{Y^{cdgiefhi}} 
      + \beta_{\eta,72}^{(4)}\,\mathcal{S}_{90}\,\tr{Y^{agbh}} \tr{Y^{cdefgihi}}
      \notag\\ &\phantom{=\,}
      + \beta_{\eta,73}^{(4)}\,\mathcal{S}_{90} \,\tr{Y^{abgh}} \tr{Y^{cdeighfi}}
      + \beta_{\eta,74}^{(4)}\,\mathcal{S}_{180}\,\tr{Y^{abgh}} \tr{Y^{cdeigifh}}
      \notag\\ &\phantom{=\,}
      + \beta_{\eta,75}^{(4)}\,\mathcal{S}_{180}\,\tr{Y^{abgh}} \tr{Y^{cdeifigh}}
      + \beta_{\eta,76}^{(4)}\,\mathcal{S}_{180}\,\tr{Y^{abgh}} \tr{Y^{cdeifghi}}
      \notag\\ &\phantom{=\,}
      + \beta_{\eta,77}^{(4)}\,\mathcal{S}_{180}\,\tr{Y^{abgh}} \tr{Y^{cdegfihi}}
      + \beta_{\eta,78}^{(4)}\,\mathcal{S}_{90}\,\tr{Y^{agbh}} \tr{Y^{cdeighfi}}
      \notag\\ &\phantom{=\,}
      + \beta_{\eta,79}^{(4)}\,\mathcal{S}_{360}\,\tr{Y^{agbh}} \tr{Y^{cdeghifi}}
      + \beta_{\eta,80}^{(4)}\,\mathcal{S}_{180}\,\tr{Y^{agbh}} \tr{Y^{cdeifigh}}
      \notag\\ &\phantom{=\,}
      + \beta_{\eta,81}^{(4)}\,\mathcal{S}_{360}\,\tr{Y^{agbh}} \tr{Y^{cdeifghi}}
      + \beta_{\eta,82}^{(4)}\,\mathcal{S}_{360}\,\tr{Y^{agbh}} \tr{Y^{cdegfihi}}
      \notag\\ &\phantom{=\,}
      + \beta_{\eta,83}^{(4)}\,\mathcal{S}_{90}\,\tr{Y^{abgh}} \tr{Y^{cgdiehfi}}
      + \beta_{\eta,84}^{(4)}\,\mathcal{S}_{180}\,\tr{Y^{abgh}} \tr{Y^{cgdheifi}}
      \notag\\ &\phantom{=\,}
      + \beta_{\eta,85}^{(4)}\,\mathcal{S}_{180}\,\tr{Y^{agbh}} \tr{Y^{cgdiehfi}}
      + \beta_{\eta,86}^{(4)}\,\mathcal{S}_{360}\,\tr{Y^{agbh}} \tr{Y^{cgdheifi}}
      \notag\\ &\phantom{=\,}
      + \beta_{\eta,87}^{(4)}\,\mathcal{S}_{180}\,\tr{Y^{aghi}} \tr{Y^{bcdhegfi}}
      + \beta_{\eta,88}^{(4)}\,\mathcal{S}_{360}\,\tr{Y^{aghi}} \tr{Y^{bcdgehfi}}
      \notag\\ &\phantom{=\,}
      + \beta_{\eta,89}^{(4)}\,\mathcal{S}_{90} \,\tr{Y^{aghi}} \tr{Y^{bcdgefhi}}
      + \beta_{\eta,90}^{(4)}\,\mathcal{S}_{180}\,\tr{Y^{aghi}} \tr{Y^{bcefdghi}}
      \notag\\ &\phantom{=\,}
      + \beta_{\eta,91}^{(4)}\,\mathcal{S}_{180} \,\tr{Y^{aghi}} \tr{Y^{bcdhefgi}}
      + \beta_{\eta,92}^{(4)}\,\mathcal{S}_{180}\,\tr{Y^{aghi}} \tr{Y^{bcefdhgi}}
      \notag\\ &\phantom{=\,}
      + \beta_{\eta,93}^{(4)}\,\mathcal{S}_{45}\,\tr{Y^{gihi}} \tr{Y^{abcgdefh}}
      + \beta_{\eta,94}^{(4)}\,\mathcal{S}_{90}\,\tr{Y^{gihi}} \tr{Y^{abcdegfh}}
      \notag\\ &\phantom{=\,}
      + \beta_{\eta,95}^{(4)}\,\mathcal{S}_{45}\,\tr{Y^{gihi}} \tr{Y^{abcdefgh}}\,,
\end{align}

\begin{align}
	\hat{\beta}_{\eta,Y^6}^{(4) abcdef} &=  
       \beta_{\eta,96}^{(4)}\,\mathcal{S}_{45}\,\tr{Y^{abcdefghhiig}} 
     + \beta_{\eta,97}^{(4)}\,\mathcal{S}_{90}\,\tr{Y^{abcdghefhiig}} 
     \notag\\ &\phantom{=\,}
     + \beta_{\eta,98}^{(4)}\,\mathcal{S}_{15}\,\tr{Y^{abghcdhiefig}} 
     + \beta_{\eta,99}^{(4)}\,\mathcal{S}_{180}\,\tr{Y^{abcdegfhgihi}} 
     \notag\\ &\phantom{=\,}
     + \beta_{\eta,100}^{(4)}\,\mathcal{S}_{180}\,\tr{Y^{abcdegfhhigi}}
     + \beta_{\eta,101}^{(4)}\,\mathcal{S}_{180}\,\tr{Y^{abcdegfghihi}}
     \notag\\ &\phantom{=\,}
     + \beta_{\eta,102}^{(4)}\,\mathcal{S}_{180}\,\tr{Y^{abcdeggifhhi}}
     + \beta_{\eta,103}^{(4)}\,\mathcal{S}_{180}\,\tr{Y^{abcdeghifhgi}}
     \notag\\ &\phantom{=\,}
     + \beta_{\eta,104}^{(4)}\,\mathcal{S}_{180}\,\tr{Y^{abcdeghifghi}}
     + \beta_{\eta,105}^{(4)}\,\mathcal{S}_{90} \,\tr{Y^{abcdeggihifh}}
     \notag\\ &\phantom{=\,}
     + \beta_{\eta,106}^{(4)}\,\mathcal{S}_{90} \,\tr{Y^{abcdeghigifh}}
     + \beta_{\eta,107}^{(4)}\,\mathcal{S}_{90} \,\tr{Y^{abcdeghihifg}}
     \notag\\ &\phantom{=\,}
     + \beta_{\eta,108}^{(4)}\,\mathcal{S}_{90} \,\tr{Y^{abcdgiegfhhi}}
     + \beta_{\eta,109}^{(4)}\,\mathcal{S}_{90} \,\tr{Y^{abcdgiehfghi}}
     \notag\\ &\phantom{=\,}
     + \beta_{\eta,110}^{(4)}\,\mathcal{S}_{90} \,\tr{Y^{abcdhiegfghi}}
     + \beta_{\eta,111}^{(4)}\,\mathcal{S}_{180}\,\tr{Y^{abcgdefhhigi}}
     \notag\\ &\phantom{=\,}
     + \beta_{\eta,112}^{(4)}\,\mathcal{S}_{180}\,\tr{Y^{abcgdefhgihi}}
     + \beta_{\eta,113}^{(4)}\,\mathcal{S}_{180}\,\tr{Y^{abcgdefghihi}}
     \notag\\ &\phantom{=\,}
     + \beta_{\eta,114}^{(4)}\,\mathcal{S}_{90} \,\tr{Y^{abcgdehifghi}}
     + \beta_{\eta,115}^{(4)}\,\mathcal{S}_{180}\,\tr{Y^{abcgdegifhhi}}
     \notag\\ &\phantom{=\,}
     + \beta_{\eta,116}^{(4)}\,\mathcal{S}_{180}\,\tr{Y^{abghcdeifigh}}
     + \beta_{\eta,117}^{(4)}\,\mathcal{S}_{180}\,\tr{Y^{abghcdegfihi}}
     \notag\\ &\phantom{=\,}
     + \beta_{\eta,118}^{(4)}\,\mathcal{S}_{180}\,\tr{Y^{abghcdeifghi}}
     + \beta_{\eta,119}^{(4)}\,\mathcal{S}_{90}\, \tr{Y^{abghcdeighfi}}
     \notag\\ &\phantom{=\,}
     + \beta_{\eta,120}^{(4)}\,\mathcal{S}_{180}\,\tr{Y^{abghcdeghifi}}
     + \beta_{\eta,121}^{(4)}\,\mathcal{S}_{90} \,\tr{Y^{abcgdgefhihi}}
     \notag\\ &\phantom{=\,}
     + \beta_{\eta,122}^{(4)}\,\mathcal{S}_{90} \,\tr{Y^{abcgdhefgihi}}
     + \beta_{\eta,123}^{(4)}\,\mathcal{S}_{90} \,\tr{Y^{abcgdhefhigi}}
     \notag\\ &\phantom{=\,}
     + \beta_{\eta,124}^{(4)}\,\mathcal{S}_{90} \,\tr{Y^{abghcidefigh}}
     + \beta_{\eta,125}^{(4)}\,\mathcal{S}_{90} \,\tr{Y^{abghcgdefihi}}
     \notag\\ &\phantom{=\,}
     + \beta_{\eta,126}^{(4)}\,\mathcal{S}_{90} \,\tr{Y^{abghcidefghi}}
     + \beta_{\eta,127}^{(4)}\,\mathcal{S}_{90} \,\tr{Y^{abghcideghfi}}
     \notag\\ &\phantom{=\,}
     + \beta_{\eta,128}^{(4)}\,\mathcal{S}_{90} \,\tr{Y^{abghcgdehifi}}
     + \beta_{\eta,129}^{(4)}\,\mathcal{S}_{90} \,\tr{Y^{abghcidegifh}}
     \notag\\ &\phantom{=\,}
     + \beta_{\eta,130}^{(4)}\,\mathcal{S}_{360}\,\tr{Y^{abghcgdheifi}}
     + \beta_{\eta,131}^{(4)}\,\mathcal{S}_{360}\,\tr{Y^{abghcgdiehfi}}
     \notag\\ &\phantom{=\,}
     + \beta_{\eta,132}^{(4)}\,\mathcal{S}_{360}\,\tr{Y^{abghcidgehfi}}
     + \beta_{\eta,133}^{(4)}\,\mathcal{S}_{360}\,\tr{Y^{abghcgdieifh}}
     \notag\\ &\phantom{=\,}
     + \beta_{\eta,134}^{(4)}\,\mathcal{S}_{360}\,\tr{Y^{abghcidgeifh}}
     + \beta_{\eta,135}^{(4)}\,\mathcal{S}_{360}\,\tr{Y^{abghcidiegfh}}
     \notag\\ &\phantom{=\,}
     + \beta_{\eta,136}^{(4)}\,\mathcal{S}_{360}\,\tr{Y^{abcgghdheifi}}
     + \beta_{\eta,137}^{(4)}\,\mathcal{S}_{360}\,\tr{Y^{abcgghdiehfi}}
     \notag\\ &\phantom{=\,}
     + \beta_{\eta,138}^{(4)}\,\mathcal{S}_{360}\,\tr{Y^{abcgghdieifh}}
     + \beta_{\eta,139}^{(4)}\,\mathcal{S}_{360}\,\tr{Y^{abcghidgehfi}}
     \notag\\ &\phantom{=\,}
     + \beta_{\eta,140}^{(4)}\,\mathcal{S}_{360}\,\tr{Y^{abcghidhegfi}}
     + \beta_{\eta,141}^{(4)}\,\mathcal{S}_{360}\,\tr{Y^{abcghidheifg}}
     \notag\\ &\phantom{=\,}
     + \beta_{\eta,142}^{(4)}\,\mathcal{S}_{180}\,\tr{Y^{abcgdhgiehfi}}
     + \beta_{\eta,143}^{(4)}\,\mathcal{S}_{180}\,\tr{Y^{abcgdhhieifg}}
     \notag\\ &\phantom{=\,}
     + \beta_{\eta,144}^{(4)}\,\mathcal{S}_{360}\,\tr{Y^{abcgdhgieifh}}
     + \beta_{\eta,145}^{(4)}\,\mathcal{S}_{360}\,\tr{Y^{abcgdhgheifi}}
     \notag\\ &\phantom{=\,}
     + \beta_{\eta,146}^{(4)}\,\mathcal{S}_{120}\,\tr{Y^{agbgchdheifi}}
     + \beta_{\eta,147}^{(4)}\,\mathcal{S}_{360}\,\tr{Y^{agbgchdieifh}}
     \notag\\ &\phantom{=\,}
     + \beta_{\eta,148}^{(4)}\,\mathcal{S}_{360}\,\tr{Y^{agbgchdiehfi}}
     + \beta_{\eta,149}^{(4)}\,\mathcal{S}_{180}\,\tr{Y^{agbhcgdiehfi}}
     \notag\\ &\phantom{=\,}
     + \beta_{\eta,150}^{(4)}\,\mathcal{S}_{60}\,\tr{Y^{agbhcidgehfi}}\,.
\end{align}

The $150$ coefficients contributing to the four-loop sextic vertex correction are found to be 
\begin{align*}
     \beta^{(4)}_{\eta,1} & = 0 \,, &
	 \beta^{(4)}_{\eta,2} & = \tfrac{1}{48} \,, &
	 \beta^{(4)}_{\eta,3} & = 0 \,, &
	 \beta^{(4)}_{\eta,4} & = -\tfrac{1}{12} \,, &
	 \beta^{(4)}_{\eta,5} & = -\tfrac{\pi^2}{32} \,, \\
	 \beta^{(4)}_{\eta,6} & = -\tfrac{\pi^2}{32} \,, &
	 \beta^{(4)}_{\eta,7} & = 0 \,, &
	 \beta^{(4)}_{\eta,8} & = -\tfrac{1}{12} \,, &
	 \beta^{(4)}_{\eta,9} & = 0 \,, &
	 \beta^{(4)}_{\eta,10} & = 0 \,, \\
	 \beta^{(4)}_{\eta,11} & = 0 \,, &
	 \beta^{(4)}_{\eta,12} & = -\tfrac{\pi^2}{32} \,, &
	 \beta^{(4)}_{\eta,13} & = -\tfrac{1}{4} \,, &
	 \beta^{(4)}_{\eta,14} & = 0 \,, &
	 \beta^{(4)}_{\eta,15} & = 0 \,, \\
	 \beta^{(4)}_{\eta,16} & = 0 \,, &
	 \beta^{(4)}_{\eta,17} & = -\tfrac{1}{12} \,, &
	 \beta^{(4)}_{\eta,18} & = 0 \,, &
	 \beta^{(4)}_{\eta,19} & = 0 \,, &
	 \beta^{(4)}_{\eta,20} & = 0 \,, \\
	 \beta^{(4)}_{\eta,21} & = 0 \,, &
	 \beta^{(4)}_{\eta,22} & = 0 \,, &
	 \beta^{(4)}_{\eta,23} & = -\tfrac{1}{4} \,, &
	 \beta^{(4)}_{\eta,24} & = -\tfrac{\pi^2}{32} \,, &
	 \beta^{(4)}_{\eta,25} & = 0 \,, \\
	 \beta^{(4)}_{\eta,26} & = 0 \,, &
	 \beta^{(4)}_{\eta,27} & = 0 \,, &
	 \beta^{(4)}_{\eta,28} & = 0 \,, &
	 \beta^{(4)}_{\eta,29} & = \tfrac{1}{8} \,, &
	 \beta^{(4)}_{\eta,30} & = -\tfrac{1}{12} \,, \\
	 \beta^{(4)}_{\eta,31} & = 0 \,, &
	 \beta^{(4)}_{\eta,32} & = \tfrac{1}{8} \,, &
	 \beta^{(4)}_{\eta,33} & = -\tfrac{\pi^2}{32} \,, &
	 \beta^{(4)}_{\eta,34} & = -\tfrac{1}{4} \,, &
	 \beta^{(4)}_{\eta,35} & = \tfrac{\pi^2}{16} \,, \\
	 \beta^{(4)}_{\eta,36} & = \tfrac{\pi^2}{16} \,, &
	 \beta^{(4)}_{\eta,37} & = 1\,, &
	 \beta^{(4)}_{\eta,38} & = \tfrac{\pi^2}{8} \,, &
	 \beta^{(4)}_{\eta,39} & = \tfrac{\pi^2}{6} \,, &
	 \beta^{(4)}_{\eta,40} & = 0 \,, \\
	 \beta^{(4)}_{\eta,41} & = 1-\tfrac{\pi^2}{8} \,, &
	 \beta^{(4)}_{\eta,42} & = 0 \,, &
	 \beta^{(4)}_{\eta,43} & = 0 \,, &
	 \beta^{(4)}_{\eta,44} & = \tfrac{1}{2} \,, &
	 \beta^{(4)}_{\eta,45} & = 2-\tfrac{\pi^2}{3} \,, \\
	 \beta^{(4)}_{\eta,46} & = 1-\tfrac{\pi^2}{8} \,, &
	 \beta^{(4)}_{\eta,47} & = 1-\tfrac{\pi^2}{8} \,, &
	 \beta^{(4)}_{\eta,48} & = 0 \,, &
	 \beta^{(4)}_{\eta,49} & = 0 \,, &
	 \beta^{(4)}_{\eta,50} & = 0 \,, \\
	 \beta^{(4)}_{\eta,51} & = -\tfrac{\pi^2}{8} \,, &
	 \beta^{(4)}_{\eta,52} & = 0 \,, &
	 \beta^{(4)}_{\eta,53} & = -\tfrac{1}{4} \,, &
	 \beta^{(4)}_{\eta,54} & = -\tfrac{1}{2} \,, &
	 \beta^{(4)}_{\eta,55} & = 0 \,, \\
	 \beta^{(4)}_{\eta,56} & = -\tfrac{1}{4} \,, &
	 \beta^{(4)}_{\eta,57} & = -\tfrac{3}{4} \,, &
	 \beta^{(4)}_{\eta,58} & = -\tfrac{1}{4} \,, &
	 \beta^{(4)}_{\eta,59} & = -\tfrac{1}{2} \,, &
	 \beta^{(4)}_{\eta,60} & = -\tfrac{\pi^2}{8} \,, \\
	 \beta^{(4)}_{\eta,61} & = -\tfrac{\pi^2}{16} \,, &
	 \beta^{(4)}_{\eta,62} & = 0 \,, &
	 \beta^{(4)}_{\eta,63} & = -\tfrac{3}{4} \,, &
	 \beta^{(4)}_{\eta,64} & = -\tfrac{\pi^2}{16} \,, &
	 \beta^{(4)}_{\eta,65} & = 0 \,, \\
	 \beta^{(4)}_{\eta,66} & = 0 \,, &
	 \beta^{(4)}_{\eta,67} & = 0 \,, &
	 \beta^{(4)}_{\eta,68} & = -\tfrac{\pi^2}{32} \,, &
	 \beta^{(4)}_{\eta,69} & = 0 \,, &
	 \beta^{(4)}_{\eta,70} & = 0 \,, \\
	 \beta^{(4)}_{\eta,71} & = 1\,, &
	 \beta^{(4)}_{\eta,72} & = \tfrac{1}{4} \,, &
	 \beta^{(4)}_{\eta,73} & = 0 \,, &
	 \beta^{(4)}_{\eta,74} & = 0 \,, &
	 \beta^{(4)}_{\eta,75} & = 0 \,, \\
	 \beta^{(4)}_{\eta,76} & = 0 \,, &
	 \beta^{(4)}_{\eta,77} & = 0 \,, &
	 \beta^{(4)}_{\eta,78} & = 0 \,, &
	 \beta^{(4)}_{\eta,79} & = 0 \,, &
	 \beta^{(4)}_{\eta,80} & = 0 \,, \\
	 \beta^{(4)}_{\eta,81} & = 0 \,, &
	 \beta^{(4)}_{\eta,82} & = \tfrac{1}{2} \,, &
	 \beta^{(4)}_{\eta,83} & = 0 \,, &
	 \beta^{(4)}_{\eta,84} & = 0 \,, &
	 \beta^{(4)}_{\eta,85} & = 1-\tfrac{\pi^2}{8} \,, \\
	 \beta^{(4)}_{\eta,86} & = 0 \,, &
	 \beta^{(4)}_{\eta,87} & = 1\,, &
	 \beta^{(4)}_{\eta,88} & = \tfrac{\pi^2}{8} \,, &
	 \beta^{(4)}_{\eta,89} & = \tfrac{\pi^2}{8} \,, &
	 \beta^{(4)}_{\eta,90} & = \tfrac{\pi^2}{16} \,, \\
	 \beta^{(4)}_{\eta,91} & = \tfrac{1}{2} \,, &
	 \beta^{(4)}_{\eta,92} & = \tfrac{3}{4} \,, &
	 \beta^{(4)}_{\eta,93} & = \tfrac{1}{3} \,, &
	 \beta^{(4)}_{\eta,94} & = \tfrac{1}{4} \,, &
	 \beta^{(4)}_{\eta,95} & = 0 \,, \\
	 \beta^{(4)}_{\eta,96} & = 0 \,, &
	 \beta^{(4)}_{\eta,97} & = \tfrac{\pi^2}{8} \,, &
	 \beta^{(4)}_{\eta,98} & = \tfrac{\pi^2}{2} \,, &
	 \beta^{(4)}_{\eta,99} & = \tfrac{3}{4} \,, &
	 \beta^{(4)}_{\eta,100} & = \tfrac{\pi^2}{16} \,, \\
	 \beta^{(4)}_{\eta,101} & = \tfrac{1}{6} \,, &
	 \beta^{(4)}_{\eta,102} & = \tfrac{1}{2} \,, &
	 \beta^{(4)}_{\eta,103} & = \tfrac{1}{4} \,, &
	 \beta^{(4)}_{\eta,104} & = \tfrac{1}{4} \,, &
	 \beta^{(4)}_{\eta,105} & = \tfrac{\pi^2}{16} \,, \\
	 \beta^{(4)}_{\eta,106} & = 0 \,, &
	 \beta^{(4)}_{\eta,107} & = \tfrac{1}{12} \,, &
	 \beta^{(4)}_{\eta,108} & = \tfrac{\pi^2}{16} \,, &
	 \beta^{(4)}_{\eta,109} & = \tfrac{\pi^2}{8} \,, &
	 \beta^{(4)}_{\eta,110} & = \tfrac{1}{4} \,, \\
	 \beta^{(4)}_{\eta,111} & = \tfrac{\pi^2}{8} \,, &
	 \beta^{(4)}_{\eta,112} & = 0 \,, &
	 \beta^{(4)}_{\eta,113} & = \tfrac{1}{3} \,, &
	 \beta^{(4)}_{\eta,114} & = 1-\tfrac{\pi^2}{8} \,, &
	 \beta^{(4)}_{\eta,115} & = 1\,, \\
	 \beta^{(4)}_{\eta,116} & = \tfrac{1}{2} \,, &
	 \beta^{(4)}_{\eta,117} & = \tfrac{\pi^2}{4} \,, &
	 \beta^{(4)}_{\eta,118} & = \tfrac{\pi^2}{4} \,, &
	 \beta^{(4)}_{\eta,119} & = 1\,, &
	 \beta^{(4)}_{\eta,120} & = 1\,, \\
	 \beta^{(4)}_{\eta,121} & = \tfrac{1}{6} \,, &
	 \beta^{(4)}_{\eta,122} & = 0 \,, &
	 \beta^{(4)}_{\eta,123} & = \tfrac{\pi^2}{4} \,, &
	 \beta^{(4)}_{\eta,124} & = 1\,, &
	 \beta^{(4)}_{\eta,125} & = \tfrac{\pi^2}{8} \,, \\
	 \beta^{(4)}_{\eta,126} & = -2+\tfrac{\pi^2}{4} \,, &
	 \beta^{(4)}_{\eta,127} & = 1-\tfrac{\pi^2}{8} \,, &
	 \beta^{(4)}_{\eta,128} & = \tfrac{1}{2} \,, &
	 \beta^{(4)}_{\eta,129} & = 2 \,, &
	 \beta^{(4)}_{\eta,130} & = \tfrac{1}{2} \,, \\
	 \beta^{(4)}_{\eta,131} & = \tfrac{\pi^2}{8} \,, &
	 \beta^{(4)}_{\eta,132} & = \tfrac{\pi^2}{12} \,, &
	 \beta^{(4)}_{\eta,133} & = \tfrac{\pi^2}{8} \,, &
	 \beta^{(4)}_{\eta,134} & = \tfrac{\pi^2}{4} \,, &
	 \beta^{(4)}_{\eta,135} & = 1\,, \\
	 \beta^{(4)}_{\eta,136} & = \tfrac{\pi^2}{16} \,, &
	 \beta^{(4)}_{\eta,137} & = \tfrac{\pi^2}{8} \,, &
	 \beta^{(4)}_{\eta,138} & = \tfrac{1}{2} \,, &
	 \beta^{(4)}_{\eta,139} & = -2+\tfrac{\pi^2}{3} \,, &
	 \beta^{(4)}_{\eta,140} & = -1+\tfrac{\pi^2}{8} \,, \\
	 \beta^{(4)}_{\eta,141} & = 1\,, &
	 \beta^{(4)}_{\eta,142} & = 2 \,, &
	 \beta^{(4)}_{\eta,143} & = \tfrac{\pi^2}{8} \,, &
	 \beta^{(4)}_{\eta,144} & = \tfrac{\pi^2}{8} \,, &
	 \beta^{(4)}_{\eta,145} & = \tfrac{1}{2} \,, \\
	 \beta^{(4)}_{\eta,146} & = \tfrac{\pi^2}{16} \,, &
	 \beta^{(4)}_{\eta,147} & = \tfrac{1}{4} \,, &
	 \beta^{(4)}_{\eta,148} & = \tfrac{\pi^2}{8} \,, &
	 \beta^{(4)}_{\eta,149} & = \tfrac{\pi^2}{6} \,, &
	 \beta^{(4)}_{\eta,150} & = 0 \,.
     \tag{\stepcounter{equation}\theequation}
\end{align*}

Our results agree with the two-loop and partial four-loop terms
of~\cite{Fraser-Taliente:2024rql}. As for the four-loop leg and Yukawa vertex
corrections, we find large agreement but also minor discrepancies
with~\cite{Jack:2016utw}.\footnote{We thank I.~Jack for helpful discussions and
for revisiting the relevant expressions in private correspondence. An ongoing
re-examination indicates minor inconsistencies in~\cite{Jack:2016utw}, which
are plausibly the reason for the remaining  discrepancies.} Note that
in~\cite{Jack:2016utw} the number of coefficients for $\hat{\beta}_Y^{(4)}$ is
smaller; the additional ones quoted in the present work 
all vanish. Both~\cite{Jack:2016utw,Fraser-Taliente:2024rql} were cross-checked against older publications such
as~\cite{Pisarski:1982vz,Avdeev:1991za,Avdeev:1992jt,Jack:2015tka,Giombi:2018qgp}.

Another important cross-check is the emergence of $\mathcal{N}=1$ and
$\mathcal{N}=2$ supersymmetry, to which the next section is dedicated.

In passing, we observe that the RGE coefficients are purely rational at
two-loop order, and contain 
$\zeta_2 \equiv \pi^2/6$ at four loops.
The number content appears
to get more complicated at higher loops---a phenomenon also found in four
dimensions. There, all coefficients are rational up to two loops, while three-
and four-loop corrections introduce $\zeta_{3}$ and $\zeta_4$,
respectively~\cite{Jack:2024sjr,Steudtner:2024teg,Steudtner:2025blh}, and five
and six loops even feature $\zeta_5$, $\zeta_6$, $\zeta_7$ and
$\zeta_{3,5}$~\cite{Baikov:2016tgj,Luthe:2016ima,Herzog:2017ohr,Luthe:2017ttg,Bednyakov:2021ojn}.
In contrast, the three-dimensional six-loop RGEs of~\cite{Hager:2002uq} predict
the appearance of $\ln\,2$, Catalan's constant, and the Dirichlet beta
function. This is a surprising result, which merits an independent
verification.

\section{Supersymmetry}\label{sec:SUSY}

We now impose $\mathcal{N} = 1$ supersymmetry on the Lagrangian~\eq{theory}.
It is generated via two supercharges $Q_\alpha$ that fulfil 
\begin{equation}
  \{Q_\alpha,\overline{Q}_\beta\} = 2 \gamma^\mu_{\alpha \beta}  P_\mu\,.
\end{equation}
 These supercharges and the corresponding
supersymmetry generators can be defined via superspace coordinates $\theta^\alpha$,
which are real, two-component Grassmann numbers
\begin{equation}
    Q_\alpha = \frac{\partial}{\partial \theta^\alpha} + i \gamma^\mu \theta_\alpha \partial_\mu\,, \qquad 
    D_\alpha =\frac{\partial}{\partial \theta^\alpha} - i \gamma^\mu \theta_\alpha \partial_\mu\,.
\end{equation}
Thus, each matter superfield 
\begin{equation}
    \Phi_A = \phi_A + \overline{\theta} \psi_A + \tfrac{1}{2} \overline{\theta}\theta \,F_A
\end{equation}
consists of real scalar fields $\phi_A$, two-component Majorana fields $\psi_A$ and real auxiliary fields $F_A$, 
again using $\overline{\theta} = \theta^T \gamma^0$.
A marginal superpotential takes the general form
\begin{equation}
W(\Phi)=\tfrac{1}{4!} \lambda^{ABCD} \Phi_A \Phi_B \Phi_C \Phi_D\,,
\end{equation}
with $\lambda^{ABCD} = \lambda^{(ABCD)}$ the real
superquartic coupling tensor. 

Integrating the superspace coordinates as well as
the auxiliary fields, we arrive at a generalisation of the three-dimensional
Wess--Zumino model~\cite{Wess:1973kz}
\begin{align}
    \mathcal{L}_\text{SUSY} &= \int \mathrm{d}^2 \theta \left[\tfrac12 D_\alpha \Phi_A D^\alpha \Phi_A + W(\Phi)\right] = \tfrac12 \partial_\mu \phi_A \partial^\mu \phi_A + \tfrac{i}2\overline{\psi}_A \gamma^\mu\partial_\mu \psi_A\notag \\
    &\phantom{=}  - \tfrac14 \lambda^{ABCD} \phi_A \phi_B \overline{\psi}_C \psi_D - \tfrac1{36} \lambda^{ABCG} \lambda^{DEFG} \phi_A \phi_B \phi_C \phi_D \phi_E \phi_F  \,.
\end{align}
Comparing this with~\eq{theory} implies
\begin{align}
     Y^{ABCD} = \lambda^{ABCD} \,,\quad \eta^{ABCDEF} = \mathcal{S}_{10} \lambda^{ABCG} \lambda^{DEFG} \,.
\end{align}
Therefore, at the loop level ${\mathcal N}=1$ supersymmetry yields
\begin{align}
    \gamma_\Phi^{AB} &\equiv \gamma_\phi^{AB} = \gamma_\psi^{AB}\,, \qquad 
    \hat{\beta}_\lambda^{ABCD} \equiv \hat{\beta}_Y^{ABCD} = \hat{\beta}_Y^{(ABCD)}\,, \notag\\
\hat{\beta}_\eta^{ABCDEF} &=  \mathcal{S}_{20} \lambda^{ABCG} \hat{\beta}_\lambda^{DEFG} + 2 \mathcal{S}_{10} \lambda^{ABCG} \gamma_\Phi^{GH} \lambda^{HDEF} \,.
\end{align}
At two-loop order, this leads to 
\begin{align}
    \gamma^{(2)}_{\phi,1} &= \gamma^{(2)}_{\psi,1}\,,\qquad
    \beta^{(2)}_{Y,1} = \beta^{(2)}_{Y,4} = 0\,,\quad 
    \beta^{(2)}_{Y,2} = \beta^{(2)}_{Y,3} = \beta^{(2)}_{Y,5} = 6\,\gamma^{(2)}_{\phi,1}\,, \notag\\
    \beta^{(2)}_{\eta,2} &= 0\,, \qquad \quad
    \tfrac12\beta^{(2)}_{\eta,1} = \tfrac16\beta^{(2)}_{\eta,3} = -\tfrac1{12}\beta^{(2)}_{\eta,4}  = -\tfrac1{24}\beta^{(2)}_{\eta,5} = \gamma^{(2)}_{\phi,1}\,,  \label{eq:susy1-2l}
\end{align}
which fixes all RGEs up to a single coefficient. At four loops, there are 214
independent relations all of which are compatible with the direct results of
this work. We list them in a separate file attached to the arXiv submission
of the present work.

With $\mathcal{N}=2$ supersymmetry, there are four supercharges that satisfy
the algebra 
 \begin{equation}
\{Q_{\alpha,i},\overline{Q}_{\beta,j}\} = 2 \gamma^\mu_{\alpha \beta} \delta_{ij} P_\mu\,.
\end{equation}
Here $i,j=1,2$ form an $\text{SO}(2) \simeq \text{U}(1)$ $R$-symmetry. Thus,
the algebra can be implemented with complex 2-component superspace coordinates
$\theta$. 
This is analogous to $\mathcal{N}=1$ supersymmetry in four spacetime
dimensions.  
Supercoordinates, superfields and all their components are complex, and we need to carefully distinguish the contractions $\overline{X} Y \equiv X^\dagger \gamma^0 Y$ and $(X Y) \equiv X^\mathrm{T} \gamma^0 Y$ between two spinors $X$ and $Y$ for the remainder of this section. The superfields 
\begin{align}
\Phi_A &= \varphi_A + \sqrt{2}\,(\theta \chi_A) + (\theta \theta) F_A + i \overline{\theta} \gamma^\mu \theta \partial_\mu \varphi_A - \tfrac{i}{\sqrt{2}} (\theta \theta) \overline{\theta} \gamma^\mu\partial_\mu \chi_A + \tfrac14 (\theta\theta) (\theta\theta)^* \partial_\mu \partial^\mu \varphi_A  
\end{align}
consist of complex scalars $\varphi_A$ and $F_A$ as well as complex
two-component Dirac fermions $\chi_A$. 
The superpotential reads 
\begin{equation}
    W(\Phi) = \tfrac{1}{4!} \lambda^{ABCD} \Phi_A \Phi_B \Phi_C \Phi_D\,,
\end{equation}
Note that the $R$-symmetry
enforces a holomorphy of the superpotential, meaning that tensors
$\lambda^{ABCD}$ only couple to superfields $\Phi_A$ but not their complex
conjugates $\Phi_A^*$. 
The Lagrangian then reads
\begin{align}\label{eq:susy-N2}
    \mathcal{L}_\text{SUSY} &= \int \mathrm{d}^2 \theta \mathrm{d}^2 \theta^*\, \Phi^*_A \Phi^{\phantom{*}}_A + \int \mathrm{d}^2 \theta \,W(\Phi) + \int \mathrm{d}^2 \theta^* \,W^*(\Phi^*) =  \partial_\mu \varphi^*_A \partial^\mu \varphi^{\phantom{*}}_A + i\overline{\chi}_A \gamma^\mu\partial_\mu \chi_A \notag \\
    & -  \left[\tfrac14\lambda^{ABCD}\varphi_{A}\varphi_{B} (\chi_C\chi_D) + \text{h.c.}\right]  - \tfrac1{36} \lambda^{ABCG}(\lambda^{DEFG})^* \, \varphi^{\phantom{*}}_{A} \varphi^{\phantom{*}}_{B} \varphi^{\phantom{*}}_{C} \varphi^*_{D} \varphi^*_{E} \varphi^*_{F}\,.
\end{align}

In order to map the $\left(\mathcal{N}=2\right)$--supersymmetric theory~\eq{susy-N2} to the language of~\eq{theory}, we decompose the complex matter fields $\varphi$ and $\chi$ into real components $\phi,\psi$. 

We first repackage the $M$ complex scalar field $\varphi_A$ and fermions $\chi_A$ into $(M \times 2)$-dimensional matrices 
\begin{equation}\label{eq:fields-repackage-tilde}
\tilde{\phi}^\alpha_{A} = (\varphi^{\phantom{*}}_A,\,\varphi_A^*)^\alpha \qquad \text{ and } \qquad \tilde{\psi}^\alpha_{A} = (\chi^{\phantom{*}}_A,\,\chi_A^*)^\alpha
\end{equation}
with $\alpha=1,2$. The interaction terms of~\eq{susy-N2} can then be rewritten as 
\begin{equation}
    \mathcal{L}_\text{int} = - \tfrac14 \tilde{Y}^{ABCD}_{\alpha\beta\gamma\delta} \,\tilde{\phi}_A^\alpha \tilde{\phi}_B^\beta (\tilde{\psi}_C^\gamma \tilde{\psi}_D^\delta) - \tfrac{1}{36} \tilde{\eta}^{ABCDEF}_{\alpha\beta\gamma\delta\epsilon\zeta} \, \tilde{\phi}_A^\alpha \tilde{\phi}_B^\beta \tilde{\phi}_C^\gamma \tilde{\phi}_D^\delta \tilde{\phi}_E^\epsilon \tilde{\phi}_F^\zeta
\end{equation}
with 
\begin{equation}\label{eq:embedded-couplings}
    \tilde{Y}^{ABCD}_{1111} = \lambda^{ABCD}\,, \qquad \tilde{Y}^{ABCD}_{2222} = (\lambda^{ABCD})^*\,, \qquad \eta^{ABCDEF}_{111222} = \lambda^{ABCG} (\lambda^{DEFG})^*\,.
\end{equation}
Next, we translate the pseudo-real fields $\tilde{\phi},\tilde{\psi}$ into real components $\phi,\psi$ as in \eq{theory}, using the embeddings
\begin{equation}\label{eq:fields-repackage}
    \phi^\alpha_{A} = (\mathrm{Re}\,\varphi_A,\,\mathrm{Im}\,\varphi_A)^\alpha \qquad \text{ and } \qquad \psi^\alpha_{A} = (\mathrm{Re}\,\chi_A,\,\mathrm{Im}\,\chi_A)^\alpha\,.
\end{equation}
Note that the scalar $(a,b,c,\dots)$ and fermionic indices ($i,j,\dots$) running from $1\dots 2M$ in \eq{theory} are here split up in two indices $A$ and $\alpha$ with the ranges $M$ and $2$.    
The bases \eq{fields-repackage-tilde} and \eq{fields-repackage} are related via the  rotation 
\begin{equation}\label{eq:X-mat}
    \tilde{\phi}^\alpha_A = X^{\alpha}_{\ \beta} \phi^\beta_A\,,\qquad \tilde{\psi}^\alpha_A = X^{\alpha}_{\ \beta} \psi^\beta_A\,, 
    \qquad \text{ where } \qquad  
    X = \frac1{\sqrt{2}} \begin{pmatrix} 1 & i\\ 1 & -i \end{pmatrix}\,.
\end{equation}
Finally, we obtain the mapping 
\begin{equation}\label{eq:mapping}
    Y^{ABCD}_{\alpha\beta\gamma\delta} = X^{\alpha}_{\ \alpha'} X^{\beta}_{\ \beta'} X^{\gamma}_{\ \gamma'} X^{\delta}_{\ \delta'} \tilde{Y}^{ABCD}_{\alpha'\beta'\gamma'\delta'}\,, \qquad
    \eta^{ABCDEF}_{\alpha\beta\gamma\delta\epsilon\zeta} = X^{\alpha}_{\ \alpha'} X^{\beta}_{\ \beta'} X^{\gamma}_{\ \gamma'} X^{\delta}_{\ \delta'} X^{\epsilon}_{\ \epsilon'} X^{\zeta}_{\ \zeta'} \tilde{\eta}^{ABCDEF}_{\alpha'\beta'\gamma'\delta'\epsilon'\zeta'}\,,
\end{equation}
which refer the couplings $Y$ and $\eta$ in \eq{theory}, again with their original indices being split up
\begin{equation}
\mathcal{L}_\text{int} = - \tfrac14 {Y}^{ABCD}_{\alpha\beta\gamma\delta} \,{\phi}_A^\alpha {\phi}_B^\beta ({\psi}_C^\gamma {\psi}_D^\delta) - \tfrac{1}{6!} {\eta}^{ABCDEF}_{\alpha\beta\gamma\delta\epsilon\zeta} \, {\phi}_A^\alpha {\phi}_B^\beta {\phi}_C^\gamma {\phi}_D^\delta {\phi}_E^\epsilon {\phi}_F^\zeta\,.
\end{equation}
While these structures appear complicated, their practical evaluation in template RGEs is straightforward: all internal contractions between two coupling tensors include a summation over greek indices of the form
\begin{equation}
    Y^{A \dots }_{\,\alpha\, \dots}\  Y^{A \dots }_{\,\alpha\, \dots} = \tilde{Y}^{A \dots }_{\,\beta \dots} \ \tilde{Y}^{A \dots }_{\,\gamma \dots} \ X^\beta_{\ \alpha} X^\gamma_{\ \alpha} = \tilde{Y}^{A \dots }_{\,\beta \dots} \ \tilde{Y}^{A \dots }_{\,\gamma \dots} \ \sigma_1^{\beta \gamma} \,.
\end{equation} 
The Pauli matrix $\sigma_1$ can be understood from the fact that propagators connect fields and their complex conjugates, e.g. $\varphi_A$ and $\varphi_A^*$ and vice versa. Inserting \eq{mapping} everywhere in our template expressions, all internal contractions of greek indices are carried over the metric $\sigma_1$. Together with \eq{embedded-couplings}, this enforces a holomorphic structure of RGEs: all $\lambda$ are only contracted with $\lambda^*$ but no other $\lambda$.
Many contractions in the template RGEs vanish in $\mathcal{N}=2$ supersymmetry because they are incompatible with this holomorphic structure.
In consequence, the superpotential parameters do not receive vertex
corrections~\cite{Salam:1974jj,Grisaru:1979wc}. Overall, the relations
\begin{align}
    \gamma_\Phi{}^{AB} &\equiv \gamma_\phi{}^{AB} = \gamma_\psi{}^{AB}\,, \qquad 
    \hat{\beta}_Y^{ABCD} = 0\,, \qquad 
    \hat{\beta}_\eta{}^{ABCDEF} =   2 \lambda^{ABCG} \gamma_\Phi{}^{HG} (\lambda^{HDEF})^*
\end{align}
must hold as long as the renormalisation scheme does not break the 
supersymmetry.
At two-loop order this implies 
\begin{align}\label{eq:susy2-2l}
    \gamma_{\phi,1}^{(2)} = \gamma_{\psi,1}^{(2)} \,, \qquad \beta_{Y,1}^{(2)} = \beta_{Y,4}^{(2)} = 0 \,,\qquad \beta_{\eta,1}^{(2)} = -\tfrac13 \beta_{\eta,3}^{(2)} -\tfrac13  \beta_{\eta,4}^{(2)} = 2\gamma_{\phi,1}^{(2)} \,, \qquad  \beta_{\eta,2}^{(2)} = 0\,.
\end{align}
It is obvious that \eq{susy2-2l} is less restrictive than its $\mathcal{N}=1$ counterpart~\eq{susy1-2l}. At
four loops, 46 conditions arise, which are all fulfilled for our results and
are listed in a separate file along with the arXiv submission of the present work.

\section{Fixed Points}\label{sec:FPs}

In this section, we investigate whether perturbatively accessible fixed points can
arise directly in $d=3$ in the context of the renormalisable theories
in~\eq{theory}.
To this end, we analyse the leading (two-loop)
contribution to the quartic Yukawa and scalar sextic $\beta$-functions
from~\Sec{templates}, which we repeat here for convenience
\begin{align}
    (4\pi)^2 \beta_Y^{ab} &=  \frac{1}{12} \mathcal{S}_2 \tr{Y^{accd}}Y^{bd} + \frac1{12} \mathcal{S}_2 Y^{abcdcd}  \notag \\ 
     &\phantom{=} + \frac12\tr{Y^{acbd}} Y^{cd}  + \frac12 Y^{cdabcd} + \frac12 \mathcal{S}_4 Y^{acbdcd}\,,\label{eq:2L-Y}\\
     (4\pi)^2 \beta_\eta^{abcdef} &= \frac16 \mathcal{S}_{10} \eta^{abcghi} \eta^{defghi} + \frac1{12} \mathcal{S}_6 \tr{Y^{aggh}} \eta^{bcdefh}  + \frac12\mathcal{S}_{15} \tr{Y^{agbh}} \eta^{cdefgh} \notag \\
     &\phantom{=} - \mathcal{S}_{90} \tr{Y^{abcdegfg}} - 2\mathcal{S}_{45} \tr{Y^{abcgdefg}} \,.\label{eq:2L-eta}
\end{align}
Schematically, RGEs take the form
\begin{equation}
    \beta(\alpha) = -B\,\alpha^2 + C\,\alpha^3 + \mathcal{O}(\alpha^4)
\end{equation}  
where the first term is determined by \eq{2L-Y} or \eq{2L-eta}. A fixed point $\alpha^*$ arises at $\beta(\alpha^*) = 0$ and exhibits a perturbative expansion $\alpha^* = B/C + \mathcal{O}(B^2)$. For perturbation theory to be reliable around the fixed point, the leading coefficient $|B|$ should be small with respect to $|C|$ and higher order coefficients.  For instance, the smallness of $|B|$ could be the product of a cancellation of terms or a large-$N$ suppression in \eq{2L-Y} and \eq{2L-eta}.

Fixed points are denominated UV or IR attractive, depending on whether they are reached in the limit of 
  high or low renormalisation scales $\mu$, respectively. For a theory with several couplings $\alpha_i$ with respective $\beta$-functions $\beta_i$, the RG flow around a fixed point reads 
\begin{equation}
    \beta_i = S_{ij} (\alpha_j - \alpha_j^*) + \mathcal{O}\left((\alpha-\alpha^*)^2\right)\,.
\end{equation}
The eigenvalues $\vartheta_i$ of the stability matrix $S_{ij}$ determine whether a fixed-point is IR ($\vartheta_i > 0$) or UV attractive ($\vartheta_i < 0$).

For purely scalar theories ($Y=0$), the existence of a UV fixed point is long
known~\cite{Townsend:1976sy,Appelquist:1981sf,Pisarski:1982vz,Hager:2002uq,Kvedaraite:2025lgi},
and under perturbative control in a large-$N$ limit. This is in contrast
to renormalisable QFTs in $d=4$, where fixed points may only arise in the presence of a 
non-Abelian gauge sector~\cite{Coleman:1973sx,Bond:2016dvk,Bond:2018oco}. 

\subsection{General Considerations}

In the following, we will search for fixed points involving fermions. 
The leading-order $\beta$ functions for $Y$ does not
involve $\eta$. Thus, a fixed-point coupling  $Y^*$ must emerge from $Y$
itself. Crucially, all terms in~\eq{2L-Y} enter with positive coefficients,
prohibiting any straightforward cancellations among them.
As a result, most QFTs exhibit a
manifestly positive $\beta_Y$, and the emergence of a perturbative fixed point
is highly non-trivial. 

Nevertheless, there is a window of opportunity, which can be recognised by
considering the quantity~\cite{Coleman:1973sx,Bond:2018oco}
\begin{align}
    (4\pi)^2 \frac{\mathrm{d}}{\mathrm{d \log \mu}}\tr{Y^{abab}} &= (4\pi)^2 \tr{Y^{ab} \beta_{Y}^{ab}} \notag\\
    &= \frac16 \tr{Y^{accb}} \tr{Y^{addb}}  + \frac1{6} \langle Y^{abcd}, Y^{abcd} \rangle  \notag \\ 
    &\phantom{=}   + \frac12 \langle Y^{ij} Y^{kl}, Y^{kl} Y^{ij} \rangle  + \frac12 \langle Y^{abcd}, Y^{cdab} \rangle + 2 \langle Y^{accb},Y^{bdda}\rangle \,,\label{eq:y*by}
\end{align} 
with $\langle A, B \rangle = \sum_{i,j} A_{ij} B_{ij}$, i.e. the sum runs over indices not explicitly
written in the brackets~$\langle \cdot,\cdot \rangle$. For the first term in
the second line we have made the fermion indices explicit but suppressed the
scalar ones. 
For the first term in the second line we have made the fermion indices explicit 
but suppressed the scalar ones.
A requirement for a QFT to contain a perturbative fixed point is that 
the absolute value of~\eq{y*by} can be made arbitrary small or even vanish.
In our case, this is signaled by a change of sign in~\eq{y*by} by any choice of $Y$.

The first two terms in
\eq{y*by} are manifestly positive while the situation is less clear for the
last three. If we fix the fermionic indices for the third and the scalar ones for
the fourth and fifth term, we observe that each can be written as a scalar
product of the form $\langle A B , B A\rangle$ where $A$, $B$ are symmetric
matrices. In case the product $B A$ is antisymmetric, the last three terms of
\eq{y*by} are negative. Matrices with these properties are part of a Clifford
algebra $\{\gamma^A,\gamma^B\} = 2 \delta^{AB} \mathbb{1}$. 

Thus, a promising candidate for negative contributions 
to the two-loop $\beta$-function is the theory  
\begin{equation}
    \mathcal{L} = \partial_\mu \phi^*_i \partial^\mu \phi_i + i \overline{\psi}{}_{i} \slashed{\partial} \psi_{i} - Y\, (\phi_i^*\gamma_{ij}^A \phi_j)(\overline{\psi}{}_{k}  \gamma_{kl}^A \psi_{l}) - \tfrac1{36} \eta (\phi^*_i \phi_i)^3
\end{equation}
with complex scalars $\phi_{i}$ and Dirac fermions $\psi_i$ both in the spinor representation of $\mathrm{SO}(n)$. Here we have the fundamental index $A=1\dots n$
as well as $i,j,k,l=1,\dots d_\gamma$ for the spinor reprentation, where $d_\gamma = 2^{\lfloor N/2\rfloor}$ and $\overline{\psi}_i = \psi_i^\dagger \gamma^0$.
The corresponding RGE
\begin{equation}\label{eq:betaYSON}
\beta_{Y} = - \left[2(n-2) + \frac{4}{3} (n-3) d_\gamma \right] \frac{Y^3}{(4\pi)^2} +
\mathcal{O}(\text{4-loop})\,, 
\end{equation} 
is indeed negative for $n \geq 3$.
Thus, we find that negative two-loop $\beta$-functions are viable, 
which motivates the existence of perturbative fixed points. 

For any fixed point to be strictly perturbative, it needs to be under the
control of a small expansion parameter, $\epsilon$. Since we search for fixed
points directly in $d=3$, $\epsilon$ must stem from the parameters of the
theory and arise within the leading order RGEs. For instance, 
\begin{equation}
    \beta_\alpha = -\epsilon\,\alpha^2 + C\,\alpha^3 + \dots   
\end{equation} 
yields a fixed-point value $\alpha^* = \epsilon/C + \mathcal{O}(\epsilon^2)$,
suppressing higher loop contributions by ever-increasing powers of $\epsilon$.
A natural source of perturbative control originates from large field
multiplicities. Commonly, two realizations can be distinguished. Large-$N$
expansions have $\epsilon \propto N^{-1}$ and become exact in the limit $N \to
\infty$. A notable example is found
in~\cite{Townsend:1976sy,Appelquist:1981sf,Pisarski:1982vz,Kvedaraite:2025lgi}.
On the other hand, Veneziano--type expansions~\cite{Veneziano:1976wm} feature
two large multiplicities $N_{1,2} \to \infty$ with a fixed ratio such that
$\epsilon = N_1/N_2 + \text{const}$, see for
instance~\cite{Banks:1981nn,Litim:2014uca}. In the following, we will discuss
some examples how large-$N$ limits can be taken for the theories at hand. 
 
A straightforward idea is to have both scalar and fermions in large
multiplicities $N_s$ and $N_f$, respectively, where $N_f \propto N_s \propto N
\to \infty$. In this large-$N$ expansion, the leading-order contributions to
$\beta_Y$ are given by an alternating bubble chain of fermions and scalars
\begin{equation}\label{eq:chain}
    \tikz[baseline=0.6em]{
        \draw[line width=0.05em] (1.0em,0) circle [radius=1.0em];
        \draw[fill=black] (0,0) circle (0.15em and 0.15em);
        \draw[dash pattern=on .2em off .2em, line width=0.05em] (-.8em,.8em) -- (0,0) -- (-.8em,-.8em); 
        \draw[fill=black] (2.0em,0) circle (0.15em and 0.15em);
        \draw[dash pattern=on .2em off .2em, line width=0.05em] (3.0em,0) circle [radius=1.0em];
        \draw[fill=black] (4.0em,0) circle (0.15em and 0.15em);
        \draw[line width=0.05em] (4.8em,.8em) -- (4em,0) -- (4.8em,-.8em);
        \node at (0.1em,0) {$\bigg[$}; 
        \node at (4.1em,0) {$\bigg]^n$}; 
    }
\end{equation}
and appear at every non-trivial loop order. Explicitly, they yield factors
$(N_f N_s Y^2)^{n}$ at $(2n)$-loop, such that a 't Hooft--rescaled
coupling~\cite{tHooft:1973alw} 
\begin{equation}\label{eq:Yhat}
    \hat{Y}^2 = N^2 Y^2
\end{equation}
has a finite $\beta$-function for $N \to \infty$. However, each diagram
of~\eq{chain} factorises into subgraphs without logarithmic divergences, thus
not contributing to the $\beta$-functions.\footnote{We remark that the
vanishing is  renormalisation-scheme independent. Although, e.g. in
cutoff-schemes, massive tadpoles exhibit a pole, they do not enter the vertex
counterterm. See~\cite{Kvedaraite:2025lgi} for a discussion in the scalar
$\phi^6$ theory.} At two and four loops, this class of diagrams corresponds to
the vanishing coefficients $\beta_{Y,1}^{(2)}$ and $\beta_{Y,10}^{(4)}$ in
\eq{chain2L} and \eq{chain4L}, respectively. The dominant contributions
therefore stem from ``next-to-leading order'' diagrams given by
inserting~\eq{chain} between two scalar, or two fermion legs, respectively
\begin{equation}\label{eq:largeNLO}
    \tikz[baseline=0.6em]{
        \draw[line width=0.05em] (-1.0em,0) circle [radius=1.0em];
        \draw[dash pattern=on .2em off .2em, line width=0.05em] (+1.0em,0) circle [radius=1.0em];
        \draw[line width=0.05em] (+3.0em,0) circle [radius=1.0em];
        \draw[fill=black] (-2em,0) circle (0.15em and 0.15em);
        \draw[fill=black] (0,0) circle (0.15em and 0.15em);
        \draw[fill=black] (+2em,0) circle (0.15em and 0.15em);
        \draw[fill=black] (+4em,0) circle (0.15em and 0.15em);
        \draw[dash pattern=on .2em off .2em, line width=0.05em] (1.em,3.em) to [bend left] (+4.0em,0em) to (+4.5em,-1.5em);
        \draw[dash pattern=on .2em off .2em, line width=0.05em] (1.em,3.em) to [bend right] (-2.0em,0em) to (-2.5em,-1.5em);
        \draw[line width=0.05em] (.2em,3.6em) -- (1.em,3.em) -- (1.8em,3.6em);
        \draw[fill=black] (1.em,3.em) circle (0.15em and 0.15em);
        \node at (0.1em,0) {$\bigg[$}; 
        \node at (4.1em,0) {$\bigg]^n$}; 
    } \qquad
    \tikz[baseline=0.6em]{
        
        \draw[dash pattern=on .2em off .2em, line width=0.05em] (-1.0em,0) circle [radius=1.0em];
        \draw[line width=0.05em] (+1.0em,0) circle [radius=1.0em];
        \draw[dash pattern=on .2em off .2em, line width=0.05em] (+3.0em,0) circle [radius=1.0em];
        \draw[fill=black] (-2em,0) circle (0.15em and 0.15em);
        \draw[fill=black] (0,0) circle (0.15em and 0.15em);
        \draw[fill=black] (+2em,0) circle (0.15em and 0.15em);
        \draw[fill=black] (+4em,0) circle (0.15em and 0.15em);
        \draw[line width=0.05em] (1.em,3.em) to [bend left] (+4.0em,0em) to (+4.5em,-1.5em);
        \draw[line width=0.05em] (1.em,3.em) to [bend right] (-2.0em,0em) to (-2.5em,-1.5em);
        \draw[dash pattern=on .2em off .2em, line width=0.05em] (.2em,3.6em) -- (1.em,3.em);
        \draw[dash pattern=on .2em off .2em, line width=0.05em] (1.8em,3.6em) -- (1.em,3.em);
        \draw[fill=black] (1.em,3.em) circle (0.15em and 0.15em);
        \node at (0.1em,0) {$\bigg[$}; 
        \node at (4.1em,0) {$\bigg]^n$}; 
    } \,,
\end{equation}
corresponding to the non-vanishing coefficients $\beta_{Y,2}^{(2)}$,
$\beta_{Y,5}^{(2)}$ and $\beta^{(4)}_{Y,16}$, $\beta^{(4)}_{Y,70}$. For
$\beta$-functions to remain finite at $N \to \infty$, we still require the
rescaling~\eq{Yhat}. This, however, produces a factor of $\epsilon= 1/N$ for
every dominant contribution~\eq{largeNLO}, suppressing the entire
$\beta$-function by $1/N$, i.e. $\beta_{\hat{Y}^2} \propto \epsilon$. We
stress that taking the strict $N \to \infty$ limit and only then searching for
fixed points is inconsistent, since the large-$N$ rescaling renders the full
$\beta$-function subleading and hence trivially vanishing. The correct
procedure is to search for fixed points within $\beta_{\hat{Y}^2}/\epsilon =
\sum_{\ell = 1}^\infty c_{2\ell} \hat{Y}^{2\ell} +
\mathcal{O}(\epsilon)$, keeping the leading $1/N$ contributions explicit before
sending $N$ to infinity~\cite{Kvedaraite:2025lgi}.

Another popular large-$N$--expansion is to only assign one large field
multiplicity, for instance $N_f \propto N\to\infty$ but not $N_s$. In this
case, the rescaling $\check{Y}^2 = N Y^2$ is sufficient. Leading-order
contributions do not merely arise through~\eq{chain}, but from many other graphs as
well, even at two loops. In the current example, the leading-order $\beta$-functions
include contributions from both leg corrections 
$\propto \gamma_{\phi,1}^{(2)}$ and the non-vanishing vertex correction 
$\propto \beta_{Y,2}^{(2)}$.

\subsection{Example: Perturbative IR Fixed Point}

Let us choose the explicit example of a QFT with global (internal) symmetry $\mathrm{SO}(3)
\times U(N)$, where complex fermions and scalars both transform under the spinor
representation of $\mathrm{SO}(3)$, but fermions are in the fundamental of
$U(N)$, while scalars are singlets. Explicitly, the Lagrangian reads 
\begin{equation}\label{eq:SO3-lag}
    \mathcal{L} = \partial_\mu \phi^*_i \partial^\mu \phi_i + i \overline{\psi}{}_{ia} \slashed{\partial} \psi_{ia} - Y\, (\phi_i^*\sigma_{ij}^A \phi_j)(\overline{\psi}{}_{ak}  \sigma_{kl}^A \psi_{al}) - \tfrac1{36} \eta (\phi^*_i \phi_i)^3 \,,
\end{equation}
with $a=1,\dots,N$ labelling the $U(N)$ flavour, $i,j,k,l=1,2$ labelling the
$\mathrm{SO}(3)$ spinor indices, $\overline{\psi}{} = \psi^\dagger \gamma^0$, and $A=1,2,3$ denoting the three Pauli
matrices of dimension $d_\gamma = 2$. Using the abbreviations 
\begin{equation}
    \epsilon = \frac{1}{N}\,,\qquad \alpha_Y = \frac{N Y^2}{(4\pi)^2}  \,,\qquad \text{and} \qquad \alpha_\eta = \frac{\eta}{(4\pi)^2}
\end{equation}
we find the RGEs 
\begin{align}\label{eq:SO3RGEs}
    \beta_{\alpha_Y} &= - 4 \epsilon \,\alpha_Y^2  + 4\left[6 + \pi^2 + \mathcal{O}(\epsilon^2)
    \right]  \alpha_Y^3 - \frac{32}{3} \epsilon \,\alpha_\eta \alpha_Y^2 + \frac{4}{9}\, \alpha_\eta^2 \alpha_Y + \mathcal{O}(\alpha^4)\,, \\
    \beta_{\alpha_\eta} &= \frac{17}{3} \alpha_\eta^2 + 48  \alpha_Y \alpha_\eta - 288 \epsilon \, \alpha_Y^2  - \left[288 \pi^2 + \mathcal{O}(\epsilon) \right] \alpha_Y^3 + \alpha_\eta \mathcal{O}(\alpha^2) + \mathcal{O}(\alpha^4) \,.
\end{align}

Note that the two-loop contribution of $\beta_{\alpha_Y}$ is negative. This is
due to fermions transforming under a spinor representation of $\mathrm{SO}(n)$,
as discussed earlier. However, the fact that the two-loop coefficient is also
suppressed $\propto \epsilon$ is a special feature of $n=3$. This is evident
from~\eq{betaYSON}, where  dominant large-$N$ contributions $\propto d_\gamma$
vanish at two loops and first reappear in the RGEs at four loops. More
precisely, there is a cancellation of terms $\propto \gamma_{\phi,1}^{(2)}$ and
$\propto \beta_{Y,2}^{(2)}$. We could have also assigned an additional
multiplicity $N_s$ to the scalar field and observe an equivalent cancellation
between  $\propto \gamma_{\psi,1}^{(2)}$ and $\propto \beta_{Y,5}^{(2)}$. The
special constellation of $\mathrm{SO}(3)$ spinors yields an IR fixed point
under perturbative control 
\begin{equation}\label{eq:SO3FP}
    \alpha_Y^* = \frac{\epsilon}{6+ \pi^2} + \mathcal{O}(\epsilon^2) \,, \qquad \alpha_\eta^* = \frac{12 (3 + \pi^2)\epsilon^2}{(6+ \pi^2)^2} + \mathcal{O}(\epsilon^3)\,.
\end{equation}

\begin{figure}
    \centering
    \includegraphics[scale=.45]{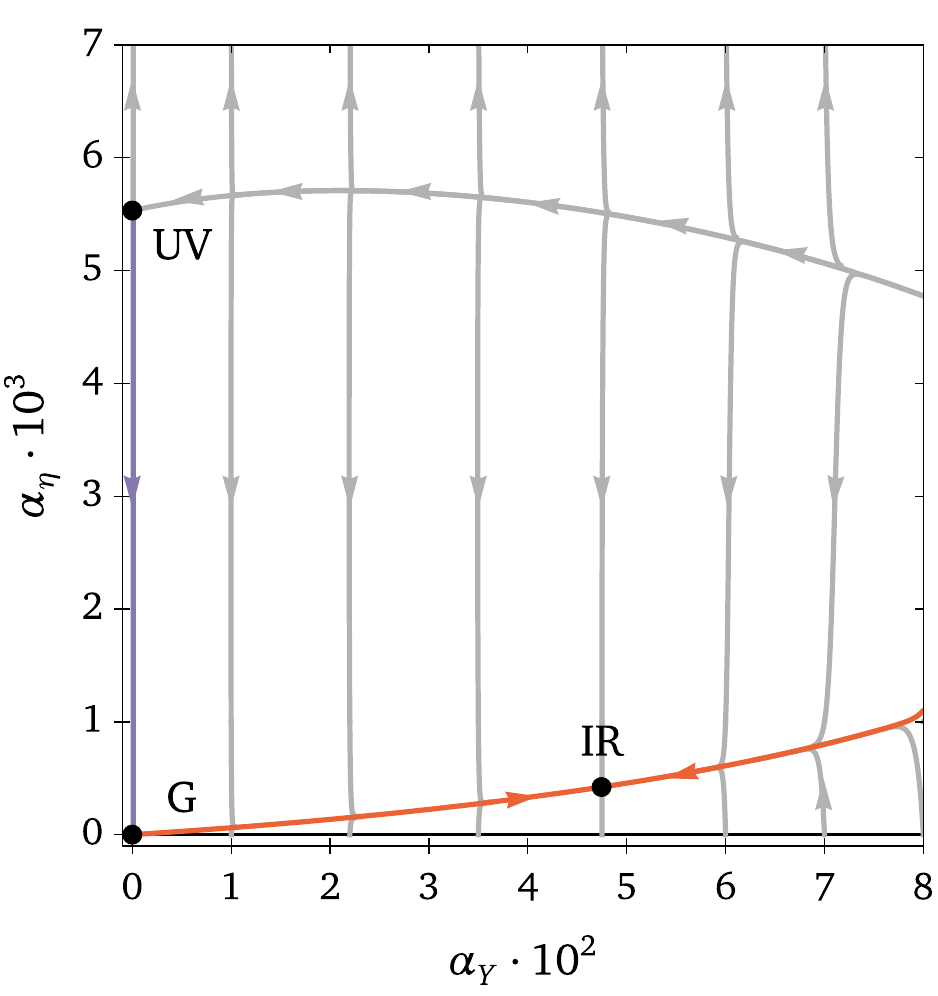}
    \caption{RG flow for the theory~\eq{SO3-lag} with $\epsilon = 10^{-1}$, showing the Gaussian fixed point ($\alpha_Y^* = \alpha_\eta^* = 0$), the infrared fixed point~\eq{SO3FP}, and the purely scalar ultraviolet fixed point ($\alpha_Y^* = 0$)~\cite{Townsend:1976sy,Appelquist:1981sf,Pisarski:1982vz,Kvedaraite:2025lgi}. Arrows point from the UV to the IR. The UV fixed point is connected to the Gaussian only through the trajectory $\alpha_Y = 0$ (violet). The RG flow along the red trajectory is slower--around the IR fixed point, it is suppressed by factor $\propto\epsilon^2$ as opposed to $\propto \epsilon^1$. }
    \label{fig:flow}
\end{figure}

Since $\alpha^*_\eta > 0$,  the scalar potential is bounded from below. The two
eigenvalues of the corresponding stability matrix $S_{m n} = (\partial
\beta_{\alpha_m} / \partial \alpha_n) \big\vert_{\alpha^{*}}$ read
\begin{equation}
    \vartheta_1 = \frac{4 \epsilon^2}{6+\pi^2} + \mathcal{O}(\epsilon^3)\,,\qquad \vartheta_2 = \frac{48 \epsilon}{6+\pi^2} + \mathcal{O}(\epsilon^2) \,,
\end{equation}
and reveal that the fixed point is totally IR attractive. The RG flow around
the fixed point is suppressed by $\propto \epsilon^2$ in one of the directions.
This slow trajectory also connects the IR fixed point with the Gaussian one
($\alpha_Y^* = \alpha_\eta^* = 0$). The situation is depicted in \fig{flow} for
the case $\epsilon=10^{-1}$, where the slow trajectory is marked red. We also
show the purely scalar UV fixed
point~\cite{Townsend:1976sy,Appelquist:1981sf,Pisarski:1982vz,Kvedaraite:2025lgi}
in this setting, which is not under strict perturbative control as the number
of scalars is not large. The UV fixed point is not directly connected to the IR
one~\eq{SO3FP}, but flows into the Gaussian along $\alpha_Y = 0$, which is
coloured violet in \fig{flow}. There are hints of even more ultraviolet fixed
points, though they lie beyond the perturbative region. 

For completeness, we also investigate if our model predicts fixed points in
$d=4$ via dimensional continuation from $d=3+\delta$. This is established by
evaluating 
\begin{equation}
    0 = 2 \delta \alpha_{Y,\eta} + \beta_{\alpha_{Y,\eta}}
\end{equation}  
using the four-loop RGEs~\eq{SO3RGEs} and evaluating fixed points as a power
series in $\delta$. As the two-loop coefficient of $\beta_{\alpha_{Y}}$ is
negative, an interacting UV fixed point emerges at $0 < \delta \ll 1$, which
needs to be tracked to $\delta \to 1$. Conversely, there is no fixed-point
solution for $\delta < 0$, excluding fixed points in two dimensions. However,
such phenomena are predicted in different three-dimensional theories, see for
instance~\cite{Fraser-Taliente:2024rql} in the context of melonic CFTs. Back
to the model at hand, we find 
\begin{equation}
    \alpha_Y^* = \frac{\delta}{2 \epsilon} + \mathcal{O}(\delta^2)\,,
\end{equation}
highlighting that the dimensional continuation is not reliable in the limit
$\epsilon \to 0$ where perturbative control is established in $d=3$, but rather
towards the upper bound $\epsilon \to 1$. We find two fixed-point solutions 
\begin{align}
    \alpha_Y^\pm &= \frac{\delta}{2 \epsilon} + \left[ \frac{50437}{5202} + \frac{173}{64} \pi^2 \mp \frac{94\sqrt{577} }{289} + \mathcal{O}(1-\epsilon) \right] \delta^2 + \mathcal{O}(\delta^3)\,,\notag \\
    \alpha_\eta^\pm &= \left[\frac{3 (\pm \sqrt{577} - 13)}{17} + \mathcal{O}(1-\epsilon) \right]\delta +  \mathcal{O}(\delta^2)\,,
\end{align}
where the coefficient $\propto \delta^2$ for $\alpha_\eta^\pm $ is known but
omitted for brevity. Only $\alpha_{Y,\eta}^+$ describes a stable fixed point as
the leading coefficient of $\alpha_\eta^-$ is negative. Computing the
eigenvalues of the stability matrix yields only one UV attractive direction
\begin{equation}
    \vartheta_1 = - 2\,\delta + \left[ \frac{100874}{2601} - \frac{376 \sqrt{577}}{289} + \frac{173}{16}\pi^2 + \mathcal{O}(1-\epsilon)\right]\delta^2 + \mathcal{O}(\delta^3)\,.
\end{equation}
A smooth limit $\delta \to 1$ is not possible, given the UV fixed point turns
marginal around $\delta \approx 0.0175$, possibly hinting at a fixed-point
merger. Although the precision is rather limited, there is no indication that
the UV fixed point survives the continuation to four dimensions.

\section{Conclusions}\label{sec:conclusions}

In this work, we derived general, renormalisation-scheme-agnostic template
expressions at four loops for all $\beta$-functions and anomalous dimensions in
any three-dimensional renormalisable theory with scalars and fermions. We
explicitly computed all coefficients of the template RGEs in the
$\overline{\text{MS}}$ scheme and checked them against existing literature. 

Our results are compatible with both $\mathcal{N}=1$ and $\mathcal{N}=2$
supersymmetry to emerge with general superpotentials. In fact, we have
extracted algebraic relations required to hold in any supersymmetry-preserving
renormalisation scheme. Both $\mathcal{N}=1$ and $\mathcal{N}=2$ can be
extended to four dimensions, where they correspond to $\mathcal{N}=\tfrac12$
and $\mathcal{N}=1$, respectively. Curiously, $\mathcal{N}=\tfrac12$
supersymmetry is violated in the $\overline{\text{MS}}$ at four-loop
order~\cite{Zerf:2017zqi,Steudtner:2025blh}, while its counterpart in three dimensions is
not. The source of the violation are terms stemming from odd spinor traces,
which always vanish in four dimensions. At four-loop in $d=3$, such terms do
not contribute, though they might play a role at higher loops. It would be
interesting to see if such terms eventually pose a problem for $\mathcal{N}=1$
supersymmetry in three dimensions.

As a direct application of our results, we have surveyed template RGEs for
perturbative fixed points in strictly three dimensions. In doing so, we
discovered a systematic way to construct theories with such critical points.
This revealed a rare example of a QFT with an interacting IR fixed point
under strict perturbative control. Our search for such theories, however, 
is not exhaustive, and other critical phenomena may yet be discovered.

A natural direction for future work is to search for a non-trivial UV fixed
point, which would render the theory asymptotically safe. Such a scenario is
not unmotivated: UV fixed points are found in purely
scalar~\cite{Townsend:1976sy,Appelquist:1981sf,Pisarski:1982vz,Kvedaraite:2025lgi}
and purely
fermionic~\cite{Rosenstein:1988pt,Gat:1990xi,deCalan:1991km,Braun:2011pp,Jakovac:2014lqa}
theories in three dimensions. In contrast, UV completion in four dimensions is
a more complicated business: at least a non-Abelian gauge sector is
required~\cite{Coleman:1973sx}, and even more restrictions apply for
non-trivial fixed points~\cite{Bond:2016dvk,Bond:2018oco,Steudtner:2020jcj}.  

Our work has revealed that the perturbative landscape of critical points in
three dimensions is quite rich. This picture will become more complete once
the catalogue of template RGEs is extended to include gauge interactions. The gauge
sector admits an exactly marginal Chern--Simons term at level $k$ amenable
to a controlled large-$k$ expansion. This, in turn, provides a systematic
setting for perturbative tests of dualities between fermionic and scalar
Chern--Simons theories enjoying rich applications, e.g. in condensed-matter
physics~\cite{Aharony:2015mjs,Seiberg:2016gmd,Aharony:2016jvv,DiPietro:2017kcd,DiPietro:2017vsp,DiPietro:2019hqe,Amoretti:2025hpi}.
In contrast, the gauge coupling is classically relevant and requires
e.g. large-$N$ expansions to be studied with perturbative methods.
A prominent example is
QED${}_{3}$~\cite{Appelquist:1981vg,Appelquist:1986qw,Appelquist:1986fd,Appelquist:1988sr,Nash:1989xx,Chester:2016ref},
sharing phenomena such as confinement, chiral symmetry breaking, and asymptotic
freedom with real-world QCD${}_{4}$.

\section*{Acknowledgements}
Y.S. acknowledges support from ANID under FONDECYT project No.~1231056 and Exploraci\'on Project No.~13250014.
E.S. and M.U. are supported by the Mercator Research Center Ruhr under Project No.~Ko-2022-0012. 
M.U. is supported by the doctoral scholarship program of the \textit{Studienstiftung des deutschen Volkes}.

\begin{appendix}

\section{Master integrals at four loops}\label{app:masters}

In this Appendix, we collect the fully massive tadpole master integrals,
required to compute the template RGEs in dimensional regularisation and the
$\overline{\mathrm{MS}}$ scheme up to four loops. First, in
App.~\ref{app:analytic}, we provide an update on~\cite{Schroder:2003kb} by
giving analytical results where available, and complementing with some
relations found by applying the integer relation algorithm
\texttt{PSLQ}~\cite{FergusonBailey1992,FergusonBaileyArno1999} to our
high-precision numerical expansions. We show the numerical results for all
master integrals in App.~\ref{app:numerics}. Lastly, in
App.~\ref{app:basisChange}, we discuss alternative choices of masters, and
connect with the basis used in the \texttt{FORM} package
\texttt{FMFT}~\cite{Pikelner:2017tgv}.

While we present analytic and numeric data in an odd dimension $d=3-2\epsilon$
here, let us note that the equivalent information for an even dimension
$d=2-2\epsilon$ had been listed in~\cite{Gracey:2016mio}. Taken together, this
suffices for expansions around any integer dimension, given the existence of
dimensional recurrences~\cite{Tarasov:1996br} that allow to analytically map
$d\leftrightarrow d+2$.

\subsection{Analytic Results in \texorpdfstring{$d=3-2\epsilon$}{d=3 - 2 epsilon}}
\label{app:analytic}

\begin{figure}[thb]
\begin{align*}
\renewcommand{\sbx}{\scalebox{0.6}}
\begin{array}{l@{\hspace{5mm}}ccccccc}
&\TLfig{1}
&\TLfig{7}
&\TLfig{51}
&\TLfig{62}
&\TLfig{63}
&\TLfig{841}
\\[5.5ex]
&\TLfig{993}
&\TLfig{952}
&\TLfig{1016}
&\TLfig{1010}
&\TLfig{1009}
&\TLfig{1020}
\\[5.5ex]
&\TLfig{1011}
&\TLfig{1022}
&\TLfig{511}
&\TLfig{841.1.3}
&\TLfig{1009.1.2}
&\TLfig{1011.1.2}
\end{array}
\end{align*}
\caption{\label{fig:masters}The master integrals up to four loops, labelled by their respective sector identifiers whose binary representation corresponds to the propagators that are present, with momenta from the set
$\{k_1,k_2,k_3,k_4,k_1-k_4,k_2-k_4,k_3-k_4,k_1-k_2,k_1-k_3,k_1-k_2-k_3\}$ in the four-loop case.
The dotted masters carry an additional pair of labels indicating which propagator carries a non-unity power.
For possible alternative choices of the latter class, see the basis change relations of App.~\ref{app:basisChange}.}
\end{figure}

We treat here the set of master integrals displayed in Fig.~\ref{fig:masters}.
All our propagators $1/(1+p_i^2)$ are Euclidean with unit mass.
We normalise each $L$-loop tadpole with the (1-loop tadpole)$^L$, leading to
\begin{align}
I_{1} &= 1
\end{align}
for the 1-loop tadpole, exactly.
For reference, in any other normalisation, the 1-loop tadpole can be inferred from 
$\int{\rm d}^dp/(1+p^2)=\pi^{d/2}\,\Gamma(1-d/2)$. 

At two loops, the only master integral $I_7$ can be represented by a
hypergeometric function~\cite{Davydychev:1992mt}, see also Sec.~6.2
of~\cite{Schroder:2005va}. For its (fast) expansion around $d=3-2\ep$
dimensions, we write 
\begin{align*}
I_{7} &= 
\frac{1-2\ep}{2\ep}\,\Big\{
\frac32
-\frac9{16}\sum_{j=1}^\infty(-\ep)^j\,g_j
-\frac{\cos(\pi\ep)}{\sqrt{{\rm sinc}(2\pi\ep)}}\,\exp\Big[\ep\ln(\tfrac{16}3)+2\ep\sum_{j=1}^\infty\ep^{2j}\,(4^j-1)\frac{\zeta(2j+1)}{2j+1}\Big]
\Big\} \,,\\
g_j &\equiv {}_{j+2}F_{j+1}(\tfrac32,1,\dots,1;2,\dots,2;\tfrac34) \,,
\end{align*}
in terms of generalized hypergeometric functions $g_j$ (which can be assigned
transcendental weight $j$, and evaluate rapidly), odd zeta values, and with
${\rm sinc}(x)\equiv\frac{\sin(x)}x$. We note that
$g_1=\frac{16}3\,\ln(\frac43)$, $g_2=\frac{16}3\,{\rm
Li}_2(\frac14)-\frac83\,\ln^2(\frac43)$,
$g_3={}_{5}F_{4}(\tfrac32,1,1,1,1;2,2,2,2;\tfrac34)$, etc.

Another analytically known result is the three-spoked wheel at $d=3$. From
Broadhurst's work~\cite{Broadhurst:1998iq,Broadhurst:1998ke} we know the
leading term of the fully massive 3d 3-loop tetrahedron $I_{63}$ to be a
difference of two Clausen integrals
\begin{align}
I_{63} &= c_{\rm tet} + {\cal O}(\ep) \,,\quad\mbox{with~~} c_{\rm tet}=\tfrac1{\sqrt2}\, {\rm Im}\Big[{\rm Li}_2(w^2)-{\rm Li}_2(w^4)\Big]\,,
\end{align}
where $w=e^{i\arcsin(1/3)}=(\sqrt{8}+i)/3$.

For two other integrals, $I_{51}$ and $I_{952}$ (which in fact constitute the
first two instances of a class of necklace integrals), it can be shown that
their expansion coefficients contain multiple zeta values (MZVs) only. The
constructive proof follows from introducing two distinct masses, and solving a
differential equation in $z=M/m$ in terms of harmonic polylogarithms, which at
$z=1$ reduce to MZVs~\cite{Vermaseren:1998uu,Blumlein:1998if,Blumlein:2009cf}. Recently, a
recursive construction of the expansion coefficients of the necklace integrals
has been given, see App.~A of~\cite{Kvedaraite:2025lgi}.

Some further terms had contributed to specific physics questions and therefore
been analytically identified before. See, e.g.~\cite{Kajantie:2003ax}, where
the 4-loop integrals $I_{841}$, $I_{993}$ and $I_{952}$ were needed up to their
constant $\ep^0$ coefficients (the $\ep^1$ terms of $I_{51}$ and $I_{7}$ had
entered as well).

Using high-precision numerical results (for reference, we show 65 digits and a
dozen $\ep$-orders in App.~\ref{app:numerics}) generated by an in-house
implementation~\cite{Luthe:2015ngq} of Laporta's difference equation
method~\cite{Laporta:2000dsw}, we have attempted to identify some
numbers by the integer relation algorithm \texttt{PSLQ}, implemented in
Mathematica~\cite{Mathematica} as {\tt FindIntegerNullVector}. We have mostly
searched at transcendental weight 2 (two $\ep^0$ terms are found with weight 3,
$I_{952}$ and $I_{1016}$), and our (rather meager) findings are documented
below. Note that some dilogarithms with different arguments are related, such
as ${\rm Li}_2(\frac14)+2{\rm Li}_2(\frac13)=\zeta(2)-\ln^2(2)-\ln^2(\frac32)$.
In the interest of compact formulae, for some integrals we pull out a rational
prefactor that renders the remaining expansion in pure-weight form.
\begin{align*}
I_{7} &= \frac{1-2\ep}{4\ep}\,\Big\{
1 - 4 \ln(\tfrac32)\,\ep + 4 \lk 3{\rm Li}_2(\tfrac13) + 2 \ln^2(\tfrac32) - \zeta_2 \rk\,\ep^2\\ 
&+ \lk \tfrac98\,g_3 - \tfrac13\,\ln^3(\tfrac{16}3) + 2 \ln(\tfrac{16}3)\zeta_2 - 4 \zeta_3 \rk\,\ep^3\\ 
&+ \lk -\tfrac98\,g_4 - \tfrac1{12}\,\ln^4(\tfrac{16}3) + \ln^2(\tfrac{16}3)\zeta_2 + \tfrac92\,\zeta_4 - 4 \ln(\tfrac{16}3)\zeta_3  \rk\,\ep^4\\
&+ \lk \tfrac98\,g_5 - \tfrac1{60}\,\ln^5(\tfrac{16}3) + \tfrac13\,\ln^3(\tfrac{16}3)\zeta_2 + \tfrac92\,\ln(\tfrac{16}3)\zeta_4 - 2 \ln^2(\tfrac{16}3)\zeta_3  +4\zeta_2\zeta_3 -12\zeta_5 \rk\,\ep^5 \\
&+ \dots \Big\}\,,\tag{\stepcounter{equation}\theequation}\\[.5em]
I_{51} &= \frac{(1-2\ep)^2}{(1-6\ep)\ep}\,\Big\{
1 - 4 \ln(2)\,\ep  + 4 \lk \ln^2(2) + \zeta_2 \rk\,\ep^2 +8 \lk -\tfrac13\,\ln^3(2) - \ln(2) \zeta_2 -\tfrac{19}4\,\zeta_3 \rk\,\ep^3 \\
&+ 8 \lk 20 {\rm Li}_4(\tfrac12) + \ln^4(2) - 4 \ln^2(2) \zeta_2 + 27  \ln(2)  \zeta_3 + 11 \zeta_4 \rk\,\ep^4\\
&+ 8 \lk 40 {\rm Li}_5(\tfrac12) -\tfrac25\, \ln^5(2) +\tfrac83\,\ln^3(2) \zeta_2 - 27 \ln^2(2)  \zeta_3 - 27 \zeta_2  \zeta_3 -22 \ln(2) \zeta_4 - \tfrac{1445}8\,\zeta_5 \rk\,\ep^5 \\
&+\dots \Big\}\,,\tag{\stepcounter{equation}\theequation}\\[.5em]
I_{62} &= \frac{(1-2\ep)^2}{\ep^2}\,\Big\{ 
\lk -2{\rm Li}_2(\tfrac13) -\tfrac12\ln^2(3) +\tfrac12\zeta_2  \rk\,\ep^2\\
&+ \lk -\tfrac{45}{64}\,g_3 +\tfrac52\,{\rm Li}_3(\tfrac13) +14{\rm Li}_2(\tfrac13)\ln(2) +20\ln^3(2) -4{\rm Li}_2(\tfrac13)\ln(3) -30\ln^2(2)\ln(3) \\
&+\tfrac{37}2\,\ln(2)\ln^2(3) -\tfrac{85}{24}\,\ln^3(3) -11\ln(2)\zeta_2 +\tfrac{19}4\,\ln(3)\zeta_2 +\tfrac{79}{24}\,\zeta_3 \rk\,\ep^3
+\dots \Big\}\,,\tag{\stepcounter{equation}\theequation}\\[.5em]
I_{63} &= (c_\text{tet}) +{\cal O}(\ep)\,,\tag{\stepcounter{equation}\theequation}\\[.5em]
I_{841} &= \frac{15}{4\ep(1+2\ep)(1-6\ep)}\,\Big\{ 
\tfrac34 + \tfrac{10}3 \ln(\tfrac25) \,\ep
+ \lk -6{\rm Li}_2(\tfrac13) +6{\rm Li}_2(\tfrac15) +6{\rm Li}_2(\tfrac16) \\
&+5\zeta_2 -3\ln^2(5) +11\ln^2(\tfrac25) +6\ln(2)\ln(3) \rk\,\ep^{2}
+\dots \Big\}\,,\tag{\stepcounter{equation}\theequation}\\[.5em]
I_{993} &= \frac{1+12\ep^2}{4\ep^2}\,\Big\{ 
\tfrac18 + \ln(\tfrac23) \,\ep
+ \lk 7{\rm Li}_2(\tfrac13) +4{\rm Li}_2(\tfrac15) -4{\rm Li}_2(\tfrac16)\\
&-5\zeta_2 +2\ln^2(3) +2\ln^2(\tfrac25) -4\ln(2)\ln(5) \rk\,\ep^{2}
+\dots \Big\}\,,\tag{\stepcounter{equation}\theequation}\\[.5em]
I_{952} &= \frac{(1-2\ep)^3}{\ep^3}\,\Big\{
\tfrac{3}{16}\,\zeta_2\,\ep^2
+\tfrac38 \lk 2 \ln(2) \zeta_2 - 7  \zeta_3 \rk\,\ep^3\\ 
&+ \tfrac34 \lk  40 {\rm Li}_4(\tfrac12) + \tfrac53\, \ln^4(2) - 8 \ln^2(2) \zeta_2 + 21  \ln(2)  \zeta_3 - \tfrac{25}2\, \zeta_4 \rk\,\ep^4\\ 
&+ \lk - 228 {\rm Li}_5(\tfrac12) - 108 {\rm Li}_4(\tfrac12)\ln(2) -\tfrac{13}5\,\ln^5(2) +10\ln^3(2)\zeta_2 -\tfrac{63}4\,\ln^2(2)\zeta_3\\
&- \tfrac{135}8\,\zeta_2\zeta_3 -\tfrac{75}2\,\ln(2)\zeta_4 +\tfrac{1023}8\,\zeta_5 \rk\,\ep^5
+\dots \Big\}\,,\tag{\stepcounter{equation}\theequation}\\[.5em]
I_{1016} &= \frac{1}{\ep^3} \,\Big\{ 
\lk -3 {\rm Li}_3(\tfrac13) +\tfrac{17}8\zeta_3 -\tfrac{3}{2}\ln(2)\zeta_2  +\tfrac14\ln^3(3) \rk\,\ep^3
+\dots \Big\}\,,\tag{\stepcounter{equation}\theequation}\\[.5em]
I_{1010} &= (c_{1010}) +{\cal O}(\ep)\,,\tag{\stepcounter{equation}\theequation}\\[.5em]
I_{1009} &= (c_{1009}) +{\cal O}(\ep)\,,\tag{\stepcounter{equation}\theequation}\\[.5em]
I_{1020} &= (c_{1020}) +{\cal O}(\ep)\,,\tag{\stepcounter{equation}\theequation}\\[.5em]
I_{1011} &= (c_{1011}) +{\cal O}(\ep)\,,\tag{\stepcounter{equation}\theequation}\\[.5em]
I_{1022} &= (c_{1022}) +{\cal O}(\ep)\,,\tag{\stepcounter{equation}\theequation}\\[.5em]
I_{511} &= (c_{511}) +{\cal O}(\ep)\,,\tag{\stepcounter{equation}\theequation}\\[.5em]
I_{841.1.3} &=  -\frac{1}{2\ep(1-4\ep)} \,\Big\{ 
\tfrac14 +\lk-\tfrac14+\ln(\tfrac25)\rk\,\ep
+\lk - {\rm Li}_2(\tfrac13) + {\rm Li}_2(\tfrac15) + {\rm Li}_2(\tfrac16) +\tfrac32\zeta_2 \\
&- \tfrac12\ln^2(5) +\tfrac52\ln^2(\tfrac25) +\ln(2)\ln(3) \rk\,\ep^2
+\dots \Big\}\,,\tag{\stepcounter{equation}\theequation}\\[.5em]
I_{1009.1.2} &=  \frac{1}{12\ep^2} \,\Big\{ 
\lk 2c_{1009} -9 {\rm Li}_2(\tfrac13) +4 {\rm Li}_2(\tfrac15) +5 {\rm Li}_2(\tfrac16) +\zeta_2\\ 
&+2\ln^2(5) -3\ln^2(3) +\tfrac52\ln^2(2) +8\ln(2)\ln(\tfrac35) \rk\,\ep^2
+\dots \Big\}\,,\tag{\stepcounter{equation}\theequation}\\[.5em]
I_{1011.1.2} &=  \frac{1}{12\ep^2} \,\Big\{ 
\lk 4c_{1011} -10 {\rm Li}_2(\tfrac13) +8 {\rm Li}_2(\tfrac15) +6 {\rm Li}_2(\tfrac16) +\tfrac12\zeta_2\\
&+4\ln^2(5) -5\ln^2(3) +3\ln^2(2) +16\ln(2)\ln(\tfrac35) \rk\,\ep^2
+\dots \Big\}\,.\tag{\stepcounter{equation}\theequation}
\end{align*}
Note that six constants $c_s$, corresponding to finite (as $d\rightarrow3$)
corner integrals of some sectors $s$ have not yet been identified analytically.
Two of them ($c_{1009}$ and $c_{1011}$) appear also in the leading terms of the
corresponding dotted master integrals, suggesting they are of weight 2.


\subsection{Numerical Results in \texorpdfstring{$d=3-2\epsilon$}{d=3 - 2 epsilon}}\label{app:numerics}

The set of integrals shown in Fig.~\ref{fig:masters}, normalised as explained
in App.~\ref{app:analytic}, is given in numerical form (we chose to display 65
digits and a dozen $\ep$-orders here) as
\begin{small}
\begin{align}
\input num3d/tex.3d.2.7.1.1.out 
\,,\\[.5em]
\input num3d/tex.3d.3.51.1.1.out 
\,,\\[.5em]
\input num3d/tex.3d.3.62.1.1.out
\,,\\[.5em]
\input num3d/tex.3d.3.63.1.1.out 
\,,\\[.5em]
\input num3d/tex.3d.4.841.1.1.out
\,,\\[.5em]
\input num3d/tex.3d.4.993.1.1.out 
\,,\\[.5em]
\input num3d/tex.3d.4.952.1.1.out
\,,\\[.5em]
\input num3d/tex.3d.4.1016.1.1.out 
\,,\\[.5em]
\input num3d/tex.3d.4.1010.1.1.out
\,,\\[.5em]
\input num3d/tex.3d.4.1009.1.1.out 
\,,\\[.5em]
\input num3d/tex.3d.4.1020.1.1.out
\,,\\[.5em]
\input num3d/tex.3d.4.1011.1.1.out 
\,,\\[.5em]
\input num3d/tex.3d.4.1022.1.1.out
\,,\\[.5em]
\input num3d/tex.3d.4.511.2.1.out 
\,,\\[.5em]
\input num3d/tex.3d.4.841.1.3.out
\,,\\[.5em]
\input num3d/tex.3d.4.1009.1.2.out 
\,,\\[.5em]
\input num3d/tex.3d.4.1011.1.2.out 
\,.
\end{align}
\end{small}


\subsection{Alternative Choices for Dotted Masters}
\label{app:basisChange}

Other possible choices (than our $I_{841.1.3}$, $I_{1009.1.2}$ and
$I_{1011.1.2}$ shown in Fig.~\ref{fig:masters}) for the dotted masters can be
inferred from linear relations among the integrals. Some relations follow from
simple dimensional analysis (by hitting the corner integral with mass
derivatives), such as
\begin{align}
(d-3)(5-2d) \diag{841} +5 \diag{841.1.3} +10 \diag{841.1.2.2.2}  &= 0 \,, \\
(d-4) \diag{1011} +2 \diag{1011.1.2}  +2 \diag{1011.3.2}  &= 0 \,, \label{eq:1011} \\
(2d-7) \diag{1009} +4  \diag{1009.1.2}  +\diag{1009.4.2} +2 \diag{1009.3.2} &= 0 \;. \label{eq:1009}
\end{align}
Other relations can be derived via IBPs, such as
\begin{align}
\label{eq:ibp1009}
(d-4)\,\diag{1009} +2\,\diag{1009.1.2} +2\,\diag{1009.4.2}  &=
\frac{3d-8}4\,\diag{993} -\frac{2(d-3)}3\,\diag{952} 
+\frac{(d-2)^2}{8(d-3)}\,\diag{7}
\nn\\
&\quad-\frac{(2d-5)(2d-7)}{20(d-3)}\,\diag{841} +\frac5{4(d-3)}\,\diag{841.1.3} \;.
\end{align}
Together with (\ref{eq:1009}), the latter relation can be used to, e.g. 
exchange our master integral choice $I_{1009.1.2}$ (cf.\
Fig.~\ref{fig:masters}) with $I_{1009.3.2}$ (dot on peripheral line) or
$I_{1009.4.2}$ (dot on central line). Using (\ref{eq:1011}) one could exchange
our master integral choice $I_{1011.1.2}$ (dot on wheel rim) with
$I_{1011.3.2}$ (dot on spoke), and this was in fact the choice that had been
made in~\cite{Schroder:2003kb}. See Fig.~\ref{fig:alternatives} for these
alternatives.

\begin{figure}[thb]
\begin{align*}
\renewcommand{\sbx}{\scalebox{0.6}}
\begin{array}{l@{\hspace{5mm}}ccccccccc}
&\TLfig{841.1.2.2.2}
&\TLfig{1009.3.2}
&\TLfig{1009.4.2}
&\TLfig{1011.3.2}
\end{array}
\end{align*}
\caption{\label{fig:alternatives}Alternative choices for dotted masters. Note that the integral $I_{1009.4.2}$ is part of the basis chosen in the package \texttt{FMFT}~\cite{Czakon:2004bu,Pikelner:2017tgv}, replacing $I_{1009.1.2}$ of the basis displayed in Fig.~\ref{fig:masters}, with~\eq{ibp1009} available for an analytic basis change if needed.}
\end{figure}

Complementing the numerical values given in App.~\ref{app:numerics}, the set of
integrals shown in Fig.~\ref{fig:alternatives}, normalised as explained in
App.~\ref{app:analytic}, reads
\begin{small}
\begin{align}
\nn
\input num3d/tex.3d.4.841.1.2.2.2.out
\,,\\[.5em]
\input num3d/tex.3d.4.1009.3.2.out 
\,,\\[.5em]
\input num3d/tex.3d.4.1009.4.2.out 
\,,\\[.5em]
\input num3d/tex.3d.4.1011.3.2.out 
\,.
\end{align}
\end{small}
\end{appendix}
\bibliography{ref.bib}

@article{Machacek:1983tz,
  author       = {Machacek, Marie E. and Vaughn, Michael T.},
  title        = {{Two-loop renormalization group equations in a general quantum field theory: (I). Wave function renormalization}},
  journal      = {Nucl. Phys.},
  volume       = {B222},
  year         = {1983},
  pages        = {83-103},
  doi          = {10.1016/0550-3213(83)90610-7},
  reportnumber = {NUB-2590, HUTP-83/A003},
  slaccitation = {%%CITATION = NUPHA,B222,83;%%}
}

@article{Machacek:1983fi,
  author       = {Machacek, Marie E. and Vaughn, Michael T.},
  title        = {{Two-loop renormalization group equations in a general quantum field theory (II). Yukawa couplings}},
  journal      = {Nucl. Phys.},
  volume       = {B236},
  year         = {1984},
  pages        = {221-232},
  doi          = {10.1016/0550-3213(84)90533-9},
  reportnumber = {NUB 2611},
  slaccitation = {%%CITATION = NUPHA,B236,221;%%}
}

@article{Machacek:1984zw,
  author       = {Machacek, Marie E. and Vaughn, Michael T.},
  title        = {{Two-loop renormalization group equations in a general quantum field theory: (III). Scalar quartic couplings}},
  journal      = {Nucl. Phys.},
  volume       = {B249},
  year         = {1985},
  pages        = {70-92},
  doi          = {10.1016/0550-3213(85)90040-9},
  reportnumber = {NUB-2653-REV, NUB-2653},
  slaccitation = {%%CITATION = NUPHA,B249,70;%%}
}

@article{Martin:1993zk,
  author        = {Martin, Stephen P. and Vaughn, Michael T.},
  title         = {{Two-loop renormalization group equations for soft supersymmetry-breaking couplings}},
  eprint        = {hep-ph/9311340},
  archiveprefix = {arXiv},
  reportnumber  = {NUB-3081-93-TH},
  doi           = {10.1103/PhysRevD.50.2282},
  journal       = {Phys. Rev. D},
  volume        = {50},
  pages         = {2282},
  year          = {1994},
  note          = {[Erratum: Phys.Rev.D 78, 039903 (2008)]}
}

@article{Luo:2002ti,
  author        = {Luo, Mingxing and Wang, Huawen and Xiao, Yong},
  title         = {{Two-loop renormalization group equations in general gauge field theories}},
  eprint        = {hep-ph/0211440},
  archiveprefix = {arXiv},
  doi           = {10.1103/PhysRevD.67.065019},
  journal       = {Phys. Rev. D},
  volume        = {67},
  pages         = {065019},
  year          = {2003}
}

@article{Schienbein:2018fsw,
  author        = {Schienbein, Ingo and Staub, Florian and Steudtner, Tom and Svirina, Kseniia},
  title         = {{Revisiting RGEs for general gauge theories}},
  eprint        = {1809.06797},
  archiveprefix = {arXiv},
  primaryclass  = {hep-ph},
  reportnumber  = {KA-TP-27-2018},
  doi           = {10.1016/j.nuclphysb.2018.12.001},
  journal       = {Nucl. Phys. B},
  volume        = {939},
  pages         = {1--48},
  year          = {2019}
}

@article{Litim:2014uca,
  author        = {Litim, Daniel F. and Sannino, Francesco},
  title         = {{Asymptotic safety guaranteed}},
  eprint        = {1406.2337},
  archiveprefix = {arXiv},
  primaryclass  = {hep-th},
  reportnumber  = {CP3-ORIGINS-2014-24, DIAS-2014-24},
  doi           = {10.1007/JHEP12(2014)178},
  journal       = {JHEP},
  volume        = {12},
  pages         = {178},
  year          = {2014}
}

@article{Litim:2015iea,
  author        = {Litim, Daniel F. and Mojaza, Matin and Sannino, Francesco},
  title         = {{Vacuum stability of asymptotically safe gauge-Yukawa theories}},
  eprint        = {1501.03061},
  archiveprefix = {arXiv},
  primaryclass  = {hep-th},
  reportnumber  = {CP3-ORIGINS-2014-049, DIAS-2014-049, NORDITA-2014-145},
  doi           = {10.1007/JHEP01(2016)081},
  journal       = {JHEP},
  volume        = {01},
  pages         = {081},
  year          = {2016}
}

@article{tHooft:1973mfk,
  author  = {'t Hooft, Gerard},
  title   = {{Dimensional regularization and the renormalization group}},
  doi     = {10.1016/0550-3213(73)90376-3},
  journal = {Nucl. Phys. B},
  volume  = {61},
  pages   = {455--468},
  year    = {1973}
}

@article{Poole:2019kcm,
  author        = {Poole, Colin and Thomsen, Anders Eller},
  title         = {{Constraints on 3- and 4-loop $\beta$-functions in a general four-dimensional Quantum Field Theory}},
  eprint        = {1906.04625},
  archiveprefix = {arXiv},
  primaryclass  = {hep-th},
  doi           = {10.1007/JHEP09(2019)055},
  journal       = {JHEP},
  volume        = {09},
  pages         = {055},
  year          = {2019}
}

@article{Chetyrkin:2013wya,
  author        = {Chetyrkin, K.G. and Zoller, M.F.},
  title         = {{$\beta$-function for the Higgs self-interaction in the Standard Model at three-loop level}},
  eprint        = {1303.2890},
  archiveprefix = {arXiv},
  primaryclass  = {hep-ph},
  reportnumber  = {TTP13-008, SFB-CPP-13-19},
  doi           = {10.1007/JHEP04(2013)091},
  journal       = {JHEP},
  volume        = {04},
  pages         = {091},
  year          = {2013},
  note          = {[Erratum: JHEP 09, 155 (2013)]}
}

@article{Chetyrkin:2012rz,
  author        = {Chetyrkin, K.G. and Zoller, M.F.},
  title         = {{Three-loop $\beta$-functions for top-Yukawa and the Higgs self-interaction in the standard model}},
  eprint        = {1205.2892},
  archiveprefix = {arXiv},
  primaryclass  = {hep-ph},
  reportnumber  = {SFB-CPP-12-23, TTP12-012},
  doi           = {10.1007/JHEP06(2012)033},
  journal       = {JHEP},
  volume        = {06},
  pages         = {033},
  year          = {2012}
}

@article{Bednyakov:2013eba,
  author        = {Bednyakov, A.V. and Pikelner, A.F. and Velizhanin, V.N.},
  title         = {{Higgs self-coupling beta-function in the Standard Model at three loops}},
  eprint        = {1303.4364},
  archiveprefix = {arXiv},
  primaryclass  = {hep-ph},
  doi           = {10.1016/j.nuclphysb.2013.07.015},
  journal       = {Nucl. Phys. B},
  volume        = {875},
  pages         = {552--565},
  year          = {2013}
}

@article{Bednyakov:2012en,
  author        = {Bednyakov, A.V. and Pikelner, A.F. and Velizhanin, V.N.},
  title         = {{Yukawa coupling beta-functions in the Standard Model at three loops}},
  eprint        = {1212.6829},
  archiveprefix = {arXiv},
  primaryclass  = {hep-ph},
  doi           = {10.1016/j.physletb.2013.04.038},
  journal       = {Phys. Lett. B},
  volume        = {722},
  pages         = {336--340},
  year          = {2013}
}

@article{Bednyakov:2012rb,
  author        = {Bednyakov, A.V. and Pikelner, A.F. and Velizhanin, V.N.},
  title         = {{Anomalous dimensions of gauge fields and gauge coupling beta-functions in the Standard Model at three loops}},
  eprint        = {1210.6873},
  archiveprefix = {arXiv},
  primaryclass  = {hep-ph},
  doi           = {10.1007/JHEP01(2013)017},
  journal       = {JHEP},
  volume        = {01},
  pages         = {017},
  year          = {2013}
}

@article{Herren:2017uxn,
  author        = {Herren, Florian and Mihaila, Luminita and Steinhauser, Matthias},
  title         = {{Gauge and Yukawa coupling beta functions of two-Higgs-doublet models to three-loop order}},
  eprint        = {1712.06614},
  archiveprefix = {arXiv},
  primaryclass  = {hep-ph},
  reportnumber  = {TTP17-046, TTP17-046},
  doi           = {10.1103/PhysRevD.97.015016},
  journal       = {Phys. Rev. D},
  volume        = {97},
  number        = {1},
  pages         = {015016},
  year          = {2018},
  note          = {[Erratum: Phys.Rev.D 101, 079903 (2020)]}
}

@article{Mihaila:2012pz,
  author        = {Mihaila, Luminita N. and Salomon, Jens and Steinhauser, Matthias},
  title         = {{Renormalization constants and beta functions for the gauge couplings of the standard model to three-loop order}},
  eprint        = {1208.3357},
  archiveprefix = {arXiv},
  primaryclass  = {hep-ph},
  reportnumber  = {SFB-CPP-12-61, TTP12-30},
  doi           = {10.1103/PhysRevD.86.096008},
  journal       = {Phys. Rev. D},
  volume        = {86},
  pages         = {096008},
  year          = {2012}
}

@article{Bednyakov:2013cpa,
  author        = {Bednyakov, A.V. and Pikelner, A.F. and Velizhanin, V.N.},
  title         = {{Three-loop Higgs self-coupling beta-function in the Standard Model with complex Yukawa matrices}},
  eprint        = {1310.3806},
  archiveprefix = {arXiv},
  primaryclass  = {hep-ph},
  reportnumber  = {HU-MATHEMATIK-2013-20, HU-EP-13-53},
  doi           = {10.1016/j.nuclphysb.2013.12.012},
  journal       = {Nucl. Phys. B},
  volume        = {879},
  pages         = {256--267},
  year          = {2014}
}

@article{Chetyrkin:2016ruf,
  author        = {Chetyrkin, K.G. and Zoller, M.F.},
  title         = {{Leading QCD-induced four-loop contributions to the $\beta$-function of the Higgs self-coupling in the SM and vacuum stability}},
  eprint        = {1604.00853},
  archiveprefix = {arXiv},
  primaryclass  = {hep-ph},
  reportnumber  = {TTP16-008, ZU-TH-3-16},
  doi           = {10.1007/JHEP06(2016)175},
  journal       = {JHEP},
  volume        = {06},
  pages         = {175},
  year          = {2016}
}

@article{Zoller:2015tha,
  author        = {Zoller, M.F.},
  title         = {{Top-Yukawa effects on the $\beta$-function of the strong coupling in the SM at four-loop level}},
  eprint        = {1508.03624},
  archiveprefix = {arXiv},
  primaryclass  = {hep-ph},
  reportnumber  = {TTP15-031},
  doi           = {10.1007/JHEP02(2016)095},
  journal       = {JHEP},
  volume        = {02},
  pages         = {095},
  year          = {2016}
}

@article{Pickering:2001aq,
  author        = {Pickering, A.G.M. and Gracey, J.A. and Jones, D.R.T.},
  title         = {{Three loop gauge $\beta$-function for the most general single gauge-coupling theory}},
  eprint        = {hep-ph/0104247},
  archiveprefix = {arXiv},
  reportnumber  = {IUHET-434, LTH-498},
  doi           = {10.1016/S0370-2693(01)00624-4},
  journal       = {Phys. Lett. B},
  volume        = {510},
  pages         = {347--354},
  year          = {2001},
  note          = {[Erratum: Phys.Lett.B 535, 377 (2002)]}
}

@article{Mihaila:2014caa,
  author  = {Mihaila, Luminita},
  title   = {{Three-loop gauge beta function in non-simple gauge groups}},
  doi     = {10.22323/1.197.0060},
  journal = {PoS},
  volume  = {RADCOR2013},
  pages   = {060},
  year    = {2013}
}

@article{Sperling:2013eva,
  author        = {Sperling, Marcus and Stöckinger, Dominik and Voigt, Alexander},
  title         = {{Renormalization of vacuum expectation values in spontaneously broken gauge theories}},
  eprint        = {1305.1548},
  archiveprefix = {arXiv},
  primaryclass  = {hep-ph},
  doi           = {10.1007/JHEP07(2013)132},
  journal       = {JHEP},
  volume        = {07},
  pages         = {132},
  year          = {2013}
}

@article{Sperling:2013xqa,
  author        = {Sperling, Marcus and Stöckinger, Dominik and Voigt, Alexander},
  title         = {{Renormalization of vacuum expectation values in spontaneously broken gauge theories: two-loop results}},
  eprint        = {1310.7629},
  archiveprefix = {arXiv},
  primaryclass  = {hep-ph},
  doi           = {10.1007/JHEP01(2014)068},
  journal       = {JHEP},
  volume        = {01},
  pages         = {068},
  year          = {2014}
}

@article{Herzog:2017ohr,
  author        = {Herzog, F. and Ruijl, B. and Ueda, T. and Vermaseren, J.A.M. and Vogt, A.},
  archiveprefix = {arXiv},
  doi           = {10.1007/JHEP02(2017)090},
  eprint        = {1701.01404},
  journal       = {JHEP},
  pages         = {090},
  primaryclass  = {hep-ph},
  reportnumber  = {NIKHEF-2017-001, LTH-1117},
  title         = {The five-loop beta function of Yang-Mills theory with fermions},
  volume        = {02},
  year          = {2017}
}

@article{Luthe:2017ttg,
  author        = {Luthe, Thomas and Maier, Andreas and Marquard, Peter and Schröder, York},
  archiveprefix = {arXiv},
  doi           = {10.1007/JHEP10(2017)166},
  eprint        = {1709.07718},
  journal       = {JHEP},
  pages         = {166},
  primaryclass  = {hep-ph},
  reportnumber  = {BI-TP-2017-13, DESY-17-142, IPPP-17-68},
  title         = {{The five-loop Beta function for a general gauge group and anomalous dimensions beyond Feynman gauge}},
  volume        = {10},
  year          = {2017}
}

@article{Czakon:2004bu,
  author        = {Czakon, M.},
  archiveprefix = {arXiv},
  doi           = {10.1016/j.nuclphysb.2005.01.012},
  eprint        = {hep-ph/0411261},
  journal       = {Nucl.Phys.B},
  pages         = {485--498},
  reportnumber  = {DESY-04-223, SFB-CPP-04-62},
  title         = {{The four-loop QCD $\beta$-function and anomalous dimensions}},
  volume        = {710},
  year          = {2005}
}

@article{Chetyrkin:1981jq,
  author  = {Chetyrkin, K.G. and Kataev, A.L. and Tkachov, F.V.},
  title   = {{Five-loop calculations in the $g \phi^4$ model and the critical index $\eta$}},
  doi     = {10.1016/0370-2693(81)90968-0},
  journal = {Phys. Lett. B},
  volume  = {99},
  pages   = {147},
  year    = {1981},
  note    = {[Erratum: Phys.Lett.B 101, 457 (1981)]}
}

@article{Gorishnii:1983gp,
  author       = {Chetyrkin, K.G. and Gorishny, S.G. and Larin, S.A. and Tkachov, F.V.},
  title        = {{Five-loop renormalization group calculations in the $g \phi^4$ theory}},
  reportnumber = {JINR-P2-83-546},
  doi          = {10.1016/0370-2693(83)90324-6},
  journal      = {Phys. Lett. B},
  volume       = {132},
  pages        = {351},
  year         = {1983}
}

@article{Kleinert:1991rg,
  author        = {Kleinert, H. and Neu, J. and Schulte-Frohlinde, N. and Chetyrkin, K.G. and Larin, S.A.},
  title         = {{Five-loop renormalization group functions of $O(n)$-symmetric $\varphi^4$-theory and $\epsilon$-expansions of critical exponents up to $\epsilon^5$}},
  eprint        = {hep-th/9503230},
  archiveprefix = {arXiv},
  doi           = {10.1016/0370-2693(91)91009-K},
  journal       = {Phys. Lett. B},
  volume        = {272},
  pages         = {39--44},
  year          = {1991},
  note          = {[Erratum: Phys.Lett.B 319, 545 (1993)]}
}

@article{Batkovich:2016jus,
  author        = {Batkovich, D.V. and Chetyrkin, K.G. and Kompaniets, M.V.},
  title         = {{Six loop analytical calculation of the field anomalous dimension and the critical exponent $\eta$ in $O(n)$-symmetric $\varphi^4$ model}},
  eprint        = {1601.01960},
  archiveprefix = {arXiv},
  primaryclass  = {hep-th},
  doi           = {10.1016/j.nuclphysb.2016.03.009},
  journal       = {Nucl. Phys. B},
  volume        = {906},
  pages         = {147--167},
  year          = {2016}
}

@article{Kompaniets:2017yct,
  author        = {Kompaniets, Mikhail V. and Panzer, Erik},
  title         = {{Minimally subtracted six loop renormalization of $O(n)$-symmetric $\phi^4$ theory and critical exponents}},
  eprint        = {1705.06483},
  archiveprefix = {arXiv},
  primaryclass  = {hep-th},
  doi           = {10.1103/PhysRevD.96.036016},
  journal       = {Phys. Rev. D},
  volume        = {96},
  number        = {3},
  pages         = {036016},
  year          = {2017}
}

@article{Wilson:1973jj,
  author  = {Wilson, Kenneth G. and Kogut, J.},
  title   = {{The renormalization group and the $\epsilon$ expansion}},
  doi     = {10.1016/0370-1573(74)90023-4},
  journal = {Phys. Rept.},
  volume  = {12},
  pages   = {75--199},
  year    = {1974}
}

@article{Wilson:1971dc,
  author  = {Wilson, Kenneth G. and Fisher, Michael E.},
  title   = {{Critical exponents in 3.99 dimensions}},
  doi     = {10.1103/PhysRevLett.28.240},
  journal = {Phys. Rev. Lett.},
  volume  = {28},
  pages   = {240--243},
  year    = {1972}
}

@article{Calabrese:2003ww,
  author        = {Calabrese, Pasquale and Parruccini, Pietro},
  title         = {{Five-loop $\epsilon$ expansion for $O(n) \times O(m)$ spin models}},
  eprint        = {cond-mat/0308037},
  archiveprefix = {arXiv},
  doi           = {10.1016/j.nuclphysb.2003.12.002},
  journal       = {Nucl. Phys. B},
  volume        = {679},
  pages         = {568--596},
  year          = {2004}
}

@article{Kompaniets:2019xez,
  author        = {Kompaniets, M.V. and Kudlis, A. and Sokolov, A.I.},
  title         = {{Six-loop $\epsilon$ expansion study of three-dimensional $O(n)\times O(m)$ spin models}},
  eprint        = {1911.01091},
  archiveprefix = {arXiv},
  primaryclass  = {cond-mat.stat-mech},
  doi           = {10.1016/j.nuclphysb.2019.114874},
  journal       = {Nucl. Phys. B},
  volume        = {950},
  pages         = {114874},
  year          = {2020}
}

@article{Adzhemyan:2019gvv,
  author        = {Adzhemyan, Loran Ts. and Ivanova, Ella V. and Kompaniets, Mikhail V. and Kudlis, Andrey. and Sokolov, Aleksandr I.},
  title         = {{Six-loop $\varepsilon$ expansion study of three-dimensional $n$-vector model with cubic anisotropy}},
  eprint        = {1901.02754},
  archiveprefix = {arXiv},
  primaryclass  = {cond-mat.stat-mech},
  doi           = {10.1016/j.nuclphysb.2019.02.001},
  journal       = {Nucl. Phys. B},
  volume        = {940},
  pages         = {332--350},
  year          = {2019}
}

@article{Kleinert:1994td,
  author        = {Kleinert, H. and Schulte-Frohlinde, V.},
  title         = {{Exact five-loop renormalization group functions of $\phi^4$ theory with $O(N)$-symmetric and cubic interactions: Critical exponents up to $\epsilon^5$}},
  eprint        = {cond-mat/9503038},
  archiveprefix = {arXiv},
  reportnumber  = {PRINT-95-096 (FREIE-U.,BERLIN)},
  doi           = {10.1016/0370-2693(94)01377-O},
  journal       = {Phys. Lett. B},
  volume        = {342},
  pages         = {284--296},
  year          = {1995}
}

@article{Bardeen:1978yd,
  author       = {Bardeen, William A. and Buras, A.J. and Duke, D.W. and Muta, T.},
  title        = {{Deep Inelastic Scattering Beyond the Leading Order in Asymptotically Free Gauge Theories}},
  reportnumber = {FERMILAB-PUB-78-42-THY, FERMILAB-PUB-78-042-T},
  doi          = {10.1103/PhysRevD.18.3998},
  journal      = {Phys. Rev. D},
  volume       = {18},
  pages        = {3998},
  year         = {1978}
}

@article{Bollini:1972bi,
  author  = {Bollini, C.G. and Giambiagi, J.J.},
  title   = {{Lowest order ``divergent graphs'' in $\nu$-dimensional space}},
  doi     = {10.1016/0370-2693(72)90483-2},
  journal = {Phys. Lett. B},
  volume  = {40},
  pages   = {566--568},
  year    = {1972}
}

@article{Bollini:1972ui,
  author  = {Bollini, C.G. and Giambiagi, J.J.},
  title   = {{Dimensional Renormalization: The Number of Dimensions as a Regularizing Parameter}},
  doi     = {10.1007/BF02895558},
  journal = {Nuovo Cim. B},
  volume  = {12},
  pages   = {20--26},
  year    = {1972}
}

@article{Zerf:2017zqi,
  author        = {Zerf, Nikolai and Mihaila, Luminita N. and Marquard, Peter and Herbut, Igor F. and Scherer, Michael M.},
  title         = {{Four-loop critical exponents for the Gross-Neveu-Yukawa models}},
  eprint        = {1709.05057},
  archiveprefix = {arXiv},
  primaryclass  = {hep-th},
  reportnumber  = {DESY-17-133},
  doi           = {10.1103/PhysRevD.96.096010},
  journal       = {Phys. Rev. D},
  volume        = {96},
  number        = {9},
  pages         = {096010},
  year          = {2017}
}

@article{Poole:2019txl,
  author        = {Poole, C. and Thomsen, A.E.},
  title         = {{Weyl Consistency Conditions and $\gamma_5$}},
  eprint        = {1901.02749},
  archiveprefix = {arXiv},
  primaryclass  = {hep-th},
  doi           = {10.1103/PhysRevLett.123.041602},
  journal       = {Phys. Rev. Lett.},
  volume        = {123},
  number        = {4},
  pages         = {041602},
  year          = {2019}
}

@article{Grisaru:1979wc,
  author       = {Grisaru, Marcus T. and Siegel, W. and Rocek, M.},
  title        = {{Improved Methods for Supergraphs}},
  reportnumber = {PRINT-79-0485 (BRANDEIS)},
  doi          = {10.1016/0550-3213(79)90344-4},
  journal      = {Nucl. Phys. B},
  volume       = {159},
  pages        = {429},
  year         = {1979}
}

@article{tHooft:1972tcz,
  author  = {'t Hooft, Gerard and Veltman, M.J.G.},
  title   = {{Regularization and Renormalization of Gauge Fields}},
  doi     = {10.1016/0550-3213(72)90279-9},
  journal = {Nucl. Phys. B},
  volume  = {44},
  pages   = {189--213},
  year    = {1972}
}

@article{Salam:1974jj,
  author       = {Salam, Abdus and Strathdee, J.A.},
  title        = {{On Superfields and Fermi-Bose Symmetry}},
  reportnumber = {IC/74/42},
  doi          = {10.1103/PhysRevD.11.1521},
  journal      = {Phys. Rev. D},
  volume       = {11},
  pages        = {1521--1535},
  year         = {1975}
}

@article{Litim:2020jvl,
  author        = {Litim, Daniel F. and Steudtner, Tom},
  title         = {{ARGES {\textendash} Advanced Renormalisation Group Equation Simplifier}},
  eprint        = {2012.12955},
  archiveprefix = {arXiv},
  primaryclass  = {hep-ph},
  reportnumber  = {DO-TH 20/16},
  doi           = {10.1016/j.cpc.2021.108021},
  journal       = {Comput. Phys. Commun.},
  volume        = {265},
  pages         = {108021},
  year          = {2021}
}

@article{Bednyakov:2021ojn,
  author        = {Bednyakov, A. and Pikelner, A.},
  title         = {{Six-loop beta functions in general scalar theory}},
  eprint        = {2102.12832},
  archiveprefix = {arXiv},
  primaryclass  = {hep-ph},
  doi           = {10.1007/JHEP04(2021)233},
  journal       = {JHEP},
  volume        = {04},
  pages         = {233},
  year          = {2021}
}

@article{Sartore:2020gou,
  author        = {Sartore, Lohan and Schienbein, Ingo},
  title         = {{PyR@TE 3}},
  eprint        = {2007.12700},
  archiveprefix = {arXiv},
  primaryclass  = {hep-ph},
  doi           = {10.1016/j.cpc.2020.107819},
  journal       = {Comput. Phys. Commun.},
  volume        = {261},
  pages         = {107819},
  year          = {2021}
}

@article{Thomsen:2021ncy,
  author        = {Thomsen, Anders Eller},
  title         = {{Introducing RGBeta: a Mathematica package for the evaluation of renormalization group $ \beta $-functions}},
  eprint        = {2101.08265},
  archiveprefix = {arXiv},
  primaryclass  = {hep-ph},
  doi           = {10.1140/epjc/s10052-021-09142-4},
  journal       = {Eur. Phys. J. C},
  volume        = {81},
  number        = {5},
  pages         = {408},
  year          = {2021}
}

@article{Staub:2013tta,
  author        = {Staub, Florian},
  title         = {{SARAH 4 : A tool for (not only SUSY) model builders}},
  eprint        = {1309.7223},
  archiveprefix = {arXiv},
  primaryclass  = {hep-ph},
  reportnumber  = {BONN-TH-2013-17},
  doi           = {10.1016/j.cpc.2014.02.018},
  journal       = {Comput. Phys. Commun.},
  volume        = {185},
  pages         = {1773--1790},
  year          = {2014}
}

@article{Davies:2021mnc,
  author        = {Davies, Joshua and Herren, Florian and Thomsen, Anders Eller},
  title         = {{General gauge-Yukawa-quartic $\beta$-functions at 4-3-2-loop order}},
  eprint        = {2110.05496},
  archiveprefix = {arXiv},
  primaryclass  = {hep-ph},
  reportnumber  = {FERMILAB-PUB-21-471-T},
  doi           = {10.1007/JHEP01(2022)051},
  journal       = {JHEP},
  volume        = {01},
  pages         = {051},
  year          = {2022}
}

@article{Steudtner:2021fzs,
  author        = {Steudtner, Tom},
  title         = {{Towards general scalar-Yukawa renormalisation group equations at three-loop order}},
  eprint        = {2101.05823},
  archiveprefix = {arXiv},
  primaryclass  = {hep-th},
  reportnumber  = {DO-TH 21/02},
  doi           = {10.1007/JHEP05(2021)060},
  journal       = {JHEP},
  volume        = {05},
  pages         = {060},
  year          = {2021}
}

@article{Wess:1973kz,
  author       = {Wess, J. and Zumino, B.},
  title        = {{A Lagrangian Model Invariant Under Supergauge Transformations}},
  reportnumber = {CERN-TH-1794},
  doi          = {10.1016/0370-2693(74)90578-4},
  journal      = {Phys. Lett. B},
  volume       = {49},
  pages        = {52},
  year         = {1974}
}

@article{Bednyakov:2021qxa,
  author        = {Bednyakov, Alexander and Pikelner, Andrey},
  title         = {{Four-Loop Gauge and Three-Loop Yukawa Beta Functions in a General Renormalizable Theory}},
  eprint        = {2105.09918},
  archiveprefix = {arXiv},
  primaryclass  = {hep-ph},
  doi           = {10.1103/PhysRevLett.127.041801},
  journal       = {Phys. Rev. Lett.},
  volume        = {127},
  number        = {4},
  pages         = {041801},
  year          = {2021}
}

@article{Herren:2021yur,
  author        = {Herren, Florian and Thomsen, Anders Eller},
  title         = {{On ambiguities and divergences in perturbative renormalization group functions}},
  eprint        = {2104.07037},
  archiveprefix = {arXiv},
  primaryclass  = {hep-th},
  reportnumber  = {FERMILAB-PUB-21-196-T},
  doi           = {10.1007/JHEP06(2021)116},
  journal       = {JHEP},
  volume        = {06},
  pages         = {116},
  year          = {2021}
}

@article{Chetyrkin:1997fm,
  author        = {Chetyrkin, Konstantin G. and Misiak, Mikolaj and Munz, Manfred},
  title         = {{Beta functions and anomalous dimensions up to three loops}},
  eprint        = {hep-ph/9711266},
  archiveprefix = {arXiv},
  reportnumber  = {MPI-PHT-97-45, TTP-97-43, ZU-TH-16-97, TUM-HEP-284-97, IFT-11-97},
  doi           = {10.1016/S0550-3213(98)00122-9},
  journal       = {Nucl. Phys. B},
  volume        = {518},
  pages         = {473--494},
  year          = {1998}
}

@article{Davydychev:1992mt,
  author       = {Davydychev, Andrei I. and Tausk, J. B.},
  title        = {{Two loop selfenergy diagrams with different masses and the momentum expansion}},
  reportnumber = {INLO-PUB-11-92},
  doi          = {10.1016/0550-3213(93)90338-P},
  journal      = {Nucl. Phys. B},
  volume       = {397},
  pages        = {123--142},
  year         = {1993}
}

@article{Nogueira:1991ex,
  author       = {Nogueira, Paulo},
  title        = {{Automatic Feynman graph generation}},
  reportnumber = {IFM-7-91},
  doi          = {10.1006/jcph.1993.1074},
  journal      = {J. Comput. Phys.},
  volume       = {105},
  pages        = {279--289},
  year         = {1993}
}

@article{Kuipers:2012rf,
  author        = {Kuipers, J. and Ueda, T. and Vermaseren, J. A. M. and Vollinga, J.},
  title         = {{FORM version 4.0}},
  eprint        = {1203.6543},
  archiveprefix = {arXiv},
  primaryclass  = {cs.SC},
  reportnumber  = {NIKHEF-2012-004, TTP12-008, SFB-CPP-12-15},
  doi           = {10.1016/j.cpc.2012.12.028},
  journal       = {Comput. Phys. Commun.},
  volume        = {184},
  pages         = {1453--1467},
  year          = {2013}
}

@article{Steudtner:FoRGEr,
  author  = {Steudtner, Tom},
  title   = {{FoRGEr}},
  journal = {{unpublished}},
  year    = {2025}
}

@article{tHooft:1973alw,
  author       = {'t Hooft, Gerard},
  editor       = {Taylor, J. C.},
  title        = {{A Planar Diagram Theory for Strong Interactions}},
  reportnumber = {CERN-TH-1786},
  doi          = {10.1016/0550-3213(74)90154-0},
  journal      = {Nucl. Phys. B},
  volume       = {72},
  pages        = {461},
  year         = {1974}
}

@article{Banks:1981nn,
  author       = {Banks, Tom and Zaks, A.},
  title        = {{On the Phase Structure of Vector-Like Gauge Theories with Massless Fermions}},
  reportnumber = {TAUP-944-81},
  doi          = {10.1016/0550-3213(82)90035-9},
  journal      = {Nucl. Phys. B},
  volume       = {196},
  pages        = {189--204},
  year         = {1982}
}

@article{Bond:2016dvk,
  author        = {Bond, Andrew D. and Litim, Daniel F.},
  title         = {{Theorems for Asymptotic Safety of Gauge Theories}},
  eprint        = {1608.00519},
  archiveprefix = {arXiv},
  primaryclass  = {hep-th},
  doi           = {10.1140/epjc/s10052-017-4976-5},
  journal       = {Eur. Phys. J. C},
  volume        = {77},
  number        = {6},
  pages         = {429},
  year          = {2017},
  note          = {[Erratum: Eur.Phys.J.C 77, 525 (2017)]}
}

@article{Bond:2018oco,
  author        = {Bond, Andrew D. and Litim, Daniel F.},
  title         = {{Price of Asymptotic Safety}},
  eprint        = {1801.08527},
  archiveprefix = {arXiv},
  primaryclass  = {hep-th},
  doi           = {10.1103/PhysRevLett.122.211601},
  journal       = {Phys. Rev. Lett.},
  volume        = {122},
  number        = {21},
  pages         = {211601},
  year          = {2019}
}

@phdthesis{Steudtner:2020jcj,
  author = {Steudtner, Tom},
  title  = {{Asymptotic safety: from perturbatively exact models to particle physics}},
  school = {Sussex U.},
  year   = {2020}
}

@article{Luthe:2016ima,
  author        = {Luthe, Thomas and Maier, Andreas and Marquard, Peter and Schr\"oder, York},
  title         = {{Towards the five-loop Beta function for a general gauge group}},
  eprint        = {1606.08662},
  archiveprefix = {arXiv},
  primaryclass  = {hep-ph},
  reportnumber  = {TTP16-025, IPPP-16-57, DESY-16-117},
  doi           = {10.1007/JHEP07(2016)127},
  journal       = {JHEP},
  volume        = {07},
  pages         = {127},
  year          = {2016}
}

@article{Baikov:2016tgj,
  author        = {Baikov, P. A. and Chetyrkin, K. G. and K\"uhn, J. H.},
  title         = {{Five-Loop Running of the QCD coupling constant}},
  eprint        = {1606.08659},
  archiveprefix = {arXiv},
  primaryclass  = {hep-ph},
  reportnumber  = {TTP16-026},
  doi           = {10.1103/PhysRevLett.118.082002},
  journal       = {Phys. Rev. Lett.},
  volume        = {118},
  number        = {8},
  pages         = {082002},
  year          = {2017}
}

@phdthesis{Luthe:2015ngq,
  author = {Luthe, Thomas},
  title  = {{Fully massive vacuum integrals at 5 loops}},
  school = {Bielefeld U.},
  year   = {2015}
}

@article{Pikelner:2017tgv,
  author        = {Pikelner, Andrey},
  title         = {{FMFT: Fully Massive Four-loop Tadpoles}},
  eprint        = {1707.01710},
  archiveprefix = {arXiv},
  primaryclass  = {hep-ph},
  reportnumber  = {DESY-17-098},
  doi           = {10.1016/j.cpc.2017.11.017},
  journal       = {Comput. Phys. Commun.},
  volume        = {224},
  pages         = {282--287},
  year          = {2018}
}

@article{Laporta:2000dsw,
  author        = {Laporta, S.},
  title         = {{High-precision calculation of multiloop Feynman integrals by difference equations}},
  eprint        = {hep-ph/0102033},
  archiveprefix = {arXiv},
  doi           = {10.1142/S0217751X00002159},
  journal       = {Int. J. Mod. Phys. A},
  volume        = {15},
  pages         = {5087--5159},
  year          = {2000}
}

@article{Schroder:2005va,
  author        = {Schroder, Y. and Vuorinen, A.},
  title         = {{High-precision epsilon expansions of single-mass-scale four-loop vacuum bubbles}},
  eprint        = {hep-ph/0503209},
  archiveprefix = {arXiv},
  reportnumber  = {BI-TP-2005-11, UW-PT-05-5},
  doi           = {10.1088/1126-6708/2005/06/051},
  journal       = {JHEP},
  volume        = {06},
  pages         = {051},
  year          = {2005}
}

@article{Veneziano:1976wm,
  author       = {Veneziano, G.},
  title        = {{Some Aspects of a Unified Approach to Gauge, Dual and Gribov Theories}},
  reportnumber = {CERN-TH-2200},
  doi          = {10.1016/0550-3213(76)90412-0},
  journal      = {Nucl. Phys. B},
  volume       = {117},
  pages        = {519--545},
  year         = {1976}
}

@article{Rosenstein:1988pt,
  author       = {Rosenstein, Baruch and Warr, Brian J. and Park, Seon H.},
  title        = {{The Four Fermi Theory Is Renormalizable in (2+1)-Dimensions}},
  reportnumber = {UTTG-22-88},
  doi          = {10.1103/PhysRevLett.62.1433},
  journal      = {Phys. Rev. Lett.},
  volume       = {62},
  pages        = {1433--1436},
  year         = {1989}
}

@article{Rosenstein:1990nm,
  author       = {Rosenstein, B. and Warr, Brian and Park, S. H.},
  title        = {{Dynamical symmetry breaking in four Fermi interaction models}},
  reportnumber = {SLAC-PUB-5349},
  doi          = {10.1016/0370-1573(91)90129-A},
  journal      = {Phys. Rept.},
  volume       = {205},
  pages        = {59--108},
  year         = {1991}
}

@article{Townsend:1976sy,
  author       = {Townsend, P. K.},
  title        = {{Consistency of the 1/n Expansion for Three-Dimensional $\phi^6$ Theory}},
  reportnumber = {Print-76-0244 (BRANDEIS)},
  doi          = {10.1016/0550-3213(77)90306-6},
  journal      = {Nucl. Phys. B},
  volume       = {118},
  pages        = {199--217},
  year         = {1977}
}

@article{deCalan:1991km,
  author  = {de Calan, C. and Faria da Veiga, P. A. and Magnen, J. and Seneor, R.},
  title   = {{Constructing the three-dimensional Gross-Neveu model with a large number of flavor components}},
  doi     = {10.1103/PhysRevLett.66.3233},
  journal = {Phys. Rev. Lett.},
  volume  = {66},
  pages   = {3233--3236},
  year    = {1991}
}

@article{Jakovac:2014lqa,
  author        = {Jakov\'ac, A. and Patk\'os, A. and P\'osfay, P.},
  title         = {{Non-Gaussian fixed points in fermionic field theories without auxiliary Bose-fields}},
  eprint        = {1406.3195},
  archiveprefix = {arXiv},
  primaryclass  = {hep-th},
  doi           = {10.1140/epjc/s10052-014-3228-1},
  journal       = {Eur. Phys. J. C},
  volume        = {75},
  number        = {1},
  pages         = {2},
  year          = {2015}
}

@article{Braun:2011pp,
  author        = {Braun, Jens},
  title         = {{Fermion Interactions and Universal Behavior in Strongly Interacting Theories}},
  eprint        = {1108.4449},
  archiveprefix = {arXiv},
  primaryclass  = {hep-ph},
  doi           = {10.1088/0954-3899/39/3/033001},
  journal       = {J. Phys. G},
  volume        = {39},
  pages         = {033001},
  year          = {2012}
}

@article{Appelquist:1981sf,
  author       = {Appelquist, Thomas and Heinz, Ulrich W.},
  title        = {{Three-dimensional O(N) theories at large distances}},
  reportnumber = {YTP-81-16},
  doi          = {10.1103/PhysRevD.24.2169},
  journal      = {Phys. Rev. D},
  volume       = {24},
  pages        = {2169},
  year         = {1981}
}

@article{Pisarski:1982vz,
  author  = {Pisarski, R. D.},
  title   = {{Fixed point structure of ($\phi^6$) in three-dimensions at large N}},
  doi     = {10.1103/PhysRevLett.48.574},
  journal = {Phys. Rev. Lett.},
  volume  = {48},
  pages   = {574--576},
  year    = {1982}
}

@article{Brod:2024zaz,
  author        = {Brod, Joachim and H{\"u}depohl, Lorenz and Stamou, Emmanuel and Steudtner, Tom},
  title         = {{MaRTIn -- Manual for the ''Massive Recursive Tensor Integration''}},
  eprint        = {2401.04033},
  archiveprefix = {arXiv},
  primaryclass  = {hep-ph},
  doi           = {10.1016/j.cpc.2024.109372},
  journal       = {Comput. Phys. Commun.},
  volume        = {306},
  pages         = {109372},
  year          = {2025}
}

@article{Jack:2024sjr,
  author        = {Jack, Ian and Osborn, Hugh and Steudtner, Tom},
  title         = {{Explorations in scalar fermion theories: {\ensuremath{\beta}}-functions, supersymmetry and fixed points}},
  eprint        = {2301.10903},
  archiveprefix = {arXiv},
  primaryclass  = {hep-th},
  reportnumber  = {DO-TH 22/06},
  doi           = {10.1007/JHEP02(2024)038},
  journal       = {JHEP},
  volume        = {02},
  pages         = {038},
  year          = {2024}
}

@article{Breitenlohner:1977hr,
  author  = {Breitenlohner, P. and Maison, D.},
  title   = {{Dimensional Renormalization and the Action Principle}},
  doi     = {10.1007/BF01609069},
  journal = {Commun. Math. Phys.},
  volume  = {52},
  pages   = {11--38},
  year    = {1977}
}

@article{Jack:1984vj,
  author       = {Jack, I. and Osborn, H.},
  title        = {{General Background Field Calculations With Fermion Fields}},
  reportnumber = {DAMTP-84-2},
  doi          = {10.1016/0550-3213(85)90088-4},
  journal      = {Nucl. Phys. B},
  volume       = {249},
  pages        = {472--506},
  year         = {1985}
}

@article{Fortin:2012hn,
  author        = {Fortin, Jean-Francois and Grinstein, Benjamin and Stergiou, Andreas},
  title         = {{Limit Cycles and Conformal Invariance}},
  eprint        = {1208.3674},
  archiveprefix = {arXiv},
  primaryclass  = {hep-th},
  reportnumber  = {UCSD-PTH-12-10, CERN-PH-TH-2012-297, SU-ITP-12-38},
  doi           = {10.1007/JHEP01(2013)184},
  journal       = {JHEP},
  volume        = {01},
  pages         = {184},
  year          = {2013}
}

@article{Steudtner:2024teg,
  author        = {Steudtner, Tom and Thomsen, Anders Eller},
  title         = {{General quartic \ensuremath{\beta}-function at three loops}},
  eprint        = {2408.05267},
  archiveprefix = {arXiv},
  primaryclass  = {hep-ph},
  doi           = {10.1007/JHEP10(2024)163},
  journal       = {JHEP},
  volume        = {10},
  pages         = {163},
  year          = {2024}
}

@misc{Mathematica,
  author = {{Wolfram Research Inc.}},
  title  = {Mathematica, {V}ersion 14.2},
  url    = {https://www.wolfram.com/mathematica},
  note   = {Champaign, IL, 2024}
}

@article{Belusca-Maito:2023wah,
  author        = {B{\'e}lusca-Ma{\"\i}to, Herm{\`e}s and Ilakovac, Amon and K{\"u}hler, Paul and Ma{\dj}or-Bo{\v{z}}inovi{\'c}, Marija and St{\"o}ckinger, Dominik and Wei{\ss}wange, Matthias},
  title         = {{Introduction to Renormalization Theory and Chiral Gauge Theories in Dimensional Regularization with Non-Anticommuting {\ensuremath{\gamma}}$_{5}$}},
  eprint        = {2303.09120},
  archiveprefix = {arXiv},
  primaryclass  = {hep-ph},
  doi           = {10.3390/sym15030622},
  journal       = {Symmetry},
  volume        = {15},
  number        = {3},
  pages         = {622},
  year          = {2023}
}

@article{Fonseca:2011sy,
  author        = {Fonseca, Renato M.},
  title         = {{Calculating the renormalisation group equations of a SUSY model with Susyno}},
  eprint        = {1106.5016},
  archiveprefix = {arXiv},
  primaryclass  = {hep-ph},
  reportnumber  = {CFTP-11-011},
  doi           = {10.1016/j.cpc.2012.05.017},
  journal       = {Comput. Phys. Commun.},
  volume        = {183},
  pages         = {2298--2306},
  year          = {2012}
}

@article{Uetrecht:2025llz,
  author        = {Uetrecht, Max and Herbut, Igor F. and Scherer, Michael M. and Stamou, Emmanuel and Steudtner, Tom},
  title         = {{Quantum multicriticality and emergent symmetry in Dirac systems with two order parameters at three-loop order}},
  eprint        = {2505.22723},
  archiveprefix = {arXiv},
  primaryclass  = {cond-mat.str-el},
  doi           = {10.1103/xrg3-yd9b},
  journal       = {Phys. Rev. B},
  volume        = {112},
  number        = {8},
  pages         = {085126},
  year          = {2025}
}

@article{Herbut:2022zzw,
  author        = {Herbut, Igor F. and Scherer, Michael M.},
  title         = {{SO(4) multicriticality of two-dimensional Dirac fermions}},
  eprint        = {2206.04073},
  archiveprefix = {arXiv},
  primaryclass  = {cond-mat.str-el},
  doi           = {10.1103/PhysRevB.106.115136},
  journal       = {Phys. Rev. B},
  volume        = {106},
  number        = {11},
  pages         = {115136},
  year          = {2022}
}

@article{Uetrecht:2023uou,
  author        = {Uetrecht, Max and Herbut, Igor F. and Stamou, Emmanuel and Scherer, Michael M.},
  title         = {{Absence of SO(4) quantum criticality in Dirac semimetals at two-loop order}},
  eprint        = {2308.12426},
  archiveprefix = {arXiv},
  primaryclass  = {cond-mat.str-el},
  reportnumber  = {DO-TH 23/13},
  doi           = {10.1103/PhysRevB.108.245130},
  journal       = {Phys. Rev. B},
  volume        = {108},
  number        = {24},
  pages         = {245130},
  year          = {2023}
}

@article{Herbut:2023xgz,
  author        = {Herbut, Igor F.},
  title         = {{Wilson-Fisher fixed points in the presence of Dirac fermions}},
  eprint        = {2304.07654},
  archiveprefix = {arXiv},
  primaryclass  = {cond-mat.str-el},
  doi           = {10.1142/S0217984924300060},
  journal       = {Mod. Phys. Lett. B},
  volume        = {38},
  number        = {34},
  pages         = {2430006},
  year          = {2024}
}

@article{Boyack:2020xpe,
  author        = {Boyack, Rufus and Yerzhakov, Hennadii and Maciejko, Joseph},
  title         = {{Quantum phase transitions in Dirac fermion systems}},
  eprint        = {2004.09414},
  archiveprefix = {arXiv},
  primaryclass  = {cond-mat.str-el},
  doi           = {10.1140/epjs/s11734-021-00069-1},
  journal       = {Eur. Phys. J. ST},
  volume        = {230},
  number        = {4},
  pages         = {979--992},
  year          = {2021}
}

@article{Herbut:2006cs,
  author        = {Herbut, Igor F.},
  title         = {{Interactions and phase transitions on graphene's honeycomb lattice}},
  eprint        = {cond-mat/0606195},
  archiveprefix = {arXiv},
  doi           = {10.1103/PhysRevLett.97.146401},
  journal       = {Phys. Rev. Lett.},
  volume        = {97},
  pages         = {146401},
  year          = {2006}
}

@article{Kvedaraite:2025lgi,
  author        = {Kvedarait{\.{e}}, Sandra and Steudtner, Tom and Uetrecht, Max},
  title         = {{Revisiting the {\ensuremath{\phi^6}} theory in three dimensions at large N}},
  eprint        = {2502.07880},
  archiveprefix = {arXiv},
  primaryclass  = {hep-th},
  doi           = {10.1103/tnh4-7lnv},
  journal       = {Phys. Rev. D},
  volume        = {112},
  number        = {5},
  pages         = {056004},
  year          = {2025}
}

@article{Steudtner:2025blh,
  author        = {Steudtner, Tom},
  title         = {{Four loop renormalisation group equations in general Gross-Neveu-Yukawa theories}},
  eprint        = {2507.18710},
  archiveprefix = {arXiv},
  primaryclass  = {hep-th},
  doi           = {10.1007/JHEP10(2025)096},
  journal       = {JHEP},
  volume        = {10},
  pages         = {096},
  year          = {2025}
}

@article{Fraser-Taliente:2024rql,
  author        = {Fraser-Taliente, Ludo and Wheater, John},
  title         = {{Melonic limits of the quartic Yukawa model and general features of melonic CFTs}},
  eprint        = {2410.09152},
  archiveprefix = {arXiv},
  primaryclass  = {hep-th},
  doi           = {10.1007/JHEP01(2025)187},
  journal       = {JHEP},
  volume        = {01},
  pages         = {187},
  year          = {2025}
}

@article{Hager:2002uq,
  author  = {Hager, J. S.},
  title   = {{Six-loop renormalization group functions of O(n)-symmetric $\phi^6$-theory and epsilon-expansions of tricritical exponents up to $\epsilon^3$}},
  doi     = {10.1088/0305-4470/35/12/301},
  journal = {J. Phys. A},
  volume  = {35},
  pages   = {2703--2711},
  year    = {2002}
}

@article{Jack:2016utw,
  author        = {Jack, I. and Poole, C.},
  title         = {{$\alpha$-function in three dimensions: Beyond the leading order}},
  eprint        = {1607.00236},
  archiveprefix = {arXiv},
  primaryclass  = {hep-th},
  reportnumber  = {LTH-1082},
  doi           = {10.1103/PhysRevD.95.025010},
  journal       = {Phys. Rev. D},
  volume        = {95},
  number        = {2},
  pages         = {025010},
  year          = {2017}
}

@article{Jack:2015tka,
  author        = {Jack, I. and Jones, D. R. T. and Poole, C.},
  title         = {{Gradient flows in three dimensions}},
  eprint        = {1505.05400},
  archiveprefix = {arXiv},
  primaryclass  = {hep-th},
  reportnumber  = {LTH1043},
  doi           = {10.1007/JHEP09(2015)061},
  journal       = {JHEP},
  volume        = {09},
  pages         = {061},
  year          = {2015}
}

@article{Avdeev:1992jt,
  author       = {Avdeev, L. V. and Kazakov, D. I. and Kondrashuk, I. N.},
  title        = {{Renormalizations in supersymmetric and nonsupersymmetric nonAbelian Chern-Simons field theories with matter}},
  reportnumber = {JINR-E2-92-270},
  doi          = {10.1016/0550-3213(93)90151-E},
  journal      = {Nucl. Phys. B},
  volume       = {391},
  pages        = {333--357},
  year         = {1993}
}

@article{Avdeev:1991za,
  author       = {Avdeev, L. V. and Grigorev, G. V. and Kazakov, D. I.},
  title        = {{Renormalizations in Abelian Chern-Simons field theories with matter}},
  reportnumber = {CERN-TH-6091-91},
  doi          = {10.1016/0550-3213(92)90659-Y},
  journal      = {Nucl. Phys. B},
  volume       = {382},
  pages        = {561--580},
  year         = {1992}
}

@article{Giombi:2018qgp,
  author        = {Giombi, Simone and Klebanov, Igor R. and Popov, Fedor and Prakash, Shiroman and Tarnopolsky, Grigory},
  title         = {{Prismatic Large $N$ Models for Bosonic Tensors}},
  eprint        = {1808.04344},
  archiveprefix = {arXiv},
  primaryclass  = {hep-th},
  reportnumber  = {PUPT-2568},
  doi           = {10.1103/PhysRevD.98.105005},
  journal       = {Phys. Rev. D},
  volume        = {98},
  number        = {10},
  pages         = {105005},
  year          = {2018}
}

@article{Coleman:1973sx,
  author  = {Coleman, Sidney R. and Gross, David J.},
  title   = {{Price of asymptotic freedom}},
  doi     = {10.1103/PhysRevLett.31.851},
  journal = {Phys. Rev. Lett.},
  volume  = {31},
  pages   = {851--854},
  year    = {1973}
}

@article{Schroder:2003kb,
  author        = {Schroder, Y. and Vuorinen, A.},
  title         = {{High precision evaluation of four loop vacuum bubbles in three-dimensions}},
  eprint        = {hep-ph/0311323},
  archiveprefix = {arXiv},
  reportnumber  = {HIP-2003-55-TH, MIT-CTP-3431},
  month         = {11},
  year          = {2003}
}

@article{Broadhurst:1998iq,
  author        = {Broadhurst, David J.},
  title         = {{A Dilogarithmic three-dimensional Ising tetrahedron}},
  eprint        = {hep-th/9805025},
  archiveprefix = {arXiv},
  reportnumber  = {OUT-4102-73},
  doi           = {10.1007/s100529900983},
  journal       = {Eur. Phys. J. C},
  volume        = {8},
  pages         = {363--366},
  year          = {1999}
}

@article{Broadhurst:1998ke,
  author        = {Broadhurst, David J.},
  title         = {{Solving differential equations for three loop diagrams: Relation to hyperbolic geometry and knot theory}},
  eprint        = {hep-th/9806174},
  archiveprefix = {arXiv},
  reportnumber  = {OUT-4102-74},
  month         = {6},
  year          = {1998}
}

@article{Vermaseren:1998uu,
  author        = {Vermaseren, J. A. M.},
  title         = {{Harmonic sums, Mellin transforms and integrals}},
  eprint        = {hep-ph/9806280},
  archiveprefix = {arXiv},
  reportnumber  = {FTUAM-98-7, NIKHEF-98-014},
  doi           = {10.1142/S0217751X99001032},
  journal       = {Int. J. Mod. Phys. A},
  volume        = {14},
  pages         = {2037--2076},
  year          = {1999}
}

@article{Blumlein:2009cf,
  author        = {Blumlein, J. and Broadhurst, D. J. and Vermaseren, J. A. M.},
  title         = {{The Multiple Zeta Value Data Mine}},
  eprint        = {0907.2557},
  archiveprefix = {arXiv},
  primaryclass  = {math-ph},
  reportnumber  = {DESY-09-003, SFB-CPP-09-65},
  doi           = {10.1016/j.cpc.2009.11.007},
  journal       = {Comput. Phys. Commun.},
  volume        = {181},
  pages         = {582--625},
  year          = {2010}
}

@article{Kajantie:2003ax,
  author        = {Kajantie, K. and Laine, M. and Rummukainen, K. and Schroder, Y.},
  title         = {{Four loop vacuum energy density of the SU(N(c)) + adjoint Higgs theory}},
  eprint        = {hep-ph/0304048},
  archiveprefix = {arXiv},
  reportnumber  = {CERN-TH-2003-076, HIP-2003-16-TH, MIT-CTP-3354},
  doi           = {10.1088/1126-6708/2003/04/036},
  journal       = {JHEP},
  volume        = {04},
  pages         = {036},
  year          = {2003}
}

@article{Aharony:2015mjs,
  author        = {Aharony, Ofer},
  title         = {{Baryons, monopoles and dualities in Chern-Simons-matter theories}},
  eprint        = {1512.00161},
  archiveprefix = {arXiv},
  primaryclass  = {hep-th},
  reportnumber  = {WIS-12-15-NOV-DPPA},
  doi           = {10.1007/JHEP02(2016)093},
  journal       = {JHEP},
  volume        = {02},
  pages         = {093},
  year          = {2016}
}

@article{Seiberg:2016gmd,
  author        = {Seiberg, Nathan and Senthil, T. and Wang, Chong and Witten, Edward},
  title         = {{A Duality Web in 2+1 Dimensions and Condensed Matter Physics}},
  eprint        = {1606.01989},
  archiveprefix = {arXiv},
  primaryclass  = {hep-th},
  doi           = {10.1016/j.aop.2016.08.007},
  journal       = {Annals Phys.},
  volume        = {374},
  pages         = {395--433},
  year          = {2016}
}

@article{DiPietro:2019hqe,
  author        = {Di Pietro, Lorenzo and Gaiotto, Davide and Lauria, Edoardo and Wu, Jingxiang},
  title         = {{3d Abelian Gauge Theories at the Boundary}},
  eprint        = {1902.09567},
  archiveprefix = {arXiv},
  primaryclass  = {hep-th},
  doi           = {10.1007/JHEP05(2019)091},
  journal       = {JHEP},
  volume        = {05},
  pages         = {091},
  year          = {2019}
}

@article{DiPietro:2017kcd,
  author        = {Di Pietro, Lorenzo and Stamou, Emmanuel},
  title         = {{Scaling dimensions in QED$_3$ from the $\epsilon$-expansion}},
  eprint        = {1708.03740},
  archiveprefix = {arXiv},
  primaryclass  = {hep-th},
  doi           = {10.1007/JHEP12(2017)054},
  journal       = {JHEP},
  volume        = {12},
  pages         = {054},
  year          = {2017}
}

@article{DiPietro:2017vsp,
  author        = {Di Pietro, Lorenzo and Stamou, Emmanuel},
  title         = {{Operator mixing in the $\boldsymbol{\epsilon}$-expansion: Scheme and evanescent-operator independence}},
  eprint        = {1708.03739},
  archiveprefix = {arXiv},
  primaryclass  = {hep-th},
  doi           = {10.1103/PhysRevD.97.065007},
  journal       = {Phys. Rev. D},
  volume        = {97},
  number        = {6},
  pages         = {065007},
  year          = {2018}
}

@article{Aharony:2016jvv,
  author        = {Aharony, Ofer and Benini, Francesco and Hsin, Po-Shen and Seiberg, Nathan},
  title         = {{Chern-Simons-matter dualities with $SO$ and $USp$ gauge groups}},
  eprint        = {1611.07874},
  archiveprefix = {arXiv},
  primaryclass  = {cond-mat.str-el},
  reportnumber  = {SISSA-62-2016-FISI},
  doi           = {10.1007/JHEP02(2017)072},
  journal       = {JHEP},
  volume        = {02},
  pages         = {072},
  year          = {2017}
}

@article{Appelquist:1986fd,
  author       = {Appelquist, Thomas W. and Bowick, Mark J. and Karabali, Dimitra and Wijewardhana, L. C. R.},
  title        = {{Spontaneous Chiral Symmetry Breaking in Three-Dimensional QED}},
  reportnumber = {YTP-85-26},
  doi          = {10.1103/PhysRevD.33.3704},
  journal      = {Phys. Rev. D},
  volume       = {33},
  pages        = {3704},
  year         = {1986}
}

@article{Appelquist:1988sr,
  author       = {Appelquist, Thomas and Nash, Daniel and Wijewardhana, L. C. R.},
  title        = {{Critical Behavior in (2+1)-Dimensional QED}},
  reportnumber = {YCTP-P2-88},
  doi          = {10.1103/PhysRevLett.60.2575},
  journal      = {Phys. Rev. Lett.},
  volume       = {60},
  pages        = {2575},
  year         = {1988}
}

@article{Chester:2016ref,
  author        = {Chester, Shai M. and Pufu, Silviu S.},
  title         = {{Anomalous dimensions of scalar operators in QED$_{3}$}},
  eprint        = {1603.05582},
  archiveprefix = {arXiv},
  primaryclass  = {hep-th},
  reportnumber  = {PUPT-2501},
  doi           = {10.1007/JHEP08(2016)069},
  journal       = {JHEP},
  volume        = {08},
  pages         = {069},
  year          = {2016}
}

@article{Nash:1989xx,
  author       = {Nash, Daniel},
  title        = {{Higher Order Corrections in (2+1)-Dimensional QED}},
  reportnumber = {YCTP-P6-89},
  doi          = {10.1103/PhysRevLett.62.3024},
  journal      = {Phys. Rev. Lett.},
  volume       = {62},
  pages        = {3024},
  year         = {1989}
}

@article{Appelquist:1981vg,
  author       = {Appelquist, Thomas and Pisarski, Robert D.},
  title        = {{High-Temperature Yang-Mills Theories and Three-Dimensional Quantum Chromodynamics}},
  reportnumber = {Print-81-0020 (YALE), YTP-81-01, COO-3075-203},
  doi          = {10.1103/PhysRevD.23.2305},
  journal      = {Phys. Rev. D},
  volume       = {23},
  pages        = {2305},
  year         = {1981}
}

@article{Appelquist:1986qw,
  author       = {Appelquist, Thomas and Bowick, Mark J. and Karabali, Dimitra and Wijewardhana, L. C. R.},
  title        = {{Spontaneous Breaking of Parity in (2+1)-dimensional {QED}}},
  reportnumber = {PRINT-86-0039 (YALE), YTP-85-32},
  doi          = {10.1103/PhysRevD.33.3774},
  journal      = {Phys. Rev. D},
  volume       = {33},
  pages        = {3774},
  year         = {1986}
}

@article{Ginsparg:1980ef,
  author       = {Ginsparg, Paul H.},
  title        = {{First Order and Second Order Phase Transitions in Gauge Theories at Finite Temperature}},
  reportnumber = {SACLAY-DPh-T 80/27},
  doi          = {10.1016/0550-3213(80)90418-6},
  journal      = {Nucl. Phys. B},
  volume       = {170},
  pages        = {388--408},
  year         = {1980}
}

@article{Farakos:1994kx,
  author        = {Farakos, K. and Kajantie, K. and Rummukainen, K. and Shaposhnikov, Mikhail E.},
  title         = {{3-D physics and the electroweak phase transition: Perturbation theory}},
  eprint        = {hep-ph/9404201},
  archiveprefix = {arXiv},
  reportnumber  = {CERN-TH-6973-94, IUHET-273},
  doi           = {10.1016/0550-3213(94)90173-2},
  journal       = {Nucl. Phys. B},
  volume        = {425},
  pages         = {67--109},
  year          = {1994}
}

@article{Kajantie:1995dw,
  author        = {Kajantie, K. and Laine, M. and Rummukainen, K. and Shaposhnikov, Mikhail E.},
  title         = {{Generic rules for high temperature dimensional reduction and their application to the standard model}},
  eprint        = {hep-ph/9508379},
  archiveprefix = {arXiv},
  reportnumber  = {CERN-TH-95-226, HU-TFT-95-50, IUHET-312},
  doi           = {10.1016/0550-3213(95)00549-8},
  journal       = {Nucl. Phys. B},
  volume        = {458},
  pages         = {90--136},
  year          = {1996}
}

@article{Chala:2025cya,
  author        = {Chala, Mikael and Dashko, Andrii and Guedes, Guilherme},
  title         = {{Running Couplings in High-Temperature Effective Field Theory}},
  eprint        = {2510.26878},
  archiveprefix = {arXiv},
  primaryclass  = {hep-ph},
  reportnumber  = {CERN-TH-2025-220},
  month         = {10},
  year          = {2025}
}

@article{Ekstedt:2024etx,
  author        = {Ekstedt, Andreas and Schicho, Philipp and Tenkanen, Tuomas V. I.},
  title         = {{Cosmological phase transitions at three loops: The final verdict on perturbation theory}},
  eprint        = {2405.18349},
  archiveprefix = {arXiv},
  primaryclass  = {hep-ph},
  reportnumber  = {HIP-2024-15/TH},
  doi           = {10.1103/PhysRevD.110.096006},
  journal       = {Phys. Rev. D},
  volume        = {110},
  number        = {9},
  pages         = {096006},
  year          = {2024}
}

@article{Aharony:2014uya,
  author        = {Aharony, Ofer and Fleischer, Daniel},
  title         = {{IR Dualities in General 3d Supersymmetric SU(N) QCD Theories}},
  eprint        = {1411.5475},
  archiveprefix = {arXiv},
  primaryclass  = {hep-th},
  reportnumber  = {WIS-09-14-NOV-DPPA},
  doi           = {10.1007/JHEP02(2015)162},
  journal       = {JHEP},
  volume        = {02},
  pages         = {162},
  year          = {2015}
}

@article{Braaten:1995cm,
  author        = {Braaten, Eric and Nieto, Agustin},
  title         = {{Effective field theory approach to high temperature thermodynamics}},
  eprint        = {hep-ph/9501375},
  archiveprefix = {arXiv},
  reportnumber  = {NUHEP-TH-95-2},
  doi           = {10.1103/PhysRevD.51.6990},
  journal       = {Phys. Rev. D},
  volume        = {51},
  pages         = {6990--7006},
  year          = {1995}
}

@article{Gat:1990xi,
  author       = {Gat, G. and Kovner, A. and Rosenstein, B. and Warr, B. J.},
  title        = {{Four Fermi Interaction in (2+1)-dimensions Beyond Leading Order in 1/$N$}},
  reportnumber = {UBCTG-1-90},
  doi          = {10.1016/0370-2693(90)90425-6},
  journal      = {Phys. Lett. B},
  volume       = {240},
  pages        = {158--162},
  year         = {1990}
}

@techreport{FergusonBailey1992,
  author      = {Helaman Rolfe Pratt Ferguson and David H. Bailey},
  title       = {{A Polynomial Time, Numerically Stable Integer Relation Algorithm}},
  institution = {NASA Ames Research Center},
  year        = {1992},
  number      = {RNR-91-032},
  url         = {http://davidhbailey.com/dhbpapers/pslq.pdf}
}

@article{FergusonBaileyArno1999,
  author  = {Helaman Rolfe Pratt Ferguson and David H. Bailey and Steve Arno},
  title   = {{Analysis of {PSLQ}, an Integer Relation Finding Algorithm}},
  journal = {Mathematics of Computation},
  year    = {1999},
  volume  = {68},
  number  = {225},
  pages   = {351--369},
  doi     = {10.1090/S0025-5718-99-00995-3}
}

@article{Herbut:2009qb,
  author        = {Herbut, Igor F. and Juricic, Vladimir and Roy, Bitan},
  title         = {{Theory of interacting electrons on the honeycomb lattice}},
  eprint        = {0811.0610},
  archiveprefix = {arXiv},
  primaryclass  = {cond-mat.str-el},
  doi           = {10.1103/PhysRevB.79.085116},
  journal       = {Phys. Rev. B},
  volume        = {79},
  pages         = {085116},
  year          = {2009}
}

@article{Hawashin:2025cua,
  author        = {Hawashin, Bilal and Scherer, Michael M. and Janssen, Lukas},
  title         = {{Gross-Neveu-XY quantum criticality in moir{\'e} Dirac materials}},
  eprint        = {2503.19963},
  archiveprefix = {arXiv},
  primaryclass  = {cond-mat.str-el},
  doi           = {10.1103/PhysRevB.111.205129},
  journal       = {Phys. Rev. B},
  volume        = {111},
  number        = {20},
  pages         = {205129},
  year          = {2025}
}

@article{Wilson:1971bg,
  author  = {Wilson, Kenneth G.},
  title   = {{Renormalization group and critical phenomena. 1. Renormalization group and the Kadanoff scaling picture}},
  doi     = {10.1103/PhysRevB.4.3174},
  journal = {Phys. Rev. B},
  volume  = {4},
  pages   = {3174--3183},
  year    = {1971}
}

@article{Wilson:1971dh,
  author  = {Wilson, Kenneth G.},
  title   = {{Renormalization group and critical phenomena. 2. Phase space cell analysis of critical behavior}},
  doi     = {10.1103/PhysRevB.4.3184},
  journal = {Phys. Rev. B},
  volume  = {4},
  pages   = {3184--3205},
  year    = {1971}
}

@article{LeGuillou:1979ixc,
  author       = {Le Guillou, J. C. and Zinn-Justin, Jean},
  title        = {{Critical Exponents from Field Theory}},
  reportnumber = {SACLAY-DPh-T 79/94},
  doi          = {10.1103/PhysRevB.21.3976},
  journal      = {Phys. Rev. B},
  volume       = {21},
  pages        = {3976--3998},
  year         = {1980}
}

@article{Tarasov:1996br,
  author        = {Tarasov, O. V.},
  title         = {{Connection between Feynman integrals having different values of the space-time dimension}},
  eprint        = {hep-th/9606018},
  archiveprefix = {arXiv},
  reportnumber  = {DESY-96-068, JINR-E2-96-62},
  doi           = {10.1103/PhysRevD.54.6479},
  journal       = {Phys. Rev. D},
  volume        = {54},
  pages         = {6479--6490},
  year          = {1996}
}

@article{Gracey:2016mio,
  author        = {Gracey, J. A. and Luthe, T. and Schroder, Y.},
  title         = {{Four loop renormalization of the Gross-Neveu model}},
  eprint        = {1609.05071},
  archiveprefix = {arXiv},
  primaryclass  = {hep-th},
  reportnumber  = {LTH-1098, TTP16-034},
  doi           = {10.1103/PhysRevD.94.125028},
  journal       = {Phys. Rev. D},
  volume        = {94},
  number        = {12},
  pages         = {125028},
  year          = {2016}
}

@article{Pannell:2025ajf,
    author = "Pannell, William H. and Ronayne, William Patrick and Stergiou, Andreas",
    title = "{Gradient RG Flow in Scalar-Fermion QFTs}",
    eprint = "2511.01971",
    archivePrefix = "arXiv",
    primaryClass = "hep-th",
    month = "11",
    year = "2025"
}

@article{Henriksson:2025vyi,
    author = "Henriksson, Johan and Kousvos, Stefanos R. and Roosmale Nepveu, Jasper",
    title = "{EFT meets CFT: Multiloop renormalization of higher-dimensional operators in general $\phi^4$ theories}",
    eprint = "2511.16740",
    archivePrefix = "arXiv",
    primaryClass = "hep-th",
    month = "11",
    year = "2025"
}

@article{Amoretti:2025hpi,
    author = "Amoretti, Andrea and Anselmi, Matteo and Brattan, Daniel K.",
    title = "{A new web of dualities from Majorana Fermions}",
    eprint = "2511.22261",
    archivePrefix = "arXiv",
    primaryClass = "hep-th",
    month = "11",
    year = "2025"
}

@article{Blumlein:1998if,
    author = "Blumlein, Johannes and Kurth, Stefan",
    title = "{Harmonic sums and Mellin transforms up to two loop order}",
    eprint = "hep-ph/9810241",
    archivePrefix = "arXiv",
    reportNumber = "DESY-98-141",
    doi = "10.1103/PhysRevD.60.014018",
    journal = "Phys. Rev. D",
    volume = "60",
    pages = "014018",
    year = "1999"
}

\nolinenumbers

\end{document}